\documentclass[iop,twocolappendix]{emulateapj}
\usepackage{amsfonts}
\usepackage{natbib}
\include{epsf}
\include{epstopdf}
\DeclareGraphicsExtensions{.jpg,.pdf}
\newcommand{\beq}{\begin{equation}}
\newcommand{\eeq}{\end{equation}}
\def\la{\hbox{\raise.35ex\rlap{$<$}\lower.6ex\hbox{$\sim$}\ }}
\def\ga{\hbox{\raise.35ex\rlap{$>$}\lower.6ex\hbox{$\sim$}\ }}

\def\BV{Brunt-V\"ais\"al\"a \ }

\def\beq{\begin{equation}}
\def\eeq{\end{equation}}
\def\beqa{\begin{eqnarray}}
\def\eeqa{\end{eqnarray}}
\def\sub#1{_{_{#1}}}
\def\order#1{{\cal O}\left({#1}\right)}

\shorttitle{Critical layers and disk turbulence}
\shortauthors{Umurhan et al.}
\begin{document}

%****** End Definitions ***************************
%
%
\title{{Critical layers and protoplanetary disk turbulence}}

\author{Orkan M. Umurhan\altaffilmark{1}, Karim Shariff and Jeffrey N. Cuzzi}
\affil{NASA Ames Research Center, Moffett Field, CA 94053, U.S.A}
\altaffiltext{1}{Also at SETI Institute, 189 Bernardo Way, Mountain View, CA 94043, U.S.A.  Email: orkan.m.umurhan@nasa.gov}

\date{}
%{Received ------------- ; accepted -------------}

% 5 {} token are mandatory
%Calculations for laboratory
% experiments show that saturation of the instability is achieved through
% modification of the background shear when P$_{{\rm m}}$ is very low
 \begin{abstract}
 { A linear analysis of the zombie vortex instability
  is performed in a stratified
 shearing sheet setting for three model barotropic shear flows. 
 %the vorticity step, the shear layer and the asymmetric jet.
 %The examination assumes that both 
 % disk-normal gravity and stratification is constant.
 %{\postrevisionbf The aim is to better understand the zombie vortex instability
 %and the subsequent nonlinear self-reproduction process it generates
 %as discussed in the literature.}  
 The linear analysis
 is done by utilizing a Green's function formulation to
 resolve the critical layers of the associated normal-mode problem.  
 %We report several results: 
 The instability is the result of a resonant interaction between
 a Rossby wave and a gravity wave which we refer to as Z-modes.  The associated critical layer is
 the location where the Doppler shifted
 frequency of a distant Rossby wave equals the local 
 \BV frequency.  
 %In the shear flow model, parameter value sweeps are done to determine 
 %when unstable Z-modes exist both separately of and in conjunction with
 %the Rossby wave instability. 
 The minimum required Rossby number
 for instability,
 $\mathtt{Ro}= 0.2$, is confirmed for parameter values
 reported in the literature.  It is also found that the shear layer supports
 the instability in the limit where stratification vanishes.  The zombie vortex instability
 %as well as the Rossby wave instability
 is examined in a jet model, finding that the instability
 can occur for $\mathtt{Ro}= 0.05$.  
 %For all model flows considered, 
 Nonlinear vorticity forcing
 due to unstable Z-modes is shown to result in the creation of a jet flow at the critical
 layer
 %The jet-flow emerges 
 emerging as the result of the competition between 
 the vertical lifting of perturbation radial vorticity and the radial transport
 of perturbation vertical vorticity. 
% with the former mechanism usually dominating the latter.  
 We find that the picture of this instability leading 
 to a form of nonlinearly driven self-replicating pattern of creation
 and destruction is warranted: a parent jet
 spawns a growing child jet at associated critical layers. A mature
 child jet creates a next
 generation of child jets at associated critical layers while simultaneously
 destroying its parent jet via
 the Rossby wave instability.  
 %We conclude with some speculation.
 }
 
  \end{abstract}

 %  \maketitle
%________________________________________________________________________

\section{Introduction}
{ 
The last few years have enjoyed a revival in the study of non-magnetic routes to turbulence
in protoplanetary disks.  
Mostly with the aid of high resolution numerical investigations,
several new instability mechanisms and processes have been identified 
and shown to lead
to some kind of sustained turbulent activity, including (but not limited to),
the baroclinic instability (Klahr \& Bodenheimer, 2003, Petersen et al. 2007a,2007b,
Lesur \& Papaloizou, 2010, 
Lyra \& Klahr, 2011,
Klahr \& Hubbard, 2014, Lyra 2014
)
the Vertical Shear instability (Goldreich \& Schubert, 1967, Fricke, 1968, 
 Brandenberg and Urpin 1998, Urpin 2003, Nelson et al. 2013, Stoll \& Kley 2014, VSI hereafter),
and recently the self-replicating ``zombie vortex instability" (Marcus et al. 2013, Marcus et al. 2015, ZVI hereafter).
This study is aimed at developing deeper intuition of the linear instability associated
with the self-reproducing {  ZVI}
\footnote{{  An earlier version of this manuscript referred to the process as the ``zombie mode instability" or ``ZMI".  We have changed its nomenclature to be consistent with other studies found in recent literature on the subject.}}
examined in Marcus et al. (2013, M13 hereafter).
\par
M13 uncover the ZVI via numerical simulations in a stably stratified rotating Cartesian box model 
known as the shearing box \citep{1965MNRAS.130..125G}.
The model is composed of a
vertical component of uniform gravity and a basic, vertically uniform,
Keplerian (azimuthal) flow field varying linearly with respect to the (nominal) radial box
coordinate.
%is represented
%as varying linearly with (cylindrical) radial distance in the box frame.
%The numerical experiment reported in M13 is the response of this shearing box
%M13 report on the numerical results of this model setting experiment they conduct 
M13 consider the physical response
of an azimuthally-aligned and relatively thin ``tube"  of surplus vertical vorticity
initiated in the center of the box.
%i.e., a radially-vertically localized vertical shear profile above
%the basic background Keplerian state.  
The tubular vortex responds to perturbations by radiating internal gravity waves which
become resonant with distant buoyant critical layers.  
The resonant interaction at the critical layer is unstable and
results in the generation of 
jet flows similarly oriented with the original tubular field.  
These ``child" jets are
found in planes parallel to 
the mid-plane containing the original ``parent" tubular surplus field. 
{ 
We note here that this class of processes, i.e., of generating a mean-flow 
through the action of a buoyant critical layer,
have long been considered relevant to geophysical flow phenomena -- for example,
as a way of transferring wave momentum into mean-flow in oceans without
generating too much mixing
\citep{Booker_Bretherton_1967}}.
\par 
M13 show that the original parent vorticity field used
to showcase this process remains {  stably constituted}
during the growth and maturation of the children.  
Typically, the vorticity
of the child jet grows in magnitude until it triggers
a secondary roll-up type of instability and then gets destroyed.
Before these first generation of jets are destroyed, however, 
they have enough time to
spawn a second generation of jets through the same resonant critical layer
interaction that brought
the first generation into existence.  
The numerical experiments in M13 illustrate 
a self-reproducing process enveloping the
domain, essentially 
 ``crystallizing" it by leaving behind a lattice pattern of dynamically active
 regions of concentrated
 vertical vorticity undergoing cycles of creation and destruction. 
 The final result resembles a sustained turbulent state.
 }
  \par
 If this self-replicating process holds up to independent scrutiny, its consequences
 are profound as far as planet formation is concerned
 because such a resulting flow state has important implications with regards to dust accumulation 
 in protoplanetary disk, magnetic-Dead-Zones of protoplanetary disks (Turner, et al., 2014). 
 For example, a scenario often quoted is one in which steady anticylonic
 vortices can attract particles over time.  But if this instability is active, while creating
 vortices it also destroys them and thereby disrupting the envisioned process of steady particle
 concentration.
 The consequence of this would be especially felt during the early stages of cold disk evolution.
 Equally important is that triggering this dynamical process requires a minimum source of perturbation
 vorticity (see also Marcus et al. 2015).  
 Identification
 of this Z-mode self-replicating process in numerical experiments necessitates high resolution
 studies because the dynamical activity is concentrated in a narrow region
 centered on the critical layers wherein the region's size depends linearly on
 the growth rate.
 A similar demand on resolution is needed 
 for the VSI, but for entirely different reasons 
 \citep{
 2013MNRAS.435.2610N,
2014A&A...572A..77S,
2015MNRAS.450...21B,
2015arXiv150501892U,Richard_etal_2016}.
 \par
 \medskip
 In this study we are interested in shedding some light on the mechanism of the linear instability
 associated with this phenomenon and, furthermore, developing some insight as to how the
 the critical layer instability leads to the creation of jets.
 The linear stability analysis of this relatively simplified shearing box setting is a
 notoriously difficult problem even in the case where there is only a purely
 Keplerian flow with no other additional vorticity surplus (e.g., Dubrulle et al. 2005). 
 A normal mode reduction of this stripped-down linearized system produces a second order differential
 equation with irregular singular points and makes generating solutions, both 
 analytically and numerically, extremely delicate and subtle.  The challenges are more
 compounded if one attempts to linearly study the same system with the addition
 of the surplus vertical vorticity field considered by M13 which varies
 both radially and vertically. In this setting, the linear stability problem becomes
 inseparable in both the radial and vertical coordinates, thereby amplifying the
 challenges faced by the analyst.\footnote{This is also a feature of the VSI; see the discussion in
 \citet{2015MNRAS.450...21B,
2015arXiv150501892U}.}
 \par
 Instead, we consider the linear normal mode response in the stably stratified shearing
 box containing {\emph{a vertically uniform}} surplus vertical vorticity field -- either as a vorticity
 step located at the origin, or a vertically uniform shear layer, or a vertically uniform deviation 
 jet flow (see Figure \ref{Velocity_profiles}). 
 { By being vertically uniform, these flows are often referred to as ``barotropic".}
  These test models are chosen because of analytical
 and numerical tractability.
   While these model flow profiles are 
 different than what was considered in M13, we find that the linear
 dynamical properties of these flow profiles are qualitatively similar to the processes
 involved in the self-reproducing mechanism reported in M13.  
 We find: 
 \begin{enumerate}
 \item  The ZVI is characterized
 primarily as the near resonance between
  the 
  Doppler shifted frequency of the Rossby wave associated with the
  distant surplus vorticity field and the
  Brunt-V\"ais\"al\"a frequency at the buoyant critical layer. 
  Without the Rossby wave the buoyant critical layer is not excited.  
  \item  The nonlinear outcome of the linear instability is to
  drive into existence a jet-like vertical vorticity field at the critical layer(s). 
  \item
  The primary mechanism that spawns the jet flow is the vertical tilting of
  {{perturbation radial vorticity}} although the radial transport of
  perturbation vertical vorticity can also be important in magnitude
  depending upon the type of shear flow under consideration.
  \item Model jet flows support both the ZVI
  and the familiar
  Rossby wave instability 
  %(Lovelace et al. 1999, Li et al. 2000, 
  % Li et al. 2001, Meheut et al. 2010, Umurhan 2010, Meheut et al. 2012,
  %RWI hereafter), 
  \citep{1999ApJ...513..805L,2001ApJ...551..874L,2000ApJ...533.1023L,
  2010A&A...521A..25U,
  2010A&A...516A..31M,2012MNRAS.422.2399M,2015MNRAS.450.1503L}
  (RWI hereafter)
  the latter of which induces
  nonlinear roll-up of the anticyclonic part of the jet.  
    \item When the amplitude of the jet's vorticity exceeds a minimum value, 
    the jet gets destroyed by the RWI.  In the Keplerian flow
  frame, the anti-cyclonic side of the jet experiences destructive roll-up
  into coherent vortices.
  \item 
  Depending upon the base flow field considered,
  the analysis indicates
  that the self-replicating ZVI can be active for surplus vorticity fields whose Rossby numbers
  are as small as 0.05 -- a figure which is smaller than previously anticipated by about
  a factor of four.  
  \end{enumerate}
  \par
  This study is organized as follows: Section 2 presents the shearing box equations assumed
  for this analysis, including the scalings leading to them
  and other assumptions and simplifications.  
  The exact nonlinear form of the vertical vorticity
  evolution equation is also shown.
  In Section 3 the linearized equations of motion are derived, assuming
  a (generalized) vertically uniform surplus vertical vorticity field.  After
  stating our assumptions,  it is shown how
  the equations of motion can be written either as a set of coupled integral equations
  or as a second order differential equation in the radial coordinate of the shearing box. 
  The system is further decomposed into a potential vorticity formulation which identifies the main
  wave mode involved in the ZVI later explored.
  Section 4 presents the model surplus vorticity fields to be tested in our analysis: the 
  vorticity step, {  the shear layer} and the asymmetric jet.  
  {  These three profiles
  are considered because the vorticity step showcases the critical layer mechanism in its purest theoretical form while the shear layer is similar to the flow considered in M13 while the jet flow represents the flow to emerge from the critical layer instability itself.} 
  \par
  Section 5 details the relevant length and time scales in the problem and identifies the independent parameters of the problem depending upon which profile is considered.
  Section 6 discusses the general solution method implemented.  Because of the aforementioned
  difficulties inherent to solutions of differential operators with explicitly appearing, non-removable, irregular
   singular points, we opt for solving the coupled integro-differential equations developed
   in Section 3.  
   %We describe the numerical method employed in developing the solutions here (with technical
   %details in Appendix A).
   \par
   Section 7 concentrates on the normal mode results for the three model shear flows considered.  
   We detail the properties of the Zombie mode instability (herein termed Z-modes) in the 
   simple vortex step profile.  We also show how the asymmetric jet profile also supports
   both the Z-mode instability and the RWI (the latter detailed in Appendix B).  
   For the latter, we analytically develop
   growth rates and conditions for marginality.  
   Section 8 examines the nonlinear forcing implied by an unstable Z-mode.  We explicitly demonstrate
   how it drives into existence a jet flow at the location of the critical layer.
   Section 9 summarizes our main results and interprets the self-replicating dynamical 
   process of jet creation/destruction
   in terms of the physical results garnered by the examination peformed
   in this study.

\section{Equations}
Following M13, we consider dynamics in the so-called shearing box which
represents dynamics in the frame of a Cartesian ``box"-section
of an accretion disk which rotates around the central star at a 
distance of $R_0$ with
a rotation rate $\Omega\sub 0$ and corresponding rotation vector given by ${\bf \hat z} \Omega\sub 0$,
, see \citet{1965MNRAS.130..125G}, which is aligned with the disk vertical direction.
We take the equivalent radial direction to be along ${\bf {\hat x}}$
and  the disk's azimuthal direction vector as ${\bf {\hat y}}$.
Length scales in the box are in units of the local vertical scale height
$H_0$.  The box equations are considered valid because the ratio
$H_0/R_0$ is generally small for cold accretion disks.
We make the additional assumption
of a Boussinesq fluid and, thereby, suppressing acoustic modes.  
The equations of motion are therefore
\beqa
(\partial_t + u\partial_x + v\partial_y + w\partial_z) u  
-2\Omega\sub 0 v &=& -\partial_x \Pi + 3\Omega\sub 0^2 x,
\label{full_x_momentum}
\\
(\partial_t + u\partial_x + v\partial_y + w\partial_z) v + 
2\Omega\sub 0 u &=& -\partial_y \Pi, 
\label{full_y_momentum}
\\
(\partial_t + u\partial_x + v\partial_y + w\partial_z) w
 &=& -\partial_z \Pi + g\Theta, 
 \label{full_z_momentum}
 \\
 (\partial_t + u\partial_x + v\partial_y + w\partial_z) \Theta
 +\beta w \Theta
 &=& 0, 
 \label{full_Theta}
 \\
\partial_x u + \partial_y v + \partial_z w &=& 0.
\label{full_incompressibility}
\eeqa
The quantities $u,v,w$ denote the components of velocity in 
the radial, azimuthal and vertical directions respectively.
In the shearing box, the $u,v$ and $w$ velocity components are
scaled by the local sound speed and this is possible only because
the box equations are for dynamics in a frame moving with the exceedingly faster
Keplerian velocity.  
To facilitate comparison with the results of M13,
we make the similar assumption that is made in that study
wherein the vertical component of
the disk gravitational field $g$ is taken to be a constant.
\par
The quantity $\Theta$ can be interchangeably interpreted as either an entropy
or a buoyancy; for this analysis we consider it as the latter, in which case
$\Theta$ is a non-dimensional quantity by definition. 
Equation (\ref{full_z_momentum}) says the entropy/buoyancy field is passively
advected by the total flow with no sources or sinks meaning to say
that the perturbations are adiabatic.  Implicitly, we
are saying that the actual cooling times of the protoplanetary disk gas 
is much longer than the growth rates
calculated later on. 
\footnote{This is in contrast with the conditions giving rise to the VSI which
requires cooling times to be very short compared to both the local rotation
time of the disk and the associated growth rates.  See recent discussion in Lin \& Youdin (2015).}
Because we are in the 
Boussinesq approximation, the pressure field $p$ is rewritten in
terms of the quantity $\Pi \equiv p/\rho_0$, as the density
$\rho_0$ is taken to be a constant in the mean state.
The quantity $\beta$, which is in units of inverse length, 
represents the vertical gradient of
the disk buoyancy (itself a dimensionless quantity) and
as with $g$ above, we assume it is constant.
As taken here, $\beta > 0$ means
that it is buoyantly stable and the model is stably stratified against buoyant
instabilities.
\par
In deriving the box equations, there is always a component
representing the radial gradient of the gravitational potential of
the central star, and this appears in the form
of the expression $ 3\Omega\sub 0^2 x$ found on the right hand side
of the $x$-momentum equation.  In the absence of time dependent
dynamics, the balance of this term with the corresponding Coriolis
term $-2\Omega\sub 0 v$ admits the basic Keplerian flow state $v_k$, i.e.
\[
-2\Omega\sub 0 v_k = 3\Omega\sub 0^2 x,
\qquad \longrightarrow \ \ 
v_k = -\frac{3}{2}\Omega\sub 0 x.
\]
However, in this study, we shall consider arbitrary plane
parallel shear flows $v_0(x)$ containing both 
$v_k$ plus some deviation/surplus shear flow $V\sub 0$, the latter of
which is similarly scaled by the local sound speed.  Such
a deviation flow is a steady solution of this system provided
the pressure field shows some radial variation in its steady state as well:
\beq
 v\sub 0(x) = v_k + V\sub 0,
\eeq
where $V\sub 0 = (\partial_x \Pi\sub 0)/(2\Omega\sub 0)$ is
the departure from the basic Keplerian flow $v_k$ .  We 
detail in Section \ref{shear_profile_choices} the two types of $V\sub 0$ we analyze herein.
We often refer to these flow fields as {\emph{barotropic flow profiles}}
since we assume there is no vertical variation of $V\sub 0$.   \par
\medskip
Although this is primarily a linear stability analysis, the nonlinear vertical
vorticity equation is useful in helping us to interpret the consequences
of the results we report here in this work.  We define the vertical component
of the fluid vorticity as
\[
\zeta \equiv \partial_x v - \partial_y u.
\]
In order to obtain an evolution equation for this quantity we 
operate on equation (\ref{full_y_momentum}) by $\partial_x$
and subtract from it the result of operating on
equation (\ref{full_x_momentum}) by $\partial_y$.  Rearranging
the result and making use of (\ref{full_incompressibility}) we
find
\beqa
 & & \ \ \ \ \ \ \ \ \ \ \Big(\partial_t + u\partial_x + v\partial_y + w\partial_z\Big) \zeta = \nonumber \\
 & & (2\Omega\sub 0 + \zeta) \partial_z w 
 -\Big[(\partial_x w) \cdot (\partial_z v) - (\partial_y w) \cdot (\partial_z u) \Big].
 \label{nonlinear_vorticity_equation}
\eeqa
What this equation says is that the vertical vorticity is advected (in a Lagrangian sense) by
the flow field and, had the right hand side of the above equation been zero, 
the local value of $\zeta$ would be preserved along the way.  
However, at any given position, $\zeta$ can change
because of the two effects appearing on the right hand side of 
equation (\ref{nonlinear_vorticity_equation}).  The first of these,
arising from the expression $(2\Omega\sub 0 + \zeta) \partial_z w $,
 is the familiar
effect of vertical stretching wherein the total vorticity contained
in a moving fluid
element can spin up or down depending upon whether or not
there is concurrent vertical stretching in the flow field itself. 
This is the usual effect known from the study of Taylor columns.  
The remaining terms can be rewritten to
represent the well-known vortex tilting effects,
\beqa
& & = -\Big[(\partial_x w) \cdot (\partial_z v) - (\partial_y w) \cdot (\partial_z u) \Big],
\nonumber \\
& & = \Big(\partial_y w - \partial_z v\Big)\partial_x w 
+
\Big(\partial_z u - \partial_x w\Big)\partial_y w, \nonumber \\
& & =\zeta\sub x\partial_x w + \zeta\sub y\partial_y w,
\label{vortex_tilting}
\eeqa
where the $x$ and $y$ directed vorticities 
are defined by 
$\zeta\sub x \equiv \partial_y w - \partial_z v$ and
$\zeta\sub y \equiv \partial_z u - \partial_x w$ respectively.
For example, the expression $\zeta_x\partial_x w$ 
describes the rate in which the $x$-directed component of the vorticity
is turned into vertical vorticity due to the shear
along the $x$-direction of the vertical velocity field $w$.  A similar
interpretation holds for the term $\zeta_y\partial_y w$.
\footnote{
This is a well-known feature in incompressible rotating flows where, in general
vector form, the vorticity equation reads
\[
\frac{d\mathbb{\omega}}{dt} = (\mathbb{\omega}\cdot\nabla) {\bf u},
\]
where $\mathbb{\omega} = 2\Omega\sub 0 {\bf {\hat z}} + \mathbf \zeta$
with $\mathbb \zeta \equiv \nabla \times {\bf u}$
with $\mathbf u$ the vector velocity field.
}
We find below that when the so-called zombie modes appear, they lead to
nonlinear generation of vertical vorticity through the combined
action of the two vortex tilting terms shown in expression (\ref{vortex_tilting}).
We find that between the two, the vortex tilting generation
of vertical vorticity is usually by the $\zeta_x\partial_x w$ term.

\section{Linearization}

\par
We linearize around a plane-parallel shear state by introducing the form
\beqa
& & \ \ \ \Big(u,v,w,\Pi,\Theta\Big)^{{\mathtt T}} \mapsto 
\left(0,v\sub 0(x), 0, \Pi\sub 0(x), 0\right)^{{\mathtt T}}
+ \nonumber \\
& & \Big(u'(x,t),v'(x,t),w'(x,t),\Pi'(x,t),\Theta'(x,t)\Big)^{{\mathtt T}} 
e^{i(\alpha y +m z)} + {\rm c.c.},
\nonumber
\eeqa
where the $\mathtt T$ superscript means transpose,  
$v\sub 0(x) = -3x\Omega\sub 0 /2 +  V\sub 0(x)$, in which 
$V\sub 0(x)$ is the aforementioned arbitrary barotropic shear profile of our choosing.
The above ansatz inserted into the equations of motion yields the following
partial differential equations for the
perturbation quantities,
%\begin{subequations}
\beqa
\Big(\partial_t + i\alpha v\sub 0\Big) u' - 2\Omega\sub 0 v'
&=& -\partial_x \Pi', \label{u_prime_eqn} \\
\Big(\partial_t + i\alpha v\sub 0\Big) v' + (2\Omega\sub 0 + v\sub {0x})  u'
&=& -i\alpha \Pi', \label{v_prime_eqn} \\
\Big(\partial_t + i\alpha v\sub 0\Big) w' 
&=& -i m \Pi' + g\Theta', \label{w_prime_eqn} \\
\Big(\partial_t + i\alpha v\sub 0\Big) \Theta'
&=& -\beta w', \label{theta_prime_eqn} \\
\partial_x u' + i\alpha v' + i m w' &=& 0. \label{incompressible_prime_eqn}
\eeqa
%\end{subequations}
The subsequent analysis exploits the incompressible nature of the disturbances.
This is done by reducing the linearized equations of motion into vorticity/dilatational form, i.e.,
by defining (respectively) the vertical vorticity perturbation and horizontal velocity divergence
according to
\beq
\zeta' \equiv \partial_x v' - i\alpha u', \qquad
D' \equiv \partial_x u' + i\alpha v'.
\label{incompressible_prime_eqn}
\eeq
The horizontal velocity fields may be written in velocity-potential/streamfunction form, i.e.,
\beq
u' = -i\alpha \psi' + \partial_x \phi', \qquad
v' = \partial_x \psi' + i\alpha \phi'
\label{u_and_v_relations}
\eeq
where $\psi'$ and $\phi$ are the streamfunction and velocity-potentials, respectively.  The
above formulation automatically satisfies the incompressibility equation
(\ref{incompressible_prime_eqn}) provided
\beq
\zeta' = (\partial_x^2 - \alpha^2)\psi', \qquad
D' \equiv -im w' = (\partial_x^2 - \alpha^2)\phi',
\label{zeta_and_D_relations}
\eeq
noting here that the vertical velocity is equated with the horizontal divergence based on the above form.  The equations of motion may now be formally reduced by one order in time derivatives
to get the following
\beqa
\Big(\partial_t + i\alpha v\sub 0\Big) \zeta' &=&
-(2\Omega\sub 0 + v\sub{0x})D' - v\sub{0xx} u', \label{linearized_zeta} \\
\Big(\partial_t + i\alpha v\sub 0\Big) D' &=& - m^2\Pi' - \theta', \label{linearized_D} \\
\Big(\partial_t + i\alpha v\sub 0\Big) \theta' &=& \beta g D', \label{linearized_theta}
\eeqa
where we have defined the quantity $\theta' \equiv im\Theta'$. 
Note that throughout this study we assume stable stratification which means
that the Brunt-V\"ais\"al\"a frequency $N_B$ is real, i.e. 
$N_B^2 \equiv g\beta > 0$.  As far as nomenclature is concerned,
we mostly dispense with using the symbol $N_B$ for the 
Brunt-V\"ais\"al\"a frequency
 and, instead, retain for its designation the expression $g\beta$.
% We opt for using $N_B$ in our discussions found in Section 7.
 \par
The above third order system (in time) is supplemented by
 the diagnostic condition
relating the perturbation pressure to the other quantities appearing, and is given as 
the solution of
\beq
\Big(\partial_x^2 - \alpha^2 - m^2\Big)\Pi' = 
2\Omega\sub 0 \zeta' + \theta' -2i\alpha v\sub{0x} u'.
\label{Pi_relationship}
\eeq
We observe that the vertical vorticity field is driven by vertical stretching ($D'$)
and radial advection of the mean vorticity gradient ($v\sub {0xx} u'$).
\par
In the form as developed here, the streamfunction and velocity potential solutions
are written in terms of Green's functions, i.e.
\beqa
& & \psi' =   \int G\sub\psi(x,x') \zeta'(x') dx',\nonumber \\
& & \phi' =   \int G\sub\phi(x,x') D'(x') dx',
\label{general_greens_solutions_phi_and_psi}
\eeqa
where $G(x,x')$ is the appropriate Green's function associated with the two-dimensional Laplace operator $\partial_x^2 - \alpha^2$,
{\emph{subject to appropriate boundary conditions (see more below)}},
\beq
\Big(\partial_x^2 - \alpha^2\Big)G\sub{\phi,\psi} = \delta(x-x'),
\eeq
where $\delta(x)$ is the Dirac delta function.
Similarly, the corresponding solution of the pressure fluctuations is
given by
\beq
\Pi' = \int G\sub\Pi(x,x')\Big\{
2\Omega\sub 0 \zeta'(x') + \theta'(x') -2i\alpha v\sub{0x}(x') u'(x') \Big\} dx'.
\label{Pi_solution}
\eeq
The Green's function $G\sub\Pi$ is the solution of
\beq
\Big(\partial_x^2 - k^2\Big)G_\Pi = \delta(x-x'), \qquad
k^2 \equiv \alpha^2 + m^2.
\label{Pi_Green's_Operator}
\eeq
The system of equations (\ref{linearized_zeta}--\ref{Pi_relationship})
together with their associated diagnostic relationships
(\ref{u_and_v_relations}) and (\ref{zeta_and_D_relations}) is an integro-differential
system which must be solved
subject to boundary conditions in the radial direction.
The Green's function strategy adopted here has been used in other disk
studies (e.g., Dubrulle \& Knobloch, 1992). 
 For this study
we report upon solutions in which all quantities exponentially decay to zero as
$x\rightarrow \pm \infty$.  
We note that for the unstable
localized modes which are the subject of this study, we have checked and
verified that the results reported
here are insensitive to whether or not the perturbations are periodic (on scale $2L$) 
or if there are no-normal flow boundary conditions imposed at $x=\pm L$.  The above
statement becomes robust so long as the horizontal domain is large enough with, typically speaking, $L \ge \pi$.  The location of truncation scale $L$ does not interfere with
the existence and/or expression of an unstable mode.

\subsection{An alternative form in terms of a perturbation potential vorticity}\label{alternative_form_of_equations}
It is worth noting that the processed linearized equations 
(\ref{linearized_zeta}--\ref{linearized_theta}) may be recast instead
in terms of a perturbation vertical {\emph {potential vorticity}}, $\Xi'$, defined by
\beq
\Xi' \equiv \zeta' + \left(\frac{2\Omega\sub 0 + v\sub{0x}}{\beta g}\right)\theta'.
\label{definition_of_PV}
\eeq
In many shear flow applications, following the behavior of the potential vorticity
is an extremely useful diagnostic \citep{1985QJRMS.111..877H}.
The perturbation equations may now be rewritten in the following alternative formulation:
\beqa
\Big(\partial_t + i\alpha v\sub 0\Big) \Xi' &=& - v\sub{0xx} u' ,
\label{linearized_alternate_Xi} \\
\Big(\partial_t + i\alpha v\sub 0\Big) \zeta' &=& -(2\Omega\sub 0 + v\sub{0x})D'
- v\sub{0xx} u' , \label{linearized_alternate_zeta} \\
\Big(\partial_t + i\alpha v\sub 0\Big) D' &=& - m^2\Pi' 
+ \frac{g\beta}{2\Omega\sub 0 + v\sub{0x}}\Big(\zeta'-\Xi'\Big),
 \label{linearized_alternate_D}
\eeqa
with
\beqa
& & \Big(\partial_x^2 - k^2\Big)\Pi' = \left[
\frac{2\Omega\sub 0(2\Omega\sub 0 + v\sub {0x}) - g\beta}{2\Omega\sub 0 + v\sub {0x}}
\right]\zeta' \nonumber \\
& & \ \ \ \ \ \ \ \ \ \ \ \ \ \ \ \ \ \ \ \ \ \ \ \ \ \ \ - 2i\alpha v\sub{0x} u' + 
\frac{g\beta}{2\Omega\sub 0 + v\sub {0x}}\Xi'.
\eeqa 
We note immediately that in the event that $v\sub{0x} $ is constant (e.g. pure
Keplerian flow with $V\sub 0 = 0$), 
the system materially conserves the perturbation potential
vorticity with the basic constant shear state.  In slightly more general terms, if
$V\sub 0(x) =  \overline \Omega_x x,$ where $\overline{\Omega_x} =$ constant (units of inverse time), then
\beq
\Big(\partial_t + i\alpha q x\Big) \Xi' = 0, \qquad
q = -\frac{3}{2}\Omega\sub 0 + \overline{\Omega}\sub x,
\eeq
which means that a potential vorticity perturbation is
advected by the composite background shear flow field $=q x$.
In this way the system becomes relatively transparent.  In the event
we consider a flow field
such that  $v\sub{0xx} = 0$, together with $\Xi' = 0$ initially, 
then the linear response is of pure inertial gravity waves.

\subsection{As a single second order differential equation
and critical layers}
Previous treatments
(e.g., Dubrulle et al. 2005) of this system 
as a normal mode problem have
instead turned equations (\ref{u_prime_eqn} -- \ref{incompressible_prime_eqn})
into a single second order differential equation.  
Indeed, combining these equations into a single one for 
the normal-mode pressure perturbation results in
\beqa
\left(\partial_x ^2 - k^2\right)\hat \Pi = 2\Omega\sub 0 \hat\zeta
+\hat \theta - 2i\alpha v_{0x}\hat u,
\eeqa
in which the following normal mode ansatz has been assumed
\beq
\Pi' = \hat\Pi(x) e^{-i\omega t},
\eeq
and similarly for the other quantities appearing.
Here, $\omega$ is the unknown complex normal mode frequency. 
Insertion of this form into the fundamental perturbation
equations
(\ref{u_prime_eqn}--\ref{incompressible_prime_eqn}) 
followed by some manipulation shows that the following
relationships hold between various quantities:
\beqa
& &  \hat \theta = \frac{g\beta m^2}{\sigma^2 - g\beta} \hat\Pi,
\qquad
\hat D = \frac{i\sigma}{g\beta} \hat\theta = 
\frac{i\sigma m^2}{\sigma^2 - g\beta}  \hat\Pi,
\nonumber \\
& & \ \ \ \ \ \ \ \ \ \ \ \  \hat u
= i\frac{2\Omega_0  \alpha \hat \Pi + \sigma \partial_x \hat \Pi}
{\omega_\varepsilon^2 - \sigma^2},
\eeqa
as well as
\beq
2\Omega\sub 0 \hat\zeta = -\frac{\omega_\varepsilon^2}{g\beta}\hat\theta
-{2\Omega\sub 0 v\sub{0xx}} i \hat u/\sigma,
\eeq
where we have
defined $\sigma(x) \equiv  \alpha v\sub 0(x) - \omega$
and $\omega_\varepsilon^2(x) \equiv 2\Omega\sub 0\big(2\Omega\sub 0 + v\sub{0x}(x)\big)$
- the latter of these is the local disk
epicyclic frequency.  Rewriting equation (\ref{Pi_relationship})
in terms of the normal mode ansatz means that
\beq
\Big(\partial_x^2 - \alpha^2 - m^2\Big)\hat\Pi = 
2\Omega\sub 0 \hat \zeta + \hat \theta -2i\alpha v\sub{0x} \hat u,
\eeq
and making use of the above relationships and some additional reduction
shows that
\beqa
& & \Big(\partial_x^2 - k^2\Big)\hat\Pi
= \frac{\omega_\varepsilon^2 - g\beta}{g\beta-\sigma^2}m^2 \hat\Pi
\nonumber \\
& & \ \ +
\frac{1}
{\sigma^2 - \omega_\varepsilon^2}
\left(2\alpha v\sub{0x}-\frac{2\Omega_0 v\sub{0xx}}{\sigma}\right)
\Big(2\Omega_0 \alpha \hat \Pi + \sigma \partial_x \hat \Pi\Big).
\label{Single_2nd_Order_ODE_in_Pi}
\eeqa
This system is subject to the same boundary conditions as outlined above.
\footnote{ Had we considered no-normal flow boundary conditions
at $x=\pm L$,
then the imposition of an impenetrable flow boundary condition
at these two locations would be the same as
imposing
\[
2\Omega_0  \alpha \hat \Pi + \sigma \partial_x \hat \Pi = 0,
\]
at $x=\pm L$ provided $\sigma^2 - \omega_\varepsilon^2 \neq 0$ at the boundaries.
}
\par
Inspection of the ODE (\ref{Single_2nd_Order_ODE_in_Pi}) 
shows that there exists the possibility of the system supporting several critical layers 
\citep{2002ihs..book.....D,2004hyst.book.....D}.
Locations where the denominators vanish are candidate irregular singular points.
Of interest to us are the points associated with expressions multiplying
the pressure $\hat\Pi$ (and, not any of its derivative expressions).
These points become irregular if the denominators or the associated
expressions can pass through zero linearly with respect to variations in $x$.
Several candidate points are identified:
The first of these is the classical one associated plane parallel shear flows
and occurs at points $x\sub{pp}$ in which Re$(\sigma) = 0$, i.e.
\[
 \alpha v_0(x\sub{pp}) - {\rm Re}(\omega)= 0.
\]
In classical plane parallel shear flow problems, such critical layers
activate only if both $v_{0xx} \neq 0$ and if viscosity is included 
(Drazin \& Reid, 1981). 
The normal modes associated with these critical layers are
called Tollmien-Schlichting waves (TS waves) and they are unstable for wide
values of Reynolds numbers.  TS-waves cease to be
normal modes in the exactly inviscid problem.  
In monotonic shear flows (flows in which the shear velocity strictly increases or decreases) there tends to be
only one such critical point $x\sub{pp}$.
\par
With the inclusion of buoyancy under stable stratification, in which $g\beta > 0$,
two more critical layers emerge at points $x\sub{bg}^{\pm}$
which we henceforth refer to as {\emph{buoyant critical layers}} to distinguish them
from the others.
These locations are associated
with the first term on the right hand side of equation (\ref{Single_2nd_Order_ODE_in_Pi}), 
where Re$(\sigma) \pm \sqrt{g\beta} = 0$, i.e.
\beq
 \alpha v_0\big(x\sub{bg}^{\pm}\big) \pm \sqrt{g\beta} - {\rm Re}(\omega)= 0.
 \label{buoyant_critical_layer_guesstimate}
\eeq
The instability identified in Section \ref{results} pertains
to the action of these buoyant critical layers.  An inspection 
of equation (\ref{Single_2nd_Order_ODE_in_Pi}) shows
that these critical layers are intrinsically a three-dimensional
phenomenon as the expression $m^2 \hat\Pi/(g\beta-\sigma^2)$ 
requires vertical perturbations ($m\neq 0$),
azimuthal perturbations ($\alpha \neq 0$) and a radial
variation in the shear flow $v\sub 0(x)$.
\par
Finally, equation (\ref{Single_2nd_Order_ODE_in_Pi}) also admits the possibility
of two more critical layers associated with those points where
the denominator of the second term on the right hand side of 
equation (\ref{Single_2nd_Order_ODE_in_Pi}) is equal to zero, that is,
Re$(\sigma) \pm \omega_\varepsilon(x) = 0$.  These points are often referred
to in the astrophysical literature as the {\emph{corotation points}} or 
{\emph{Lindblad resonances}}
 of a disk (e.g., Papaloizou \& Pringle, 1984). 
While at first glance
it may seem that the corotation points ought to be important to the dynamics investigated here,
we find in our results that they in fact play little role.  The main reason
appears to be because the expressions found in the numerator
associated with this term, i.e. $2\Omega_0 \alpha \hat \Pi + \sigma \partial_x \hat \Pi$, always gets nearly as small as the value
of the denominator in this region. 
This would cease to be the case if, for example, self-gravitational physics were included in the analysis.
 While further elucidation to clarify the inactivity of
 the Lindblad resonances are surely in order, because these layers play no role in our results
 these matters are not considered forthwith.  
 \begin{figure*}
\begin{center}
\leavevmode
\includegraphics[width=\textwidth]{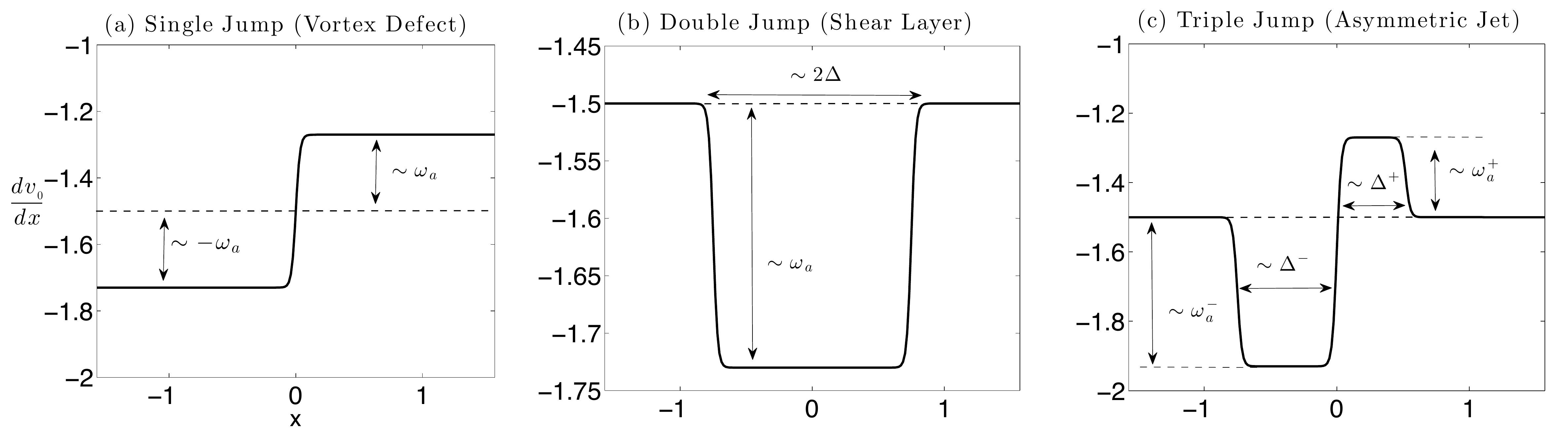}
\end{center}
\caption{The total shear profile ($dv\sub 0/dx = -3\Omega\sub 0/2 + dV\sub 0/dx$) of the two
barotropic velocity fields detailed
in Section \ref{shear_profile_choices}:  (a) The ``vortex step", 
(b) the ``shear layer",
and (c)  the ``asymmetric jet".
The shear is scaled in units of $\Omega\sub 0$ while the radial length $x$ is
scaled to the box scale $L$. 
Note that all shear fields shown include the background Keplerian state ($=-3\Omega\sub 0/2$)
which is indicated for reference by the dashed horizontal line
in panel (a).  
%In panel (a) we have chosen $\omega\sub a/\Omega\sub 0 = 0.23$
%and $\epsilon = 0.05$, while for the
%jet profile we have adopted separation parameters  $\Delta^+ = 0.50$ and
%$\Delta^- = 0.75$ with $\omega_a^+ = 0.23$ and $\omega_a^- = -0.43$.  
%Note that larger values of $\epsilon$ correspond to more rounded edges in the
%transition zones where the shear changes the most.
%Recall that $\omega\sub a/2\Omega\sub 0$ roughly measures $\mathtt {Ro}$ of the Keplerian deviation 
%flow field.  Note that in the asymmetric jet flow field,  $V\sub 0(x)$ diminishes 
%to zero sufficiently far from $x=0$.  
The deviation from Keplerian shear profile shown
in panel (b) is anti-cyclonic while
in panel (c) it is anti-cyclonic in the region $-\Delta^-<x< 0$ provided
$\omega\sub a^- < 0$
and it is cyclonic for $0<x<\Delta^+$ provided $\omega\sub a^+ > 0$.  }
\label{Velocity_profiles}
\end{figure*}

\section{Choice of barotropic velocity profiles, some nomenclature}\label{shear_profile_choices}
We consider three barotropic shear profiles $V\sub 0(x)$ that are departures from pure Keplerian
flow (Figure \ref{Velocity_profiles}):\par
\bigskip
\noindent
\emph{The vorticity step.}
This single vorticity jump profile (Figure \ref{Velocity_profiles}a) , and its derivatives, are given by
\beqa
& & V\sub 0 = \omega\sub a \epsilon \ln\left[\cosh\left(\frac{x}{\epsilon}\right)\right].
%\nonumber \\
%& & V\sub {0x} = \omega\sub a\tanh\left(\frac{x}{\epsilon}\right),
%\nonumber \\
%& & V\sub {0xx} = \frac{\omega\sub a}{\epsilon}{\rm sech}^2\left(\frac{x}{\epsilon}\right).
\label{step_profile}
\eeqa
The model profile is governed by two parameters, $\omega\sub a$ (units of inverse time) and 
$\epsilon$ (units of length). The latter quantity controls how sharply the shear transition occurs 
(around $x$) while the former dictates the shear profile for values of $|x|\gg \epsilon$, i.e. $V\sub 0 \approx \omega\sub a |x|$ and $V\sub {0x} \approx \omega\sub a {\rm sgn}(x)$ as $x\rightarrow \pm\infty$.  
The useful feature of this type of model (and for the others described hereafter) is that as $\epsilon \rightarrow 0$ the profile 
resembles a piecewise linear velocity field.  In this sense, as $\epsilon \rightarrow 0$
 $V\sub {0xx} \approx 2\omega\sub a \delta(x)$.   \par
\par
\bigskip
\noindent
{\emph{The shear layer.}}
The double jump in vorticity is a facsimile of the classic Rayleigh shear profile (Rayleigh, 1880, Drazin \& Reid, 1981) in which
constant opposite velocity layers sandwich a uniform shear layer of thickness $2\Delta$ centered
at $x=0$.  Thus we have
\beqa
V\sub 0 &=& \frac{\omega\sub a \epsilon}{2}  \left\{\ln\left[\cosh\left(\frac{x-\Delta}{\epsilon}\right)\right]
-\ln\left[\cosh\left(\frac{x+\Delta}{\epsilon}\right)\right] \right\}. 
%\nonumber \\
%V\sub {0x} &=& \frac{\omega\sub a}{2} \left[
%\tanh\left(\frac{x-\Delta}{\epsilon}\right) - \tanh\left(\frac{x+\Delta}{\epsilon}\right)
%\right],
%\nonumber \\
%V\sub {0xx} &=& \frac{\omega\sub a}{2\epsilon}\left[
%{\rm sech}^2\left(\frac{x-\Delta}{\epsilon}\right)
%-{\rm sech}^2\left(\frac{x+\Delta}{\epsilon}\right)
%\right].
\eeqa
\par
\medskip
\par
\bigskip
\noindent
{\emph{The asymmetric jet.}}
In the absence of a Keplerian shear the triple jump profile described below 
will physically resemble that of a jet. 
However, taken in aggregate with the background Keplerian
shear, the composite flow describes a shear with a weak
jet-like undulation atop of it.  Nevertheless, we refer to this as ``jet" in so far
as the deviation flow  $V\sub 0$ resembles one.
The flow $V\sub 0$ involves three
steps in the mean vorticity profile located
at positions $-\Delta^-, 0$ and $\Delta^+$.  The vorticity of $V_0$
in both the regions $x>\Delta^+$ and $x<-\Delta^-$ are zero
while the vorticity is given (approximately) to be $\omega_a^-$ for the
region $-\Delta^-< x < 0$ and 
$\omega_a^+$ for the
region $0<-\Delta^+< x$.
Thus we have

\beqa
%V\sub 0 &=& \frac{\omega\sub a}{2} \epsilon \Bigg\{\ln\left[\cosh\left(\frac{x-\Delta}{\epsilon}\right)\right]
%-2\ln\left[\cosh\left(\frac{x}{\epsilon}\right)\right] \nonumber \\
%& & +\ln\left[\cosh\left(\frac{x+\Delta}{\epsilon}\right)\right] \Bigg\},\nonumber \\
V\sub 0 &=& \frac{\omega\sub a^-}{2} \epsilon \Bigg\{\ln\left[\cosh\left(\frac{x+\Delta^-}{\epsilon}\right)\right]
-\ln\left[\cosh\left(\frac{x}{\epsilon}\right)\right]
\Bigg\}
 \nonumber \\
& & 
-\frac{\omega\sub a^+}{2} \epsilon \Bigg\{\ln\left[\cosh\left(\frac{x-\Delta^+}{\epsilon}\right)\right]
-\ln\left[\cosh\left(\frac{x}{\epsilon}\right)\right]
\Bigg\}. 
%\nonumber \\
%%V\sub {0x} &=& \frac{\omega\sub a}{2} \left[
%%\tanh\left(\frac{x-\Delta}{\epsilon}\right) - 2\tanh\left(\frac{x}{\epsilon}\right)
%% + \tanh\left(\frac{x+\Delta}{\epsilon}\right)
%%\right],
%%\nonumber \\
%V\sub {0x} &=& \frac{\omega\sub a^-}{2} \left[
%\tanh\left(\frac{x+\Delta_-}{\epsilon}\right) - \tanh\left(\frac{x}{\epsilon}\right)
%\right]\nonumber \\
%& &  -\frac{\omega\sub a^+}{2} \left[
%\tanh\left(\frac{x-\Delta^+}{\epsilon}\right) - \tanh\left(\frac{x}{\epsilon}\right)
%\right].\nonumber \\
%V\sub {0xx} &=& \frac{\omega\sub a^-}{2\epsilon}\left[
%{\rm sech}^2\left(\frac{x+\Delta^-}{\epsilon}\right)
%-{\rm sech}^2\left(\frac{x}{\epsilon}\right)\right] \nonumber \\
%& & -\frac{\omega\sub a^+}{2\epsilon}\left[
%{\rm sech}^2\left(\frac{x-\Delta^+}{\epsilon}\right)
%-{\rm sech}^2\left(\frac{x}{\epsilon}\right)\right].
\label{jet_profile}
\eeqa
Most of 
the variation of this mean velocity field is confined to within $-\Delta^-<x<\Delta^+$ centered
$x=0$ and once one has moved sufficiently far from this region the flow returns to being
largely Keplerian.
For $\omega\sub a^+ = \omega\sub a^-$ and
$\Delta^+ = \Delta^-$, in the limit $\epsilon\rightarrow 0$ the
resulting symmetric profile recovers the so-called triangular jet (Drazin, 2002).
\par
\bigskip
We reference the nature of the velocity profile in terms of deviations from the Keplerian state according to the following convention:  if ${\displaystyle {dV\sub 0}/{dx}}  < 0$ then we say
that the profile (or the part under consideration) is {\emph{anticyclonic}} with respect to the Keplerian shear, while
if ${\displaystyle {dV\sub 0}/{dx}}  > 0$ then we say that profile (or part under consideration)
is similarly {\emph{cyclonic}}.
 Often times, we will consider the Rossby number, defined by
 \[ \mathtt{Ro} \equiv {\displaystyle 
 \frac{1}{2\Omega_0}\frac{dV\sub 0}{dx}},
 \]
 which is the same definition as used by M13 and Marcus et al. (2015).
 We use $\mathtt{Ro}$ to quantify the change of the vorticity in part or in the whole
 of a profile being tested.
 For example, the jump in the vorticity across the single step vorticity defect 
 is $\omega\sub a$ while the effective
 Rossby number characterizing deviations of the shear flow about
 the Keplerian state is $\mathtt{Ro} = \order{|\omega\sub a|/2\Omega\sub 0}$.  
 Generally speaking, jumps in the vorticity relate to the $\mathtt{Ro}$
 in the way indicated
 and we think of them in terms of this equivalence hereafter.
 \par\medskip
 { 
 We perform a compreshensive analysis of these three profiles for the following reasons:
 The vorticity step showcases the essential mechanism of the critical layer mechanism.  We consider the shear layer because it is the barotropic analog of the flow profile considered in M13.  The analysis of the asymmetric jet is included because jets are flows that emerge from the critical layer instability.}

\section{Relevant length and time scales, and dependence of results on problem parameters}
The linearized equations and the flow fields we have adopted to study are
presented in dimensional units mainly in order to facilitate comparison between
these theoretical results and numerical results reported in the literature (namely, M13). The shearing box equations are expressed
in terms of length scales proportional to the local scale height of the disk
(e.g., Umurhan \& Regev, 2004).  
In the absence of any other superimposed flow fields there are no other natural length scales in the problem. Given the aforementioned flow fields
described in the previous section, there are now two length scales are introduced
to the problem, namely, $\Delta$ (or $\Delta^{\pm}$) and $\epsilon$ respectively describing the width(s) of the shear layer (jet flow) and the length scale of their corresponding transition zones.
\par
Similarly, the shearing box equations are generally expressed in time scales proportional to the inverse rotation rate of the disk, $\Omega\sub 0^{-1}$.
The plain Coriolis effects (in the absence of any $V_0$) are expressed in
equation (\ref{Single_2nd_Order_ODE_in_Pi}) as the
 square of the epicyclic frequency $\omega\sub\varepsilon^2$, which is in units of $\Omega\sub 0^2$.  In terms of the shearing box equations utilized in this work, the \BV frequency ($\sqrt{g\beta}$) is an independent time scale appearing in the system.  The addition of the flow fields $V\sub 0$, discussed in Section
 \ref{shear_profile_choices}, introduces additional timescales:
$1/\omega\sub a$ associated with the vorticity step flow and shear layer models or, the two timescales $1/\omega\sub a^{\pm}$ for
the jet flow model.
\par
Had we chosen to do so, the governing equations
(like, for example, equation \ref{Single_2nd_Order_ODE_in_Pi})
 would have been non-dimensionalized according to $\epsilon$ (spatial scales)
and $\Omega\sub 0^{-1}$ (temporal scales).  The
resulting equations would then transparently
exhibit their dependence upon the non-dimensional parameters: $\alpha\epsilon$, $m\epsilon$, $g\beta/\Omega\sub 0^2$ and $\omega\sub a/\Omega\sub 0$.  For the
vorticity-step model profile calculation, these four parameters
completely characterize the system's solutions.  The results of shear layer problem depend upon 
five parameters, i.e., the four
parameters of the vorticity-step profile problem with the addition of $\Delta/\epsilon$, describing the ratio width of the shear layer to the size of its transition zone.  The results
for the jet
profile are described by seven parameters: 
$\alpha\epsilon$, $m\epsilon$, $g\beta/\Omega\sub 0^2$, $\Delta^\pm /\epsilon$ and $\omega\sub a^\pm/\Omega\sub 0$. 
\par We have checked and verified that the numerical results we develop (and describe
hereafter), reproduces invariant solutions for invariant values of the aforementioned parameters.  
For example, for the vorticity-step problem we verify
that the eigenvalues we find depend strictly on $m\epsilon$, $\alpha\epsilon$ 
as well as $g\beta/\Omega\sub 0^2$ and $\omega\sub a/\Omega\sub 0$ only.  A variation in
$\epsilon$ accompanied by adjustments in $\alpha$ and $m$ that leave $m\epsilon$ and $\alpha\epsilon$ invariant
leave the results invariant up to numerical accuracy of the computational algorithm.
\par
\medskip
In more realistic disk models, 
the \BV frequency is a function of vertical position.  Indeed, the vertical component of gravity
has the approximate local dependence $g\sim \Omega\sub 0^2 z$. Buoyancy, which is a non-dimensional quantity, 
is best gauged by the vertical entropy profile which, in turn, is strongly dependent on the 
global disk model under consideration. Provided the
entropy structure is stably stratified and symmetric with respect to the disk midplane, we
suppose that the vertical gradient of buoyancy $\beta \sim z/(HH_\beta)$, in which $H_\beta$ is the vertical variation scale
of the (non-dimensionalized) entropy while $H$ is the usual pressure scale-height.
Supposing the model dependent $H_\beta \sim \order H$, then the square of the \BV frequency
takes on the range of values given by the relationship: $g\beta \approx \Omega\sub 0^2(z/H)^2$ .  
Considering that the shearing box equations are formally valid to 
a few scale heights above the disk midplane, and  since our model formulation assumes
constant values of $g\beta$, we make sure to consider values of $0<g\beta \le 6 \Omega\sub 0^2$, 
in which
the upper bound corresponding to about 2.5 disk scale heights
while the lower bound corresponds to locations in the vicinity of the disk midplane.

\pagebreak

\section{Solution method}\label{results}
We solve the coupled integral equations (\ref{linearized_zeta}--\ref{linearized_theta}) assuming normal mode
perturbations of the form
\beq
\left(
\begin{array}{c}
\zeta'(x,t) \\
D'(x,t) \\
\theta'(x,t)
\end{array}
\right)
=
\left(
\begin{array}{c}
\hat\zeta(x)\\
\hat D(x) \\
\hat \theta(x)
\end{array}
\right)e^{-i{\omega t}} + {\rm c.c.}
\label{nm_form}
\eeq
together with the corresponding Green's function solutions for the diagnostic
relationships $\phi,\psi'$ and $\Pi'$ found in
equations (\ref{general_greens_solutions_phi_and_psi}) and (\ref{Pi_Green's_Operator}).
As noted earlier, we seek solutions that show exponential decay as $|x| \gg 1$
which necessarily precludes simple wave normal mode solutions that have no attenuation
in the far limit.\par
The system is a series of three coupled integral equations.  The variables,
$\zeta, D$ and $\theta$, are discretized on either a uniform or Gaussian grid $x_j$ 
of $N$ points, 
but for the purposes of this report we quote the results developed using
a uniform grid only.
\footnote {All solutions obtained and reported herein are equivalently obtained using either discretization.}
Thus each variable is numerically represented as a column vector corresponding to its values on the grid, 
i.e. $\hat\zeta \mapsto \zeta_j, \hat D \mapsto D_j$ and $\hat\theta\mapsto\theta_j$.
The Green's functions are turned into matrix operators so that, for instance, the
stream function ($\hat\psi \mapsto \psi_j$) is written as a matrix operation
relating it to the vertical vorticity, that is to say
\[
\psi_j = {\mathbf G}^{(\psi)}\sub{jn} \otimes \zeta\sub{n}, \qquad
{\bf G}^{(\psi)}\sub{jn} \equiv  -\frac{dx}{2\alpha} e^{-\alpha|x_j-x_n|},
\]
where $dx$ is the grid spacing and the symbol $\otimes$ is the matrix multiplication operation.
The corresponding
derivative of $d\hat\psi/dx \mapsto \big(d\psi\big)\sub j$ is written as
\[
\Big(d\psi\Big)\sub j = {\bf dG}^{(\psi)}\sub{jn} \otimes \zeta\sub{n}, 
\]
in which
\[
\displaystyle
{\bf dG}^{(\psi)}\sub{jn} \equiv  \frac{dx}{2} {\rm sgn}(x_j-x_n) e^{-\alpha|x_j-x_n|}.
\]
A similar set of matrix operations are defined and implemented for the potential function
$\hat\phi \mapsto \phi_j$ and the pressure field $\hat\Pi \mapsto \Pi_j$.
{{In this construction, the exponential decay of solutions as $|x| \rightarrow \infty$
is ensured.}}
The complete set of equations (\ref{linearized_zeta}--\ref{linearized_theta}) is converted into
a single matrix form
\beq
\frac{\partial {\mathbf V}}{\partial t} = \mathbf M \otimes {\mathbf V}
\eeq
with
\beq
{\mathbf V} = \Big(\zeta_1, \cdots, \zeta_j, \cdots,\zeta_N,D_1, \cdots, D_N,\theta_1,\cdots,\theta_N\Big)^{\mathtt T}.
\eeq
Assumption of the normal mode form in equation (\ref{nm_form}) turns the above
into a single matrix problem to determine the unknown
eigenvalues $-i\omega$. 
$\mathbf M$ is constructed following the method described in
\citet{2010A&A...521A..25U}.
\par
We
then go through two stages to obtain a solution. {\emph{Stage 1}} uses standard matrix inversion methods
to establish both eigenvalues and the corresponding eigenfunctions. Because
the method is computationally expensive, we often use this method to determine the approximate
solution on a coarse grid and then we refine this same solution through {\emph{Stage 2}}: 
which interpolates the solution onto a finer grid (either 2 or 4 times) and then
solves the discretized system matrix operator system using
a standard Newton-Raphson-Kantorovich solution technique. \par
The benefit of this approach is that Stage 1 produces all of the normal modes permitted
by the system and Stage 2 helps to identify which of the numerically determined solutions
are spurious and which are robust.  Spurious solutions are identified as those candidate normal mode solutions
whose eigenvalues show no convergence with increased resolution. Z-modes are found to be
particularly tricky to obtain reliably, generally requiring anywhere from 700-1500 grid points
of radial resolution on domains ranging from  $\pi$ to $2\pi$ (also see section \ref{viscosity_solutions}).

\begin{figure*}
\begin{center}
\leavevmode
\includegraphics[width=16cm]{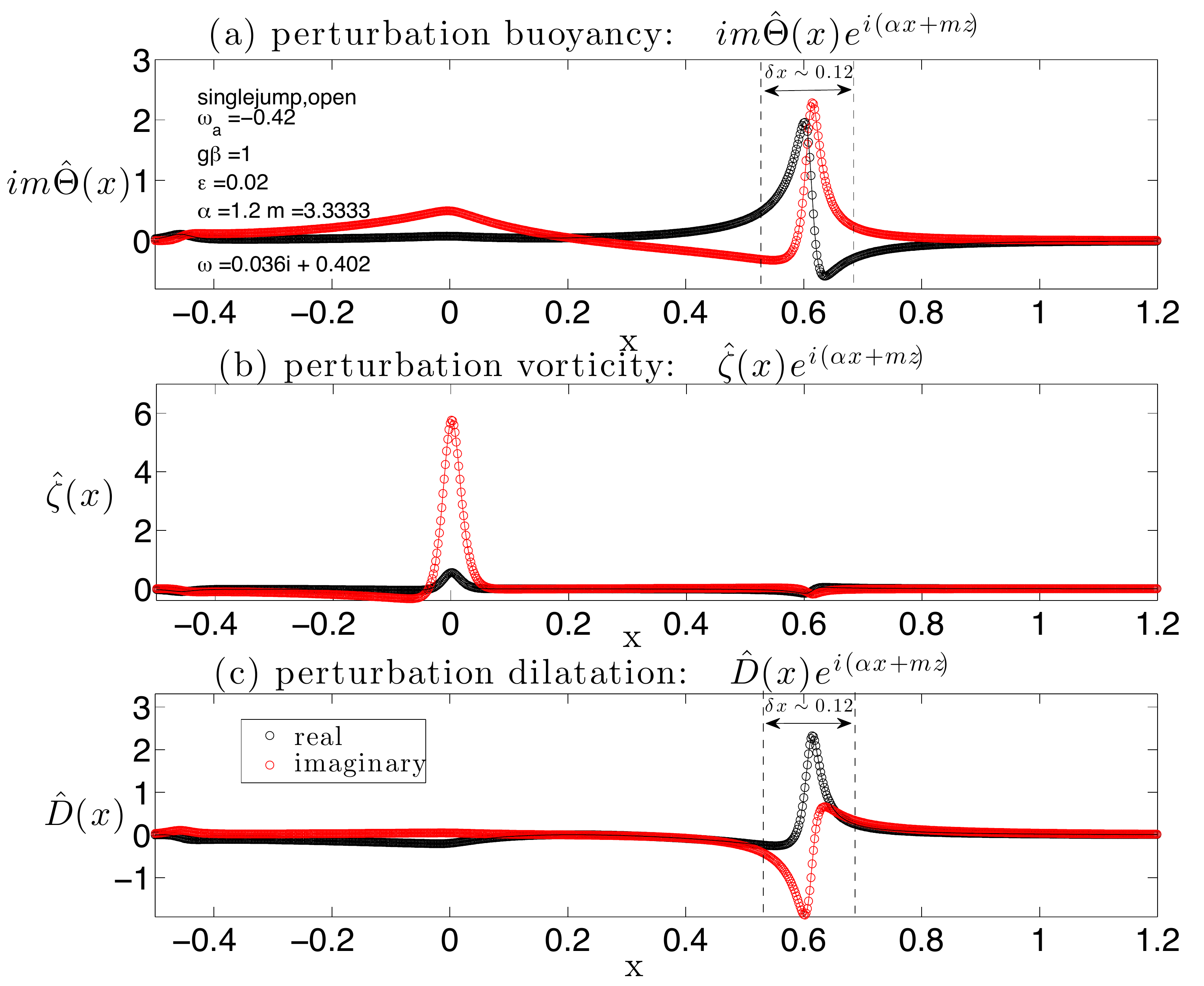}
\end{center}
\caption{Array of Z-mode eigenmodes for the vorticity step problem: $g\beta = 1.0\Omega\sub 0^2$, 
$\omega\sub a = -0.42\Omega\sub 0$, $\alpha = 1.2$
and $m = 3.33$ with $\epsilon = 1/30$. 
This sample flow field has an equivalent deviation flow field with Ro$=0.21$.
The panels show the perturbation fields for (a) the buoyancy
(b) the vorticity, (c) the dilatation.  
The basic state vorticity gradient appears imprinted in the perturbation
vorticity field around $x=0$. 
The induced critical layers appearing at $x\sub{bg}^\pm \approx -0.45, 0.62$ 
are most prominent in the buoyancy and dilatation fields.  Numerical details:
linear grid used is $N= 3064$ with solutions shown by open circles with a fitted curve connecting them.
Solutions calculated on the domain
$-\pi<x<\pi$, however only the active parts are shown in the figure.}
\label{General_profiles_1j}
\end{figure*}

 \begin{figure}
\begin{center}
\leavevmode
\includegraphics[width=8.75cm]{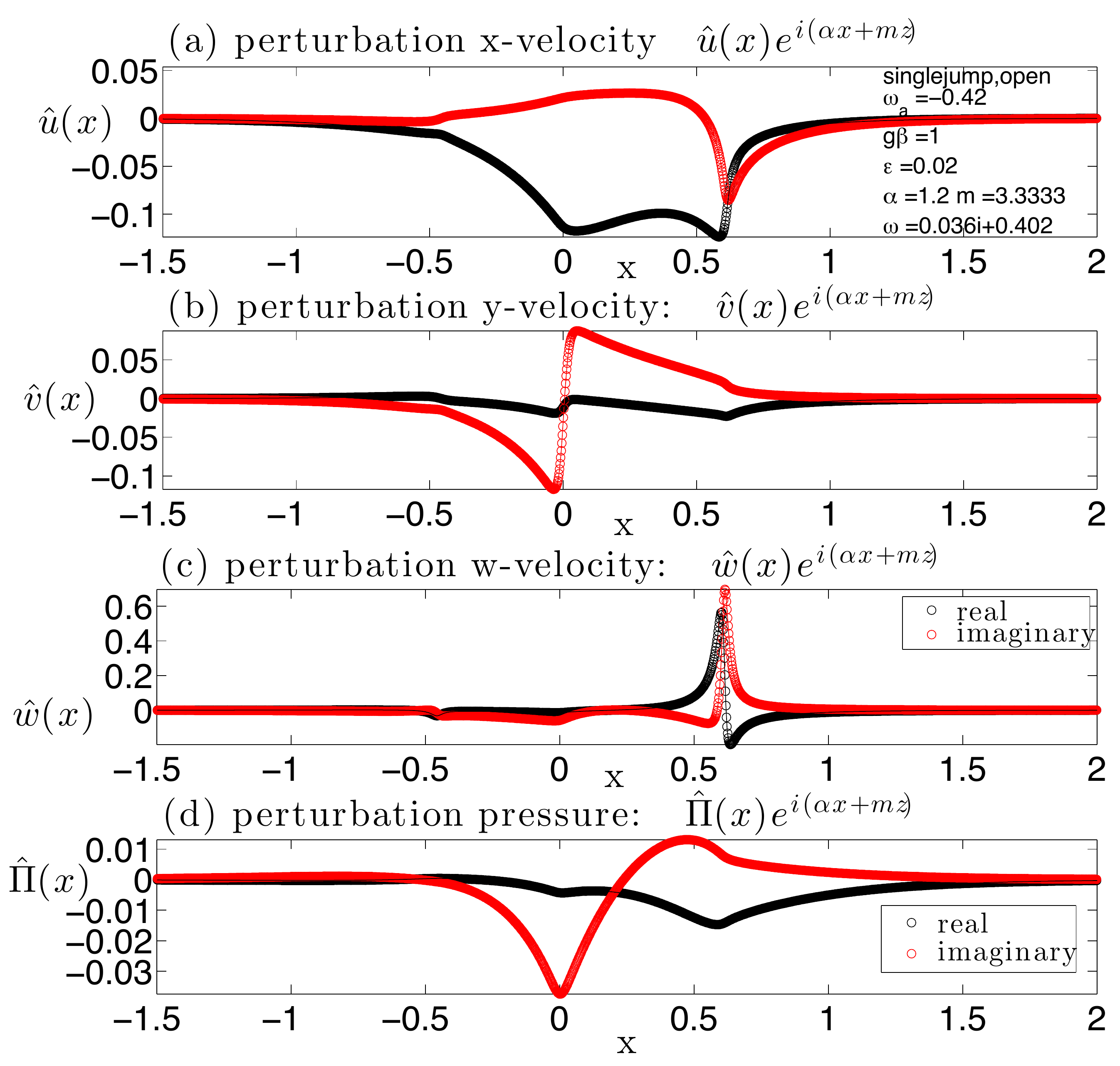}
\end{center}
\caption{Like the previous figure but showing the fields (a) $u'$, (b) $v'$,(b) $w'$ and (d) $\Pi'$.  All 
quantities show substantial drop-off as $x$ approaches the left and right boundaries emphasizing the strong localized nature of these disturbances. We emphasize that the solutions are calculated on the domain
$|x| < \pi$ but that for purposes of clarity only the inner portions are displayed 
since all solutions show exponential decay with increased $|x|$.}
\label{Velocity_profiles_1j}
\end{figure}

\section{Linear Theory Results}\label{results}

We have scanned for solutions in the event that the basic flow is
a pure Keplerian velocity field, i.e. for $V(x) = 0$. We find no converged 
continuous normal modes.  The reason appears
to us clear: the structure of
the alternative formulation of the equations, i.e. (\ref{linearized_alternate_Xi}--\ref{linearized_alternate_D}) together with $\Xi' = 0$ 
indicates that the system 
supports shear modified inertial-gravity waves.   No normal modes are
expected since the radial extent of the system is
infinite and, as such, is unable to support a ``global" supported mode.
Normal modes are potentially possible only if other boundary conditions
are adopted (not done here).  Of course, as an initial value problem with a given
initial disturbance, this system would respond by shedding inertia-gravity waves which
are non-normal mode solutions of (\ref{linearized_alternate_Xi}--\ref{linearized_alternate_D}).
These waves would propagate
out to $x\rightarrow \pm \infty$, but they do not qualify as normal modes in this case.
\par
\medskip
Normal modes do exist for $V\sub 0 (x) \neq 0$.  In this case, there 
are two kinds of modes supported which we henceforth refer to
as ``Rossby modes"(R-modes) and ``zombie modes" (Z-modes).  The R-modes
are the three-dimensional continuation of the classical
two-dimensional shear modes examined in the literature
(e.g. the Rayleigh shear layer, the RWI,the triangular jet,  etc.).  R-modes
are unstable when two or more Rossby waves
(sometimes known as ``Rossby edge waves"), 
each being associated with local extrema in the radial vorticity gradient
of the basic shear flow,
become resonantly phase-locked due to their mutual interaction, i.e., the
counter propagating Rossby wave mechanism 
\citep{1994JFM...276..327B,1999QJRMS.125.2835H,2010A&A...521A..25U}.   
There are no unstable R-modes in the single vorticity step flow
field because it can support only a single Rossby wave precluding the possibility
of resonant wave-wave interaction. 
\par
On the other hand, the Z-modes are different from the R-modes in that instability
in these modes involves the resonant interaction between a single Rossby wave
and a buoyant critical layer(s) nearby.    
In the following subsections we examine the properties of Z-modes
for the three model shear flows.

 \begin{figure}
\begin{center}
\leavevmode
\includegraphics[width=8.75cm]{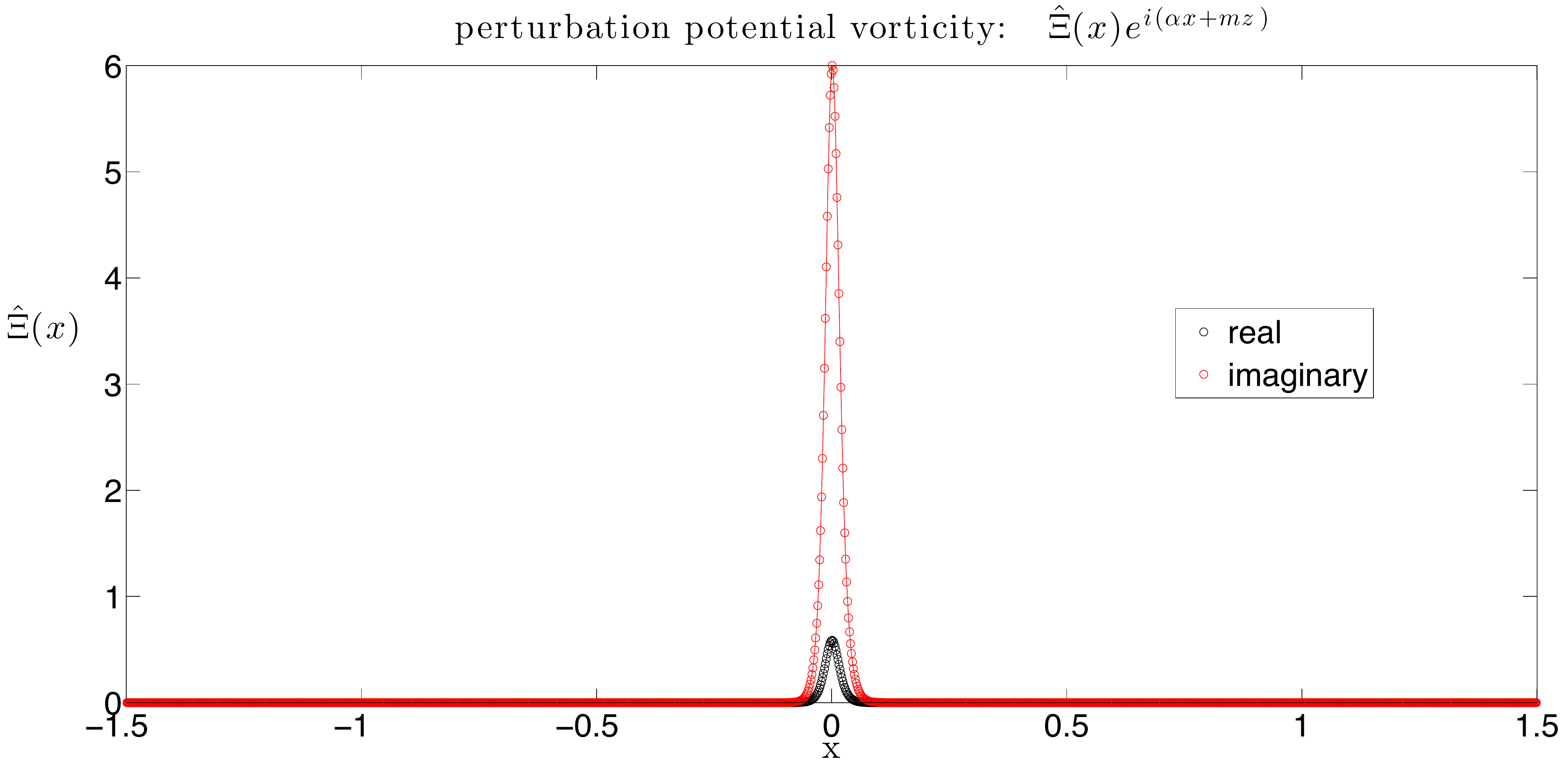}
\end{center}
\caption{Potential vorticity for the problem relating to the previous two figures.
The potential vorticity $\Xi'$ is based on its definition in equation (\ref{definition_of_PV}).
Most power in this quantity is contained mainly in the region surrounding $x=0$ corresponding
to the maximum of the mean vorticity gradient. There is no discernible power near the
critical points $x\sub{bg}^\pm$ - as expected.}
\label{Vorticity_profile_1j}
\end{figure}

\subsection{\emph{Z-modes in the vorticity step model}}\label{naked_Z-modes}
\subsubsection{Structure of the basic carrier wave}
\begin{figure*}
\begin{center}
\leavevmode
\includegraphics[width=\textwidth]{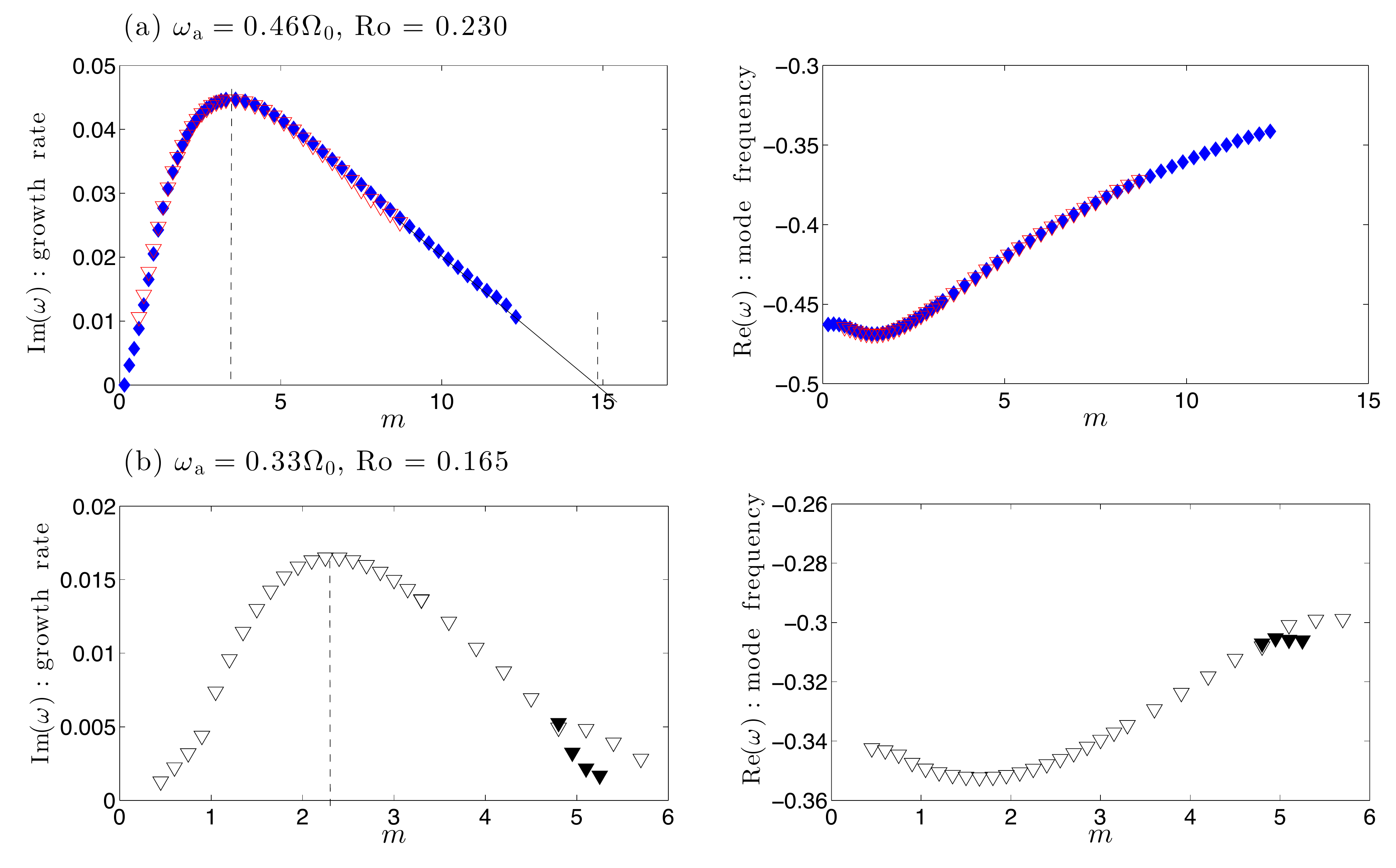}
\end{center}
\caption{Z-mode growth rates and frequencies as a function of vertical wavenumber $m$ for two different values
of $\omega_a$ in the vorticity step profile (all 
rates expressed in units of $\Omega\sub 0$): 
(a) $\omega_a = 0.46$ (b) $\omega_a = 0.33$. 
Other parameters shared by both examples: $g\beta = 1.2, \alpha = 1.1$ with $\epsilon = 0.05$.
Peak growth rates and their corresponding values of $m$ are denoted with dashed vertical lines.
General pattern indicated is that as $\omega_a$ gets smaller the wavenumber range (in $m$) and maximum
growth rate diminish.
The growth rates in panel (a) were determined on a domain $-L<x<L$ where $L=\pi$ (blue-filled diamonds)
and $L = 2\pi$ (upside-down red triangles).  Each was done with the same number of Chebyshev grid
points $N=1461$.  Note that as $|$Im$(\omega)|$ gets sufficiently small ($\sim 2L/N$) reliably converged
solutions become more difficult to determine, necessitating more resolution. 
The fitted curve (black line), extrapolating the growth rate as $m$ increases,
shows that $m_{{\rm max}}\approx 14.5$.
 In panel (b) the calculation
 was done on $L=\pi$ but for two different resolutions.  The filled black triangles are for $N=2301$
while $N=1461$ is shown with open black triangles.
}
\label{Growth rates_profiles_1j}
\end{figure*}

The velocity shear field given in equation (\ref{step_profile}) can 
support a single localized Rossby edge wave that propagates along the azimuthal
direction.  The core of the wave is localized around $x=0$ as this is the location
where the radial gradient of the basic state shear profile is the greatest.
In the limit where $m\epsilon\rightarrow 0$, together with $m \ll \alpha,$
this disturbance can be thought of as a vertically uniform azimuthally propagating edge wave 
and it was demonstrated in \citet{2010A&A...521A..25U} to have a real frequency 
given by $\omega = \omega_a$ in the effective limit $\alpha\epsilon \rightarrow 0$
($\alpha$ fixed, $\epsilon \rightarrow 0$).  The frequency
response of the unstable modes reported here are very nearly equal to this value, i.e.
Re$(\omega) \approx \omega_a$ -- and this is especially true for modes with values of $m \le 2$ (see Figure \ref{Growth rates_profiles_1j}).  
{\emph{We consider this Rossby edge wave to be the basic carrier mode of the
instability associated with Z-modes.}}
%\footnote{As we noted earlier, this is not an instance of the Rossby Wave Instability
%since that mechanism is primarily two-dimensional.  We verified that this was not
%the case for had it been so since we verified that there exists no corresponding
%normal mode (stable or not) in the exact limit of $m=0$.}.\par
\par
With this insight, we can  predict the approximate locations of the various
critical layers in the limit where both $m$ is relatively small and 
$\epsilon \ll 1$.\footnote{We also note that in this sense we
consider $\epsilon$ to be sufficiently small if its value is below
any numerically resolvable grid scale length.}  For the classical
shear critical layer we have  $ \alpha v\sub{0}(x\sub{pp}) - {\rm Re}(\omega) = 0$,
which after a little manipulation becomes
\beq
x\sub{pp}  \approx \left(\frac{1}{\alpha}\right)
\left({\displaystyle\frac{\omega\sub a}
{-{3\Omega\sub 0}/{2} +  
\omega\sub a{\rm sgn}\left(x\sub{pp}\right)}}\right).
\eeq
Similarly, the critical layers associated with gravity effects, i.e.
those points 
$\alpha v\sub{0}(x\sub{bg}^\pm) \pm \sqrt{g\beta} - {\rm Re}(\omega)= 0$
are given by
\beq
x\sub{bg}^\pm
\approx \left(\frac{1}{\alpha}\right)
\left(\frac{\omega\sub a\pm {\sqrt{g\beta}}}
{-{3\Omega\sub 0}/{2} +  
\omega\sub a{\rm sgn}\left(x\sub{bg}^\pm\right)}\right).
\label{critical_layer_positions_bg}
\eeq 
The above are rough guides -- in actuality the position of the critical
layers will differ from the above approximate form when 
$m \gg 1$, since $\omega$ of the Rossby wave is no longer expected
to be given by $\omega_a$.  
Nonetheless, when the correct value of Re$(\omega)$ appropriate
for the given Rossby wave is input,  we obtain
the correct critical layer position as expected.

\par
In Figure \ref{General_profiles_1j} we show a fairly typical result
involving unstable modes of the system.  
%Here these 
%are shown for parameter values
%$\omega_a = -0.42\Omega\sub 0$ together $\epsilon = 1/30$ and $g\beta = 1.0\Omega_0^2$.
%This step profile has a Ro$=0.21$ and the shape indicates a decrease of the
%vorticity as one crosses $x=0$ from below.
%The radial domain is $|x|< L$ where $L=\pi$.  The perturbation wave numbers
%shown are $m \approx 3.3$ and $\alpha = 1.2$ with corresponding
%wavelengths $\lambda_m = 2\pi/m \approx 1.89$ and 
%$\lambda_\alpha = 2\pi/\alpha \approx 4.70 \approx 0.75\pi$.
The three panels show the quantities $\theta',\zeta'$ and $D'$.
The vorticity shows strong power near $x=0$. This arises from
the radial transport of the mean vorticity gradient term $u' \cdot v\sub{0xx}$
and indicates that the perturbation is structurally that of a 
Rossby wave centered at that position.  According to equation (\ref{critical_layer_positions_bg})
and given the parameters of the system, 
the positions of the two buoyant gravity wave critical layers
are $x\sub{bg}^- \approx -0.448$ and $x\sub{bg}^+ \approx 0.616$.
The importance of these locations are self-evident in 
both the buoyancy and dilatation fields, demonstrating
amplified power in the vicinity of those critical layers. \par

\par
 The corresponding perturbation velocity fields 
 and pressure fluctuations are shown in Figure \ref{Velocity_profiles_1j},
 indicating the strong localized nature of the disturbances where
 the amplitude of the perturbations decays to zero rapidly as
 $|x| \gg \epsilon$ (at least, several 100 times $\epsilon$).  
 It is generally for this reason we are confident that
 these responses (and overall physical effects) 
 are independent of the assumed radial boundary conditions so long
 as the boundaries are set far enough away.  We have scanned the response
 of this system using both impenetrable walls and periodic conditions
 in the radial direction and we find that the response is essentially the same
 so long as the boundaries themselves are set far enough apart. In principle
 this means $L$ being sufficiently greater than $\pi$, but in practice we find that even
 slightly larger than $L = \pi$ is good enough.
 Observable signatures of the critical layers at $x=x\sub{bg}^\pm$ 
 are also evident in
 these fields, however, it is most prominent in the vertical velocity.
 This has profound implications for nonlinear driving (Section \ref{jets_begetting_jets}).

\par

The perturbation potential vorticity,
is displayed in Figure \ref{Vorticity_profile_1j}.
Consistent with our analysis of Section \ref{alternative_form_of_equations}, 
we see that  $\Xi'$ has power purely in the region of $x=0$ and no
discernible power in the regions around the critical points $x\sub{bg}^\pm$.
This is rationalized because the potential vorticity is (linearly) forced 
only by the advection of the mean
vorticity gradient which has appreciable power only around $x=0$. 
For us, this observation lends further credence to our claim that the main
perturbation structure is that of a Rossby wave and that it is this wave
that triggers the critical layers at $x\sub{bg}^\pm$.
\subsubsection{Critical layer scaling}

The sizes of the critical layers scale approximately as $\sim \sqrt{{\rm Im}(\omega)/\Omega\sub 0}/m$.
This fact can be seen from an elementary boundary layer scaling analysis of 
Eq. (\ref{Single_2nd_Order_ODE_in_Pi}) in the vicinity of $x=x\sub{{\rm bg}}$
.  
[For an analogous critical layer analysis see the thorough
examination of a barotropic jet instability performed by Balmforth and Piccolo (2001)].
The critical layer arises 
from the first term on the RHS of equation Eq. (\ref{Single_2nd_Order_ODE_in_Pi}),
and when the layer is activated, this term balances the second $x$ derivative of
$\hat \Pi$:
\beq
\partial_x^2 \hat \Pi \approx 
\frac{\omega\sub\varepsilon^2 - g\beta}{\Big(\sqrt{g\beta} - \alpha v\sub 0(x) + \omega\Big)
\Big(\sqrt{g\beta} + \alpha v\sub 0(x) - \omega\Big)} m^2 \hat \Pi.
\label{Boundary_Layer_Eqn}
\eeq
in which we have explicitly factored the denominator of the term on the RHS of the above expression.
For a given solution value $\omega$, at the critical
layer we have according to equation (\ref{buoyant_critical_layer_guesstimate}),
\[
\sqrt {g\beta} \pm \Big[\alpha v\sub 0(x\sub{{\rm bg}}) - {\rm Re}(\omega)\Big] = 0,
\]
A first order Taylor series expansion of the term $\alpha v\sub 0(x)$ near 
$x = x\sub{{\rm bg}}$ produces
\[
\alpha v\sub 0(x) \approx
\alpha v\sub 0(x\sub{{\rm bg}}) + \alpha \frac{d v\sub 0}{dx}\bigg|_{x\sub{{\rm bg}}}
\left(x-x\sub{{\rm bg}}\right)
\]
Considering the assumption $|{\rm Im}(\omega)| \ll |{\rm Re}(\omega)||$
(noting that all of the growth rates determined in this study are usually a factor
of 10 smaller than $\Omega\sub 0$),
putting all of the above approximate forms into Eq. (\ref{Boundary_Layer_Eqn})
reveals
\beq
\partial_x^2 \hat \Pi \approx 
\frac{\left[\omega\sub\varepsilon^2(x\sub{{\rm bg}}) - g\beta\right]m^2}{2\sqrt{g\beta}
\left[\alpha \frac{d v\sub 0}{dx}\bigg|_{x\sub{{\rm bg}}}
\left(x-x\sub{{\rm bg}}\right) - i {\rm {Im}}(\omega)\right]} \hat \Pi.
\label{Boundary_Layer_Eqn_further}
\eeq
For Im$(\omega) \neq 0$, the rate of change of the second derivative
of $\hat \Pi$ is scaled by the value of the  coefficient on the RHS of
Eq. (\ref{Boundary_Layer_Eqn_further}) evaluated at $x=x\sub{{\rm bg}}$.
\footnote{The technical reason for this
is that the branchcut of both the above ordinary differential equation and 
its associated
solution lies along one
of the two imaginary axis on the complex plane of the function 
$\chi \equiv x-x\sub{{\rm bg}}$.}
Thus, we have in the region very near $x\approx x\sub{{\rm bg}}$,
\beq
\partial_x^2 \hat\Pi \approx 
i\frac{1}{\Gamma^2}
\hat \Pi,\qquad
\Gamma^2 \equiv
{\frac
{2\sqrt{g\beta} \cdot {\rm {Im}}(\omega)}
{\left[\omega\sub\varepsilon^2(x\sub{{\rm bg}}) - g\beta\right]m^2}.
}
\label{boundary_length_scales}
\eeq
The magnitude of $\Gamma$ sets the approximate variation scale of the boundary layer
region.  For the solutions shown in figures (\ref{General_profiles_1j}--\ref{Velocity_profiles_1j}),
we have indicated with hatched vertical lines the regions in which
the critical layer is most obvious (see also Figure \ref{nonlinear_vorticity_forcing_1j}).  The width of the region, which we designate by $\delta x$, should
be approximately twice the value of $|\Gamma|$ predicted by the above analysis
(\ref{boundary_length_scales}).  For the critical layer 
near $x=0.62$, we predict values of $\Gamma \sim 0.0575$ based on the input parameters,
while we see that the
width of the buoyant critical layer zone in these figures is about $\delta x \sim 0.12$.
\par
The boundary layer scaling for the buoyant critical layers
 developed here is generally valid for the results of the other two model
 flows discussed in Sections \ref{shear_layer} and \ref{asymmetric_jet}.
\par
We have also verified that the main trend predicted by (\ref{boundary_length_scales}) holds, namely that
for all other parameters equal, the boundary layer regions 
shrinks in proportion to $\sqrt{|{\rm Im}(\omega)|}/m$.  This last observation explains why it becomes
increasingly more difficult to ascertain resolved solutions as $m$ increases.

\par

\subsubsection{Survey of growth rates}

In Figure \ref{Growth rates_profiles_1j} we show growth rates
as a function of the vertical wavenumber $m$.  The plots are fairly typical
of the responses of the system when unstable.  Keeping the azimuthal wavenumber
fixed around $\alpha \approx 1.2$, we see that the unstable vertical wave numbers will
span a finite range with a corresponding peak growth rate.  Normal modes appear
in stable/unstable pairs. In this single
step vorticity model, growth occurs irrespective of the sign of $\omega\sub a$.
For example, for values of $|\omega_\alpha| \approx 0.43 \Omega\sub 0$ (Ro $\approx 0.23$) together
with $g\beta = 1.2\Omega\sub 0^2$, the maximum growth rate occurs around $m=3.5$ and
has a growth ${\rm Im}(\omega) \approx 0.045\Omega\sub 0$ indicating that for this value of
Ro, the e-folding time of growth is 3-4 rotation times, i.e., $\sim \order{1/2\pi{\rm Im}(\omega)}$.
\par
The maximum growth rate and corresponding vertical wavenumber of maximal growth decreases
as $|\omega\sub a|$ decreases.  Furthermore, for $g\beta = 1.0\Omega\sub 0^2$
we find that if $|\omega\sub a|<0.2\Omega\sub 0$
there appears to be no instability at all.  This criterion roughly corresponds to the Ro of the
vorticity step
necessarily being less than $0.1$ for the system to be stable.  
This is determined based on our highest resolution examinations.  Lower values of Ro may
also be unstable, but they are not currently resolvable in our searches.

\subsubsection{Robustness of numerically calculated solutions}\label{viscosity_solutions}
\begin{figure*}
\begin{center}
\leavevmode
\includegraphics[width=15.75cm]{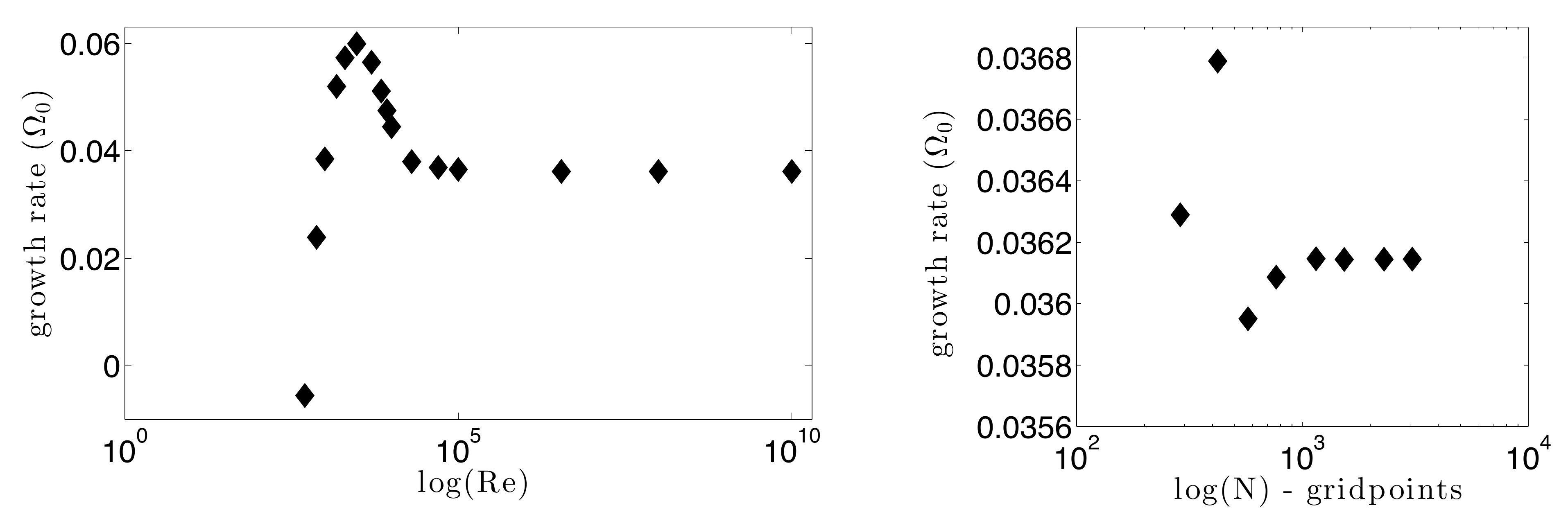}
\end{center}
\caption{Right panel: growth rates as a function of Re.  Left panel: growth
rates as a function of grid resolution.  These tests have
been performed for the solutions
showcased in Fig. \ref{General_profiles_1j}.  Numerical resolution
for solutions shown in left panel is $N=1915$. Growth rates appear to 
achieve a maximum at finite values of Re, here near 1000.}
\label{viscous_runs_checker_plotter}
\end{figure*}

{ 
During the evaluation stage of this work, an anonymous referee suggested
we consider adding a viscous operator to the equations of motion 
in order to test the robustness
of the numerical method employed in this study.  The main concern
is to identify any possible numerical artifact(s) that may be
introduced by this integral equation method because of the 
mathematical singularity posed by critical layer -- the inclusion
of viscosity acts to regularize solutions.
 As such, we have
introduced the terms ${\rm Re}^{-1}\nabla^2 \zeta', 
{\rm Re}^{-1}\nabla^2 D'$ and ${\rm Re}^{-1}\nabla^2 \theta'$
respectively to the right hand sides of equations
(\ref{linearized_zeta}-\ref{linearized_theta}).  The Reynolds number
is defined in the shearing box approximation as
\beq
{\rm Re} \equiv \Omega\sub 0^2 H/\nu,
\eeq
where $\nu$ is the effective viscosity of the medium.  The result
is now a mixed coupled set of integro-differential equations.
\footnote{While a scaling of Re on the viscous operator 
is natural for the velocity components, we interpret
the introduction of ${\rm Re}^{-1}\nabla^2 \theta'$
to the RHS of equation (\ref{linearized_theta}) as physically representing
a constant heat diffusion with a thermal diffusivity $\kappa$, in which
the Prandtl number, defined as Pr $\equiv\nu/\kappa$, is set to 1.
Indeed in the general case, the correct form of the diffusivity operator appearing
for the temperature equation would appear as ${\rm Pr}^{-1}\cdot{\rm Re}^{-1}\nabla^2 \theta'$.}
\par

Figure \ref{viscous_runs_checker_plotter} shows the behavior of
the growth rates showcased in figure
\ref{General_profiles_1j} as a function of both Re and increased numerical resolution.
The left panel of figure \ref{viscous_runs_checker_plotter} 
shows the trend in the growth rates as a function of Re.  Interestingly,
the growth rates actually show an increase as Re is lowered, indicating a peak
value $\approx 0.06 \Omega\sub 0$ at Re$\approx 1000$.  As Re is lowered below
about 600 we see that Im$(\omega) < 0$.  The structure of the eigenmodes
within the critical layer zone shows widening and smoothing (not shown here).
The reasons why the growth rate of the instability appears maximize at a
finite value of Re remains to be understood.  Nonetheless, based on the survey reported here,
we are confident that we have not introduced unwanted artifacts in the
generation of our solutions. \footnote{A recent study (Marcus et al. 2016) released during
the review phase of this work reports
similar robustness results: a viscous operator was included 
to test the numerical stability of the linear analysis of the ZVI
within the setting of the shear layer and they too find that the modes
are unstable even in the presence of viscosity.}
\par
The right panel of figure \ref{viscous_runs_checker_plotter} 
shows the behavior of the growth rates as a function of numerical
resolution for the same system in the Re $\rightarrow \infty$ limit, recalling
the equations were solved for within the domain $|x|<\pi$.  The growth rates
show clear numerical convergence for values of $N>1000$ which corresponds to a
grid spacing of $\Delta x \approx 0.006$.  For the solution shown, the size of
the critical layer is $\delta x \approx 0.12$ which means that the critical layer
zone is resolved by at least 20 points.  Note that for the highest resolved solutions
considered ($N\approx 3000$, Figs.\ref{General_profiles_1j}-\ref{Vorticity_profile_1j})
the critical layer at near $x=0.62$ is resolved by more than 50 points.
We are therefore confident that these numerically generated
solutions have been sufficiently resolved.

}

\subsection{\emph{Shear Layer}}\label{shear_layer}
\subsubsection{R-modes}
{{
In this subsection as well as in Section \ref{RWI_in_asymmetric_jet} we study the evolution of perturbations
with no vertical dependence ($m=0$).  In that case the horizontal divergence
$D' = 0$ and we only have to solve the normal
mode version of equation (\ref{linearized_alternate_zeta})
\beq
\Big[-i\omega + i\alpha\left(-1.5\Omega\sub 0 x + V\sub{0}(x)\right)\Big]\hat\zeta = 
V_{0xx} \hat u
\eeq
where the vertical vorticity and corresponding horizontal velocity
fields are related to the normal mode streamfunction $\hat\psi(x)$ more simply via
\beq
\zeta = \frac{d^2\hat\psi}{dx^2} - \alpha^2 \hat\psi,
\qquad \hat v = -\frac{d\hat \psi}{dx}, \ \ \ \hat u = -i\alpha \hat \psi.
\eeq
This is essentially an analysis of the RWI except for a system with two local extrema
in the pressure profile (whereas the usual RWI is an analysis of a configuration
with only one local pressure extremum).
This analysis of the barotropic shear layer model gets 
simplified if we implement the strategy used in \citet{2010A&A...521A..25U}, wherein
 $\epsilon \rightarrow 0$ is assumed. In this limit,
the continuous shear layer flow profile becomes  
an analytically tractable problem comprising of three piecewise linear shear profiles. 
The solution procedure follows the steps
 detailed in Appendix \ref{assymetric_jet_analysis} relating
 to the same type of analysis done on the asymmetric jet (examined in the next
 section).
\footnote{It ought to be noted that there are subtleties introduced because of the use
of piecewise linear profiles, for instance, it is sometimes the case that piecewise
linear profiles predict modes that do necessarily have counterparts in analysis
of flow profiles that are infinitely differentiable.  The normal modes generated herein, through the use of 
these piecewise linear
representations, all have counterpart modes in their corresponding
infinitely differentiable flow profiles.}
For the shear layer considered here it is a slight generalization of Rayleigh's shear layer
analysis (Rayleigh, 1880) with the additional superposition of the background Keplerian
shear profile.
The resulting frequency response is given by
\beq
4\omega^2 = -\omega_a^2 e^{-4\alpha \Delta} + \Big[3\alpha\Delta\Omega\sub 0+(1-2\alpha\Delta)\omega_a\Big]^2.
\eeq
Inspection shows that this system becomes unstable when $\omega$ goes through zero.  The physical
condition for instability is the phase locking of two counterpropagating Rossby waves 
\citep{1994JFM...276..327B,1999QJRMS.125.2835H}.  By setting $\omega = 0$ we find the condition for marginal growth to be given by
\beq
\frac{\omega\sub{R}^{\pm}}{\Omega\sub 0} = {\displaystyle\frac{3 \alpha\Delta}{\displaystyle 2\alpha\Delta -1 \pm e^{-2\alpha\Delta}}},
\label{omega-sub_RWI_shear_layer}
\eeq
where $\omega\sub{R}^\pm$ is the shear layer amplitude corresponding to marginality.
We do not consider the positive roots, $\omega\sub{R}^+$, because these generally 
correspond to composite flows that are strongly cyclonic, i.e., flows in which $dv/dx = -3\Omega\sub 0/2 + \omega\sub{R}^+ > 0$ while 
the negative root, on the other hand, often corresponds to anti-cylonic shear profiles. 
\footnote{This is generally true upon examination of the marginal condition in Eq. (\ref{omega-sub_RWI_shear_layer}).}
\par
If we consider the setting examined in M13, we suppose that the smallest azimuthal wavenumber is the one fitting the box considered therein.  The corresponding azimuthal length scale in their box units is approximately $
L = (3/2)\pi$.  This corresponds to a fundamental azimuthal wavenumber $\alpha = 2\pi/L = 4/3$.  Similarly, the width of the shear line-charge is approximately $0.28$ (also in their code units).  
Translating this into our setup, the total width is $2\Delta$, 
thus we adopt a value of $\Delta = 0.14$.  In this case, the corresponding critical value of $\omega_a$ for the R-mode
is given from
Eq. (\ref{omega-sub_RWI_shear_layer})
 by $\omega\sub R^- \approx -0.426\Omega\sub 0$, with a corresponding critical Rossby number
 of 
${\tt Ro} = |\omega\sub R^-/2\Omega\sub 0| \approx 0.213$.  Values of $0>\omega_a > \omega\sub R^-$ should then be
stable to the RWI.  We find that the value of $\omega\sub R^-$
predicted in Eq. (\ref{omega-sub_RWI_shear_layer}) generally predicts more negative values
of $\omega\sub R^-$ that are borne out in our models.  The reason for this is that we
adopt non-zero values of $\epsilon$ ($= 1/50$) for all the models reported in this section.  The discrepency is generally 2-5 \% the predicted values.
\footnote{We have separately checked 
(but have not included in this manuscript) and verified that as $\epsilon \rightarrow 0$ the model value
$\omega\sub R^-$ converges to the one predicted for $\epsilon = 0$. 
 Since smaller values of the $\epsilon$ parameter 
requires more grid points to resolve, we have adopted the value chosen here because it minimizes
the number of computations required to assess a stable numerical solution while remaining 
close to the idealized $\epsilon = 0$ model developed earlier.} 

\subsubsection{Z-modes}\label{z-modes}
The results of the vortex-step profile indicates
that Z-modes ought to be present, especially for conditions
in which $\omega_a > \omega\sub R^-$ and we accordingly
 scan the response of the system for these parameter values 
 and assess the instability in $\omega_a$.  The results reported
in M13 suggest that the Z-mode is strong and expressed for values of ${\tt Ro} \approx 0.2$, which is slightly
lower than the critical value of the Rossby number determined above for the activation of R-modes 
(i.e., ${\tt Ro} \approx 0.213$).  Indeed, we find that the Z-mode is recovered in this model
in the absence of R-modes.  We describe this in the following.\par
Provided the \BV frequency $\sqrt{g\beta}$ exceeds
some minimum value, Z-modes manifest themselves
for shear layer amplitudes, $\omega_a < \omega\sub Z$, where 
$\omega\sub Z = \omega\sub Z(\alpha\Delta,g\beta)$ is the critical value of the jet's vorticity amplitude
in order for Z-modes to appear.  
As indicated,  $\omega\sub Z$ is a function of both
the square of the \BV frequency and the relative measure of the jet-width to the  azimuthal 
perturbation length scale.
Furthermore, for a given values of $\alpha\Delta$, there 
are values of $g\beta > N^2\sub{\rm{crit}}(\alpha\Delta)$ for which Z-modes 
{\emph{with buoyant critical layers lying outside of the primary
shear layer}}
are present in the absence
of the RWI, that is to say, ``naked" Z-modes with external critical layers are observed for 
shear flow amplitudes falling within the range
$\omega\sub R^-<\omega_a < \omega\sub Z$.

 \begin{figure}
\begin{center}
\leavevmode
\includegraphics[width=8.3cm]{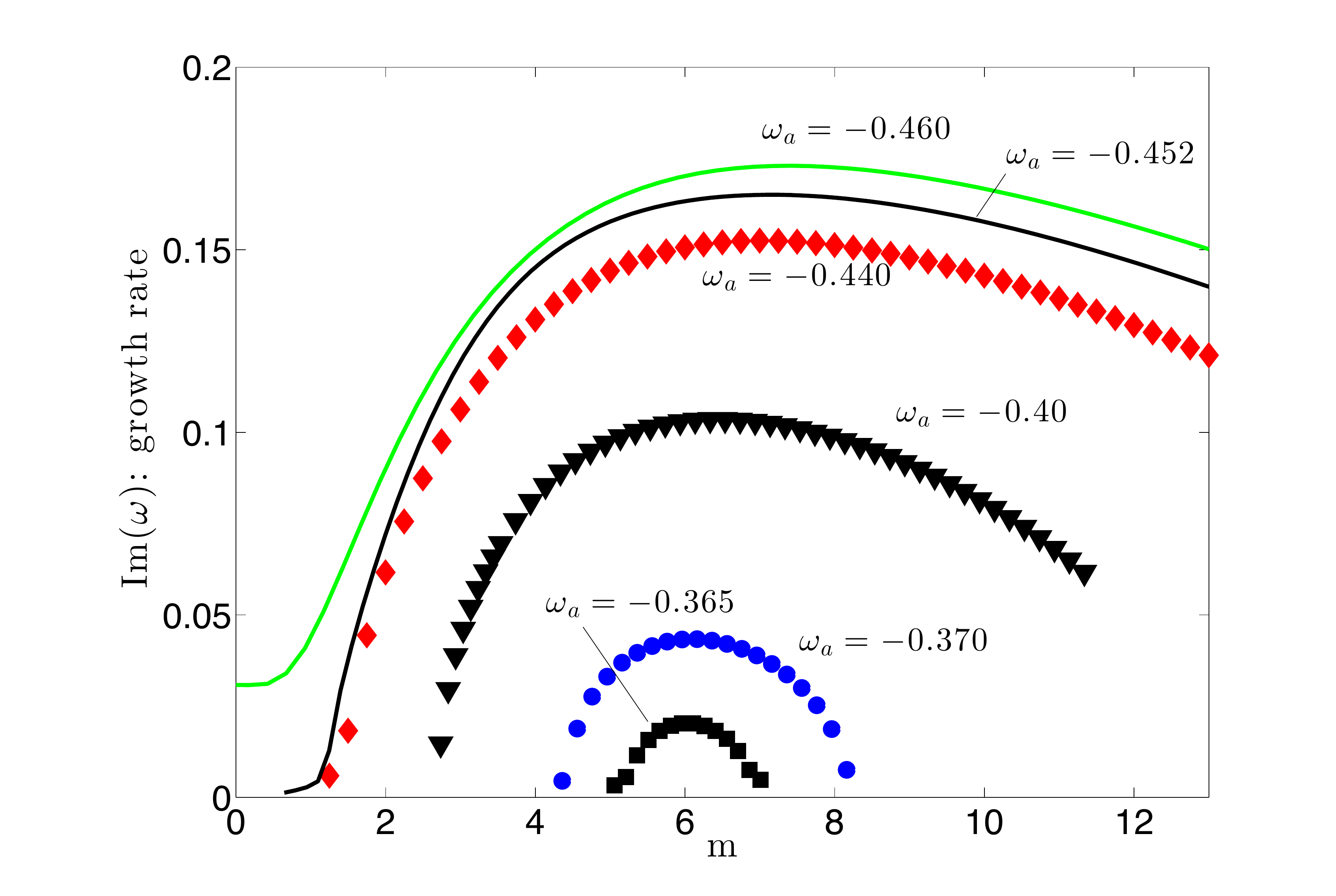}
\end{center}
\caption{Growth rates (in units of $\Omega\sub 0$) versus vertical wavenumber $m$ for
the shear layer profile:
$\epsilon = 0.02, g\beta = 4\Omega\sub 0^2, \Delta = 0.142$ and $\alpha = 1.4$.
Several different values of the shear layer amplitude are shown in plot scaled by $\Omega\sub 0$. 
%$\omega_a = 0.365\Omega\sub 0$
%(boxes), $\omega_a = -0.370\Omega\sub 0$ (cirlces),
%$\omega_a = -0.400\Omega\sub 0$ (triangles),
%$\omega_a = -0.440\Omega\sub 0$ (diamonds),
%$\omega_a = -0.452\Omega\sub 0$ (black line),
%$\omega_a = -0.460\Omega\sub 0$ (green line).  
%Recall that the scale ${\tt Ro} = |\omega_a|/(2\Omega\sub 0)$.
In the model flows used for these plots, the onset of RWI occurs for
 $\omega\sub a = \omega\sub R^{-} \approx
-0.4525 \Omega\sub 0$.  
All basic flows shown are anticyclonic compared to the background shear.
}
\label{Growth_Rates_Shear_layer}
\end{figure}

 \begin{figure}
\begin{center}
\leavevmode
\includegraphics[width=8.3cm]{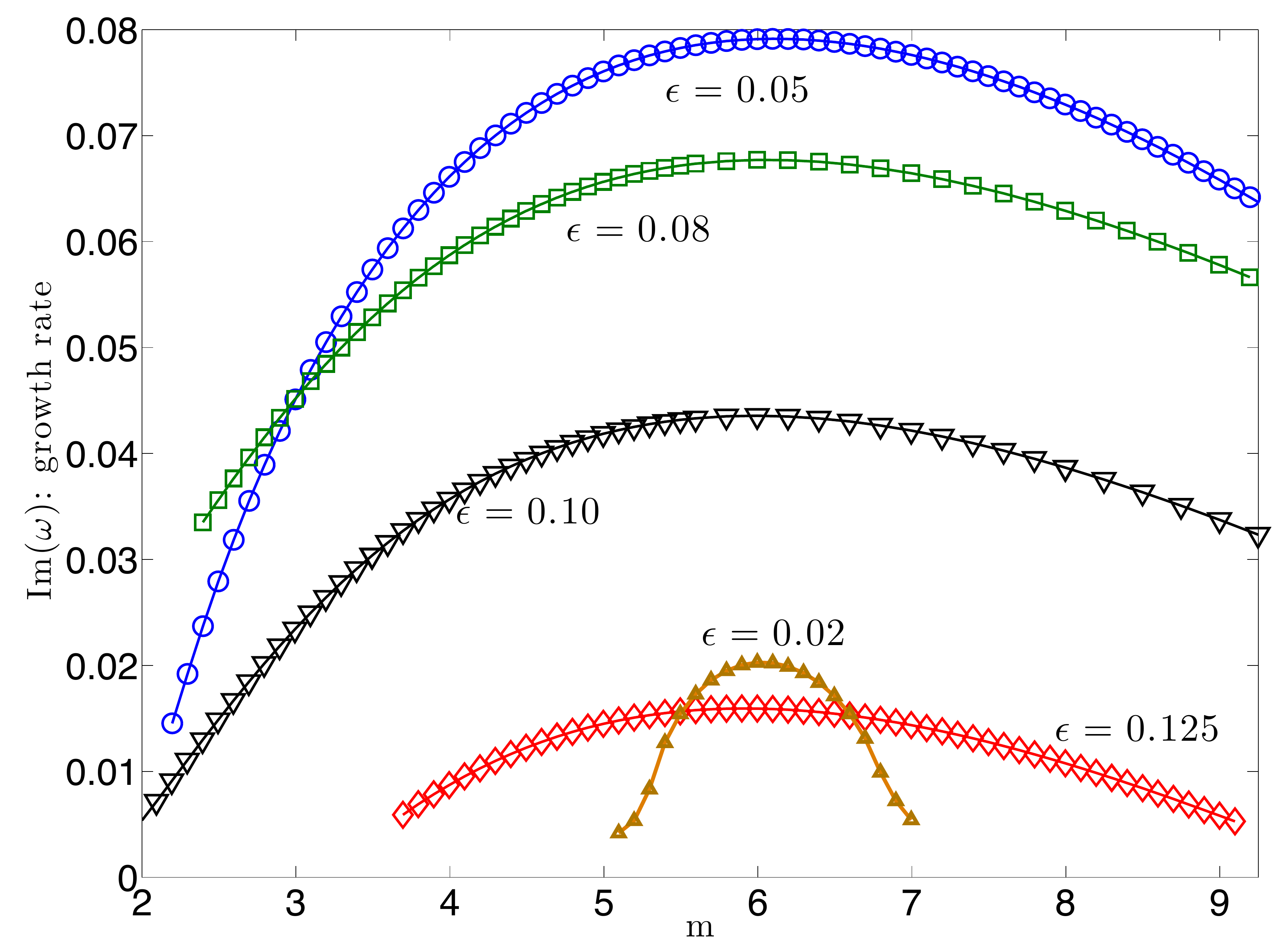}
\end{center}
\caption{Growth rates (in units of $\Omega\sub 0$) versus vertical wavenumber $m$ for the shear layer profile:
$\omega_a = -0.365, g\beta/\Omega\sub 0^2 = 4, \Delta = 0.142$ and $\alpha = 1.4$.
Shown on figure are several different values of $\epsilon$.  
For all other parameters held fixed, the figure indicates
that there is an optimal value of $\epsilon$ for maximal growth -- 
for the parameter combination considered here $\epsilon \sim 0.05$.
}
\label{Growth_Rates_Shear_layer_2}
\end{figure}

\par
Given the symmetry of the shear layer, we find that all Z-modes are unstable with zero
oscillatory part in their growth rates.  In Figure \ref{Growth_Rates_Shear_layer} we
show growth rates versus the vertical wavenumbers $m$
for given values of $\alpha\Delta$ and $g\beta$.  In this figure, 
several profiles are shown for differing values 
of the shear amplitude $\omega_a$.  In the particular example displayed, the critical
value of $\omega\sub Z$ is greater than $\omega\sub R^-$, which means that
for values of $\omega\sub R^-<\omega_a < \omega\sub Z$ Z-modes are manifested
in the models and are naked indeed.  For values of $\omega_a<\omega\sub R^-$  both
Z-modes and R-modes are present.
\par
In Figure \ref{Growth_Rates_Shear_layer_2} we show growth rates versus
$m$ for fixed values of $\alpha$, $\Delta$ and $\omega\sub a$ for several
values of the transition scale $\epsilon$. This figure indicates
that for all other parameters held fixed, there exists a value of $\epsilon$
-- and hence, the ratio $\Delta/\epsilon$ -- for which growth is
optimal.  We also note that the effect persists even as $\epsilon$ gets to be nearly the same
order as $\Delta$ (i.e, $\Delta/\epsilon \sim 1$), although the growth rates are generally far weaker
than they are for optimal conditions. In the example shown 
in Figure \ref{Growth_Rates_Shear_layer_2}, those optimal conditions 
are $\Delta/\epsilon \sim 3$.

\par
Given that $\omega$ has zero real part, 
one might approximate the critical value of $\sqrt{g\beta}$ in which
$x\sub{\rm bg}$ coincides with $\Delta$.  Setting $\omega = 0$
and replacing $x\sub{\rm bg}$ by $\Delta$ in 
Eq. (\ref{buoyant_critical_layer_guesstimate}), followed by focusing on the
positive root 
we find the following approximate rule-of-thumb,
\beq
N\sub{\rm{crit}} \equiv \left(\sqrt{g\beta}\right)_c
\approx
\alpha  \Delta \left(-\frac{3}{2}\Omega\sub 0 + \omega_a\right)
\label{approx_Nc}
\eeq
where $N\sub{\rm{crit}}$ is the critical value.  The pattern in the data suggest that
this value is in the vicinity of $\omega_a = \omega\sub R^{-}$. This is better
borne out by the data in what follows. 
\par
\par
For example
M13 adopted a value of the \BV \ frequency $=2\Omega\sub 0$.  In our parameter settings this means
choosing $g\beta = 4\Omega\sub 0^2$.  
In the following example, 
we consider a model result similar to theirs by adopting a value $\alpha = 1.4$ (slightly larger
than their value of $\alpha = 4/3$). According to Eq. (\ref{omega-sub_RWI_shear_layer}) 
this corresponds to $\omega\sub R^- \approx -0.4655$ while we find in our numerical
model flows, with $\epsilon = 1/50$, $\omega\sub R^-({\rm model}) \approx -0.4525$
(see previous subsection). 
\par
Furthermore,
using these values in (\ref{approx_Nc}) we see that 
the critical value $N^2\sub{\rm{crit}}$ is approximately $ 0.15 \Omega\sub 0^2$.
An examination of the properties of the eigenmodes is consistent
with this picture: for values of $g\beta > N^2\sub{\rm{crit}}$ Z-mode critical
layers appear outside the shear layer. For
shear layer amplitudes satisfying $\omega\sub R^-<\omega_a < \omega\sub Z$,  
Z-modes are naked.
\par
In Figure \ref{omega_Z_Shear_Layer} we display the critical values $\omega_a = \omega\sub Z$
as a function of $g\beta$ for fixed values of the product $\alpha\Delta$. The value
of $\omega\sub Z$ is a minimum in the near vicinity of  $g\beta = N^2\sub{\rm{crit}} \approx 0.15
\Omega\sub 0^2$.  For values of $g\beta > N_c^2$ the critical layers appear well outside
the shear layer (Figure \ref{omega_Z_Shear_Layer_solutions}a).  
For values of $g\beta < N^2\sub{\rm{crit}}$ the critical layers appear within
the shear layer (Figure \ref{omega_Z_Shear_Layer_solutions}b).  
 Once again, assuming $\omega = 0$
and assuming $\alpha, \omega_a$ fixed, it follows from 
Eq. (\ref{buoyant_critical_layer_guesstimate}) that lowering $g\beta$ means
shifting the critical layer $x\sub{bg}$ toward $x=0$.
It is also remarkable that the instability continues on into the limit
$g\beta \rightarrow 0$ which suggests that an analytical boundary layer
analysis is feasible in this limit (not done here).
\par
Lastly, in Figure \ref{marginal_m_shear_Z_Shear_Layer} we show
the vertical wavenumber $m$ as a function of the \BV frequency
$g\beta$ at which the unstable Z-mode first appears (at $\omega\sub a = \omega\sub Z$).
This value of $m$ generally corresponds to the fastest growing mode as
$\omega\sub a < \omega\sub Z$.  This critical vertical wavenumber begins
to increase significantly once $g\beta$ increases past $N_c^2$.
%The location
%of the critical value of $g\beta = N_c^2$ is identified as the location where
%the $\omega\sub Z$ curve meets the (fixed) value of $\omega\sub R^{-}$.  We have
%verified that
%the location of the critical layers also obeys the relationship
%expressed in Eq. (\ref{buoyant_critical_layer_guesstimate}).  Given the observation
%that instability occurs for Re$(\omega) = 0$, it follows that the location
%of $x\sub {bg}$ approximately happens where
%\beq
%\alpha x\sub{bg}\left(-\frac{3}{2}\Omega\sub 0 + \omega_a\right) + \sqrt{g\beta} = 0.
%\eeq

 \begin{figure}
\begin{center}
\leavevmode
\includegraphics[width=9.2cm]{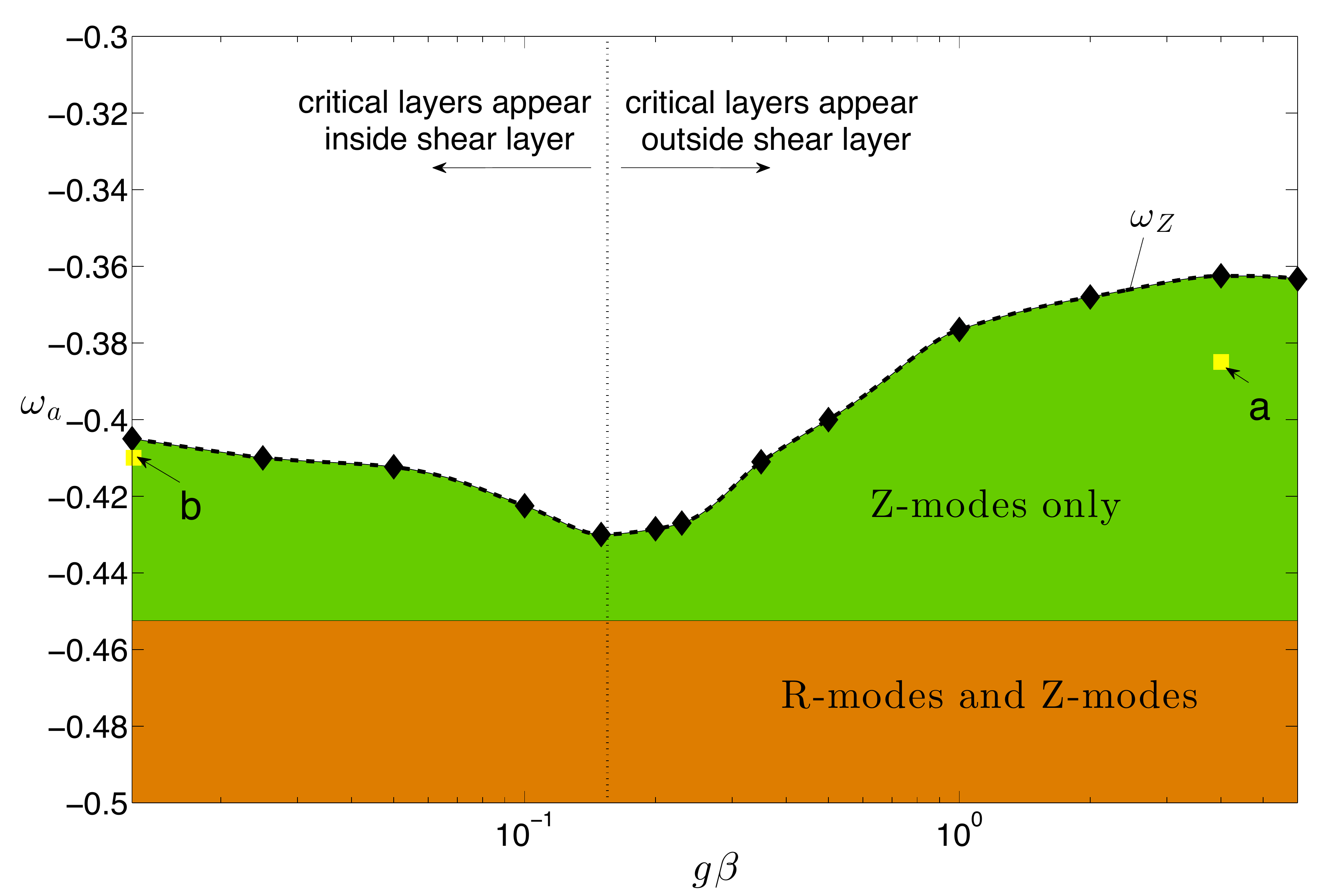}
\end{center}
\caption{Value of $\omega\sub Z$ versus $g\beta$ for fixed value of $\alpha = 1.4$ and $\Delta = 0.14$ 
(box size $L=2\pi$, $g\beta$ scaled by $\Omega\sub 0^2$, all rates in units of $\Omega\sub 0$).  
Diamonds denote numerical values determined while the dashed line is a fit
to those values.  Numerically determined value of $\omega\sub R^- \approx - 0.453$ denotes transition
into R-mode instability.  Values of $\omega_a$ corresponding to naked Z-modes designated by
the green shaded region while values of $\omega_a$ for which both Z-modes and R-modes are
expressed denoted by the orange shaded region.  Vertical dotted line represents the value of
of $g\beta = N^2\sub{{\rm crit}} \approx 0.15$ in which the critical layers correspond to the edges of the shear layer.}
\label{omega_Z_Shear_Layer}
\end{figure}

 \begin{figure*}
\begin{center}
\leavevmode
\includegraphics[width=\textwidth]{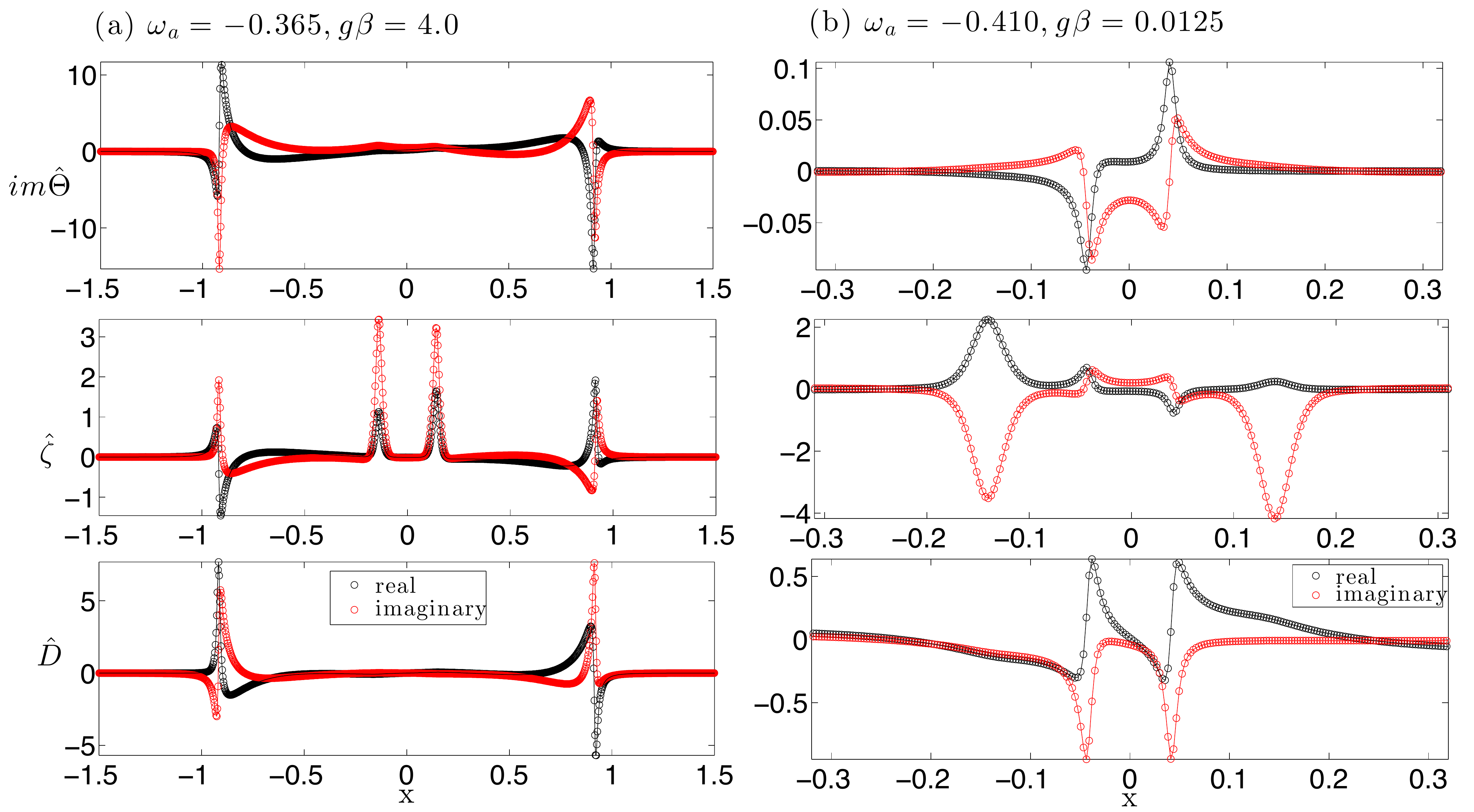}
\end{center}
\caption{Perturbation eigenfunction profiles $\hat\Theta, \hat \zeta$ and $\hat D$ in two
shear layer models.  Panel (a): $\omega_a =-0.365, g\beta = 4.0, m = 5.9$ with growth
rate -Im$(\omega)\approx 0.02$.
Panel (b):
$\omega_a =-0.410, g\beta = 0.0125, m = 2.1$ with growth
rate -Im$(\omega)\approx 0.018$.
In both panels $\alpha = 1.4$ and $\Delta = 0.14$ together
with $\epsilon = 0.02$.  The locations of these two
models in parameter space are indicated by yellow squares labelled ``a" and ``b" in 
Fig. \ref{omega_Z_Shear_Layer}.
The buoyant critical layers and their locations are predicted
according to Eq. (\ref{buoyant_critical_layer_guesstimate}) with $\omega$ set to zero: for
panel (a) $x\sub{bg} = \pm 0.91$ while for panel (b) $x\sub{bg} \approx \pm 0.042$ and these
are most prominent in both the $\hat\Theta$ and $\hat D$ fields.  The imprint of the
edges of the jet at $x=\pm 0.14$ is apparent in the $\hat\zeta$ field shown in both panels.
Note the critical layers appear inside the shear layer for the model shown in Panel (b).
All values of $\omega_a$
and $\omega$ quoted in figures are in units of $\Omega\sub 0$, while
units of $g\beta$ are $\Omega\sub 0^2$.
Solutions developed on domain $|x|<\pi$ and for values of $m$ approximately corresponding
to the fastest growth rates.
}
\label{omega_Z_Shear_Layer_solutions}
\end{figure*}

}}

 \begin{figure}
\begin{center}
\leavevmode
\includegraphics[width=8.9cm]{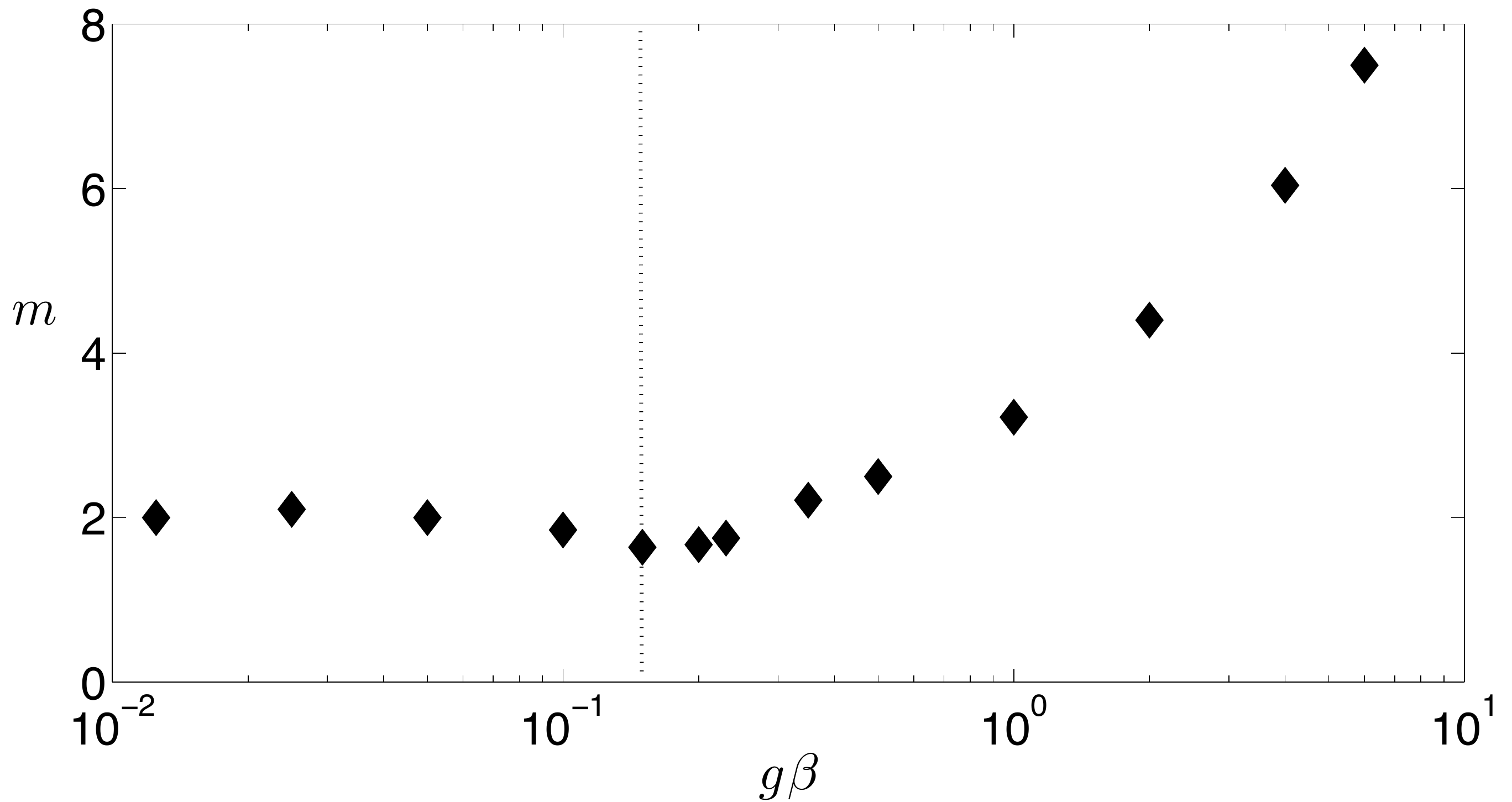}
\end{center}
\caption{The values of the vertical wavenumber $m$ corresponding to the
marginal instability transition in the vicinity of $\omega\sub a = \omega\sub Z$.
Values are shown as a function of $g\beta$ for the
values of $\alpha$ and $\Delta$ given in Figure \ref{omega_Z_Shear_Layer}.  
Values of wavenumber $m$ in azimuthal box scale units of $L=2\pi$.
$g\beta$ in units of $\Omega\sub 0^2$.  The vertical
dotted line corresponds to the value
of $g\beta = N_c^2$ (see previous figure).
}
\label{marginal_m_shear_Z_Shear_Layer}
\end{figure}

\subsection{\emph{Asymmetric Jet}}\label{asymmetric_jet}
\subsubsection{\emph{R-modes in the asymmetric jet model - the RWI}}\label{RWI_in_asymmetric_jet}

\par

 \begin{figure}
\begin{center}
\leavevmode
\includegraphics[width=9.1cm]{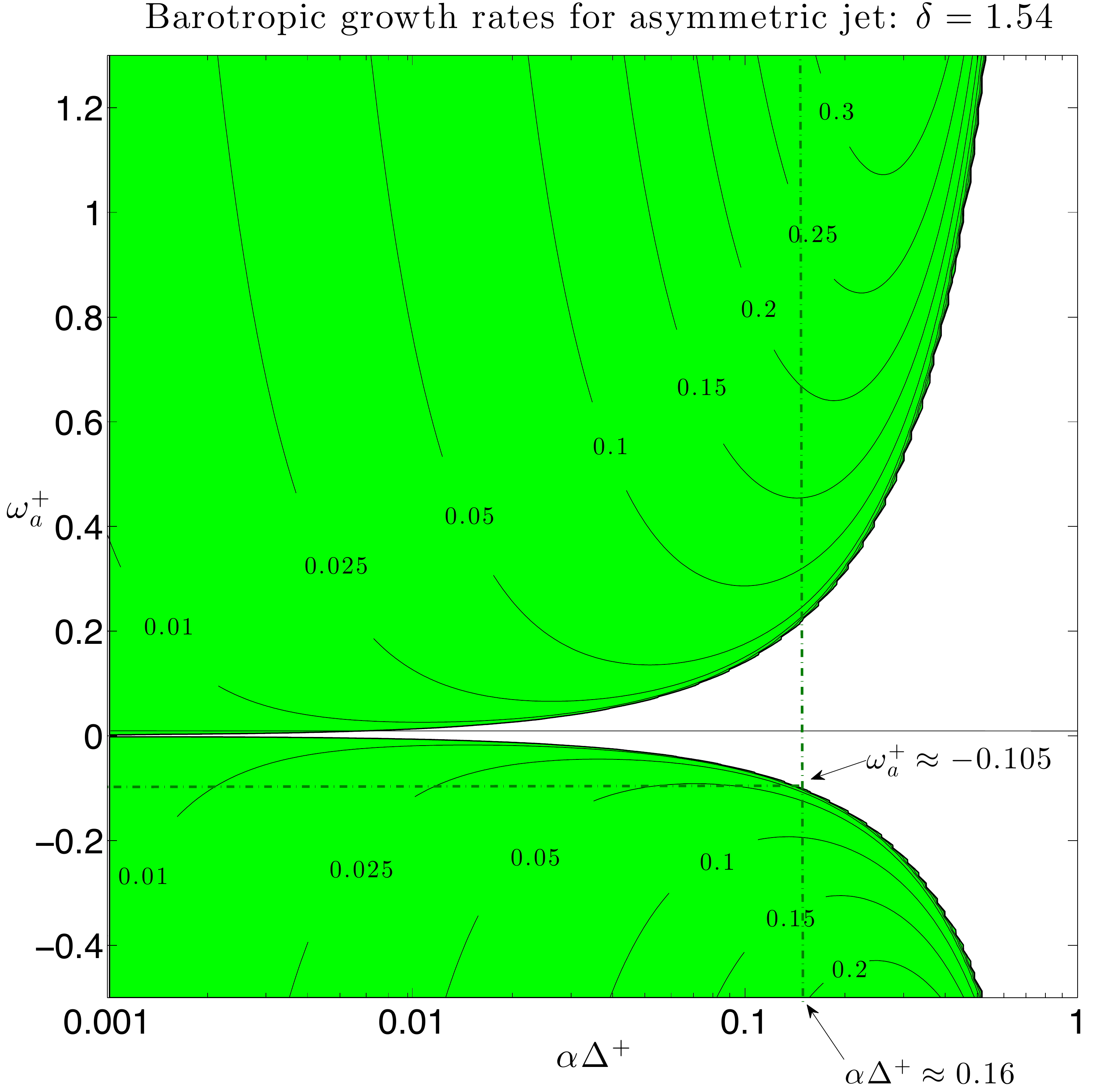}
\end{center}
\caption{Contour plots of growth rates of the barotropic asymmetric jet profile
on the $\alpha \Delta^+$ - $\omega_a^+$ plane.  Other properties
of jet are fixed in which: $\Delta^- = \delta \Delta^+$ together
with $\omega_a^- = -\omega_a^+ \Delta^+/\Delta_- = -\omega_a^+/\delta$ (see text).
This particular plot corresponds to $\delta \approx 1.54$.  Shaded regions correspond
to the RWI of this barotropic jet.  Growth rates are in units of $\Omega_0$.
Vertical hatched line corresponding to $\alpha\Delta^+ \approx 0.16$ indicating
the parameter values later examined in Section \ref{zmodes_asymmetric_jet}.
}
\label{Barotropic_Growth_rates_profiles_asymmetric_jump}
\end{figure}

In this and in the next subsection we consider asymmetric jet profiles that have zero
total integrated deviation vorticity.  In other words, in considering the flow
field given in equation (\ref{jet_profile}) we constrain our parameter variations
such that
$
\omega_a^+\Delta^+  + \omega_a^-\Delta^- = 0.
$
In our considerations we keep the ratio $\Delta^-/\Delta^+ \equiv \delta$ fixed
which means that $\omega_a^-$ is always set according to
\beq
\omega_a^- = -\omega_a^+\frac{\Delta^+}{\Delta^-} = -\frac{\omega_a^+}{\delta}.
\label{restricted_omega_minus}
\eeq
Since the only length scales in this system are given width parameters $\Delta^\pm$,
the general response is a function of $\alpha \Delta^+$ and $\omega^+$
for fixed values of $\delta$.  The analysis of the system produces the following
cubic equation for the value of the eigenvalue $\omega$
\beq
\omega^3 + a\omega^2 + b\omega + c = 0,
\label{eigenvalue_for_barotropic_jet}
\eeq
in which the coefficients $a=a(\alpha\Delta^+,\omega^+,\delta)$, 
$b=b(\alpha\Delta^+,\omega^+,\delta)$ and 
$c=c(\alpha\Delta^+,\omega^+,\delta)$ are real and whose
values are detailed in Appendix \ref{assymetric_jet_analysis}.  Cubic equations
like the one above with real coefficients have two possible
kinds of solutions either (i) three real distinct values of $\omega$ or (ii)
one real and a complex conjugate pair of solutions $\omega$.  The physical correspondence
is similar to the analysis and conclusions drawn from the simpler RWI setup in \citet{2010A&A...521A..25U}: 
when separated
Rossby waves interact with one another at a distance, they can become unstable
if their wave speeds become equal and opposite to one another in some reference frame.
The same interpretation therefore holds for this jet system as well.
\par
Figure \ref{Barotropic_Growth_rates_profiles_asymmetric_jump} depicts
a typical profile of the stability boundaries and growth rates associated
with this system for a fixed value of $\delta\approx 1.54$.  The pattern of the
solutions indicate similarity to the single pressure extremum case analyzed in 
\citet{2010A&A...521A..25U}. For the case where $-0.5\Omega\sub 0 < \omega_a^+ < 1.5\Omega\sub 0$ is a maximum value of $\alpha\Delta^+
\approx 0.53$
beyond which there is no instability.  As $\alpha\Delta^+ \rightarrow 0$ the critical value of $\omega_a^+$ for instability
also approaches zero which essentially means that all values of $\omega_a^+$, positive or negative, have some potential
for instability so long as the jet widths are thin enough.  This latter property differs from the analogous
property in the classical RWI problem wherein only anticylonic values of the deviation shear profile (i.e. $\omega_a < 0$)
lead to instability.
Finally, for a given value of $\omega_a^+$ there always exists a wavenumber corresponding to the fastest
growth rate indicated by a perusal of the contours shown in Figure \ref{Barotropic_Growth_rates_profiles_asymmetric_jump}.

 \begin{figure}
\begin{center}
\leavevmode
\includegraphics[width=8.8cm]{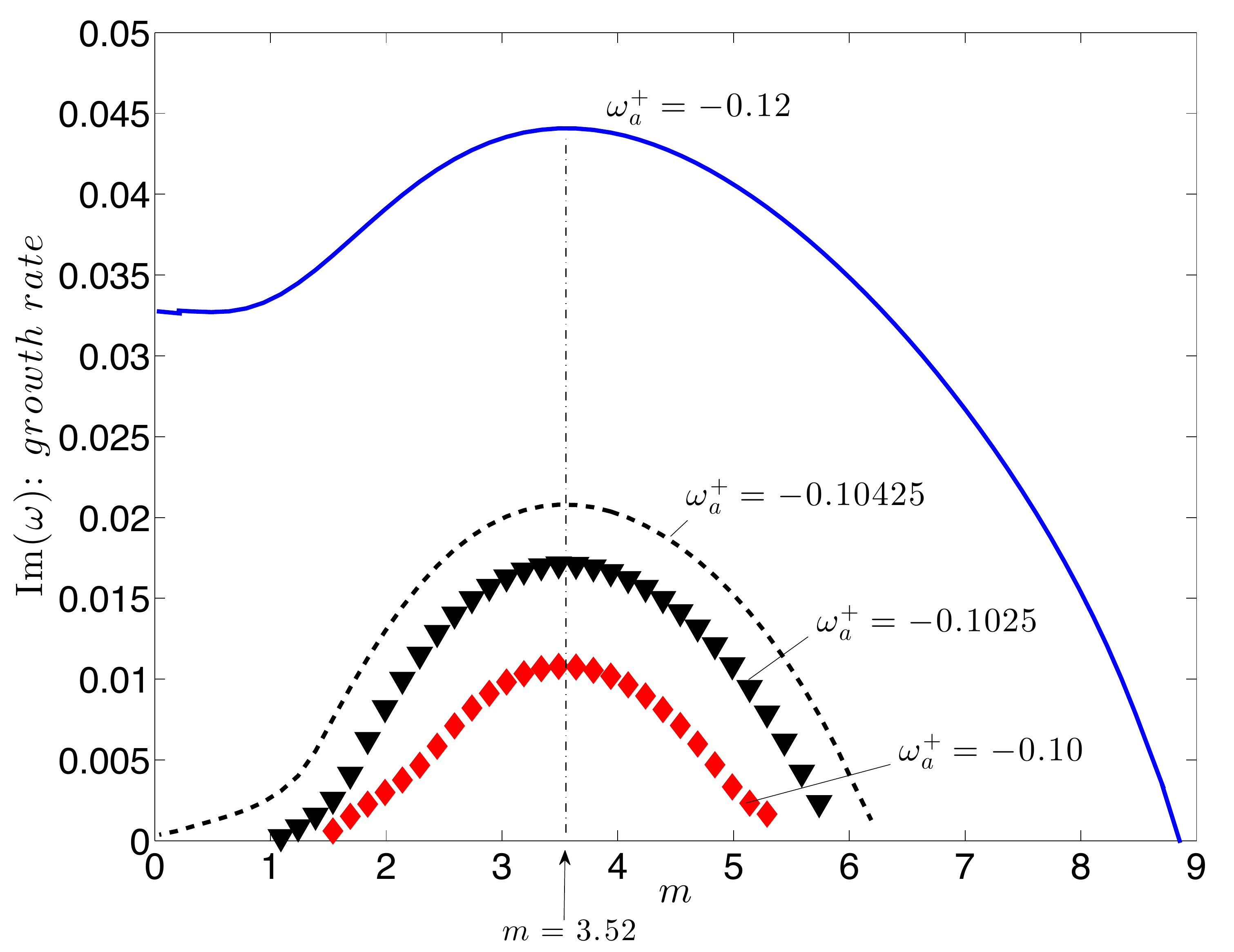}
\end{center}
\caption{Growth rates (in units of $\Omega\sub 0$) in asymmetric jet model with $\epsilon = 0.02$,
$g\beta = 4\Omega\sub 0^2$,
$\Delta^+ = 0.13$, $\Delta^- = 0.20$ (i.e. $\delta \approx 1.54$), 
and $\alpha = 1.3$
for four different values of 
$\omega_a^+$ with $\omega_a^-$ fixed and equal to
$-\omega_a^+\Delta^+/\Delta^-$
%(solid line) $\omega_a^+ = -0.12$, $\omega_a^- = 0.078$,
%(diamonds)  $\omega_a^+ = -0.1$, $\omega_a^- = 0.065$,
%(upside down triangles)   $\omega_a^+ = -0.1025$ ($\omega_a^- = 0.068$),
%(dashed curve) $\omega_a^+ = \omega^+\sub{a{{\rm c1}}} = -0.10425$  $\omega_a^+ = 0.070$ 
(values of $\omega_a^+$ in units of $\Omega\sub 0$).  
The dashed curve
signifies the critical value of $\omega_a^+$ in which
the RWI begins to also get expressed
(and corresponds to the critical
criterion indicated by where the two hatched lines meet in Figure \ref{Barotropic_Growth_rates_profiles_asymmetric_jump}).  
%For values of
%$\omega^+\sub{a{{\rm c1}}} < \omega_a^+ < 0$ only the Z-modes are unstable,
%while for $\omega_a^+ < \omega\sub{a{{\rm c1}}}$ both instabilities are active.  
In all
curves shown, the fastest growth rate corresponds to $m\approx 3.5$.
%and is
%driven primarily by the Z-mode critical layer process.  
%Also note the second critical value
%$\omega^+\sub{a{{\rm c2}}} \approx 0.22$ for the same value $\alpha \Delta^+ = 0.16$.
}
\label{Growth rates_profiles_asymmetric_jump}
\end{figure}

\subsubsection{\emph{Z-modes}}\label{zmodes_asymmetric_jet}
As might be anticipated, the asymmetric jet model also supports Z-mode instability.
A representative survey of the results is found in the growth rates as a function
of vertical wavenumber shown in Figure \ref{Growth rates_profiles_asymmetric_jump}.
As in the last section, the asymmetric models are ones in which the total integrated
deviation vorticity is zero, so that the flow profile is dictated only by values
of $\omega_a^+, \Delta^+, \Delta^-/\Delta^+ = \delta$ with $\omega_a^-$ given
by (\ref{restricted_omega_minus}).  We consider horizontal wavenumber values
of $\alpha = 1.3$ which is approximately the smallest non-zero wavenumber 
appropriate for the numerical experiments of M13.\par
Figure \ref{Growth rates_profiles_asymmetric_jump} shows the growth rates for
four different values of $\omega_a^+$ holding $\Delta^+, \delta$ and $\alpha$ fixed.  
The results show that if the jet amplitude $\omega_a^+$ lies between
$0$ and $\omega_{a{{\rm c1}}}$ with  $\omega_{a{{\rm c1}}} \approx -0.104\Omega\sub 0$, then only a Z-mode instability is possible. This corresponds to a value of Ro slightly larger than 0.05.
In this case, there is no growth for $m=0$ and the fastest
growing mode occurs for $m\approx 3.5$
-  which is similar to the periodic
pattern seen in the emerging crystallization profile of Figure 2d of M13.
However right at $\omega_a^+ = \omega^+_{a{{\rm c}}}$ (the dashed curve 
in Figure \ref{Growth rates_profiles_asymmetric_jump})
the $m=0$ state also becomes marginal.  Coincidentally
this critical value of $\omega^+_{a{{\rm c1}}}$ is approximately the same marginal condition
for the RWI and this is borne out by an inspection of the marginal boundary
shown in Figure \ref{Barotropic_Growth_rates_profiles_asymmetric_jump}
(see the location where the two hatched lines meet).  When 
the magnitude of the jet's amplitude exceeds this marginal
value, i.e. for $\omega_a^+ < \omega^+_{a{{\rm c1}}}$, then the
the growth curves show growth also for vertically uniform perturbations indicating
the concurrence of the Z-mode instability and the RWI.  The Z-mode instability
is clearly evident as the maximum growth rate still sits around the $m\approx 3.5$
value, however, its presence is diminished by the growing importance of the RWI
as the magnitude of  $\omega_a^+$ increases well past the
corresponding magnitude of $\omega^+_{a{{\rm c1}}}$ (see the solid curve in that same figure).
\par
 In conclusion we find that, irrespective of the sign of $\omega^+_a$, 
 an asymmetric jet profile can support both
 the  RWI and the Z-mode instability provided the jet's amplitude is sufficiently strong, in other
 words, jets are unstable and likely undergo nonlinear destructive transformation
 if $|\omega^\pm_a|$ is large enough.  We examine the basis for this
 expectation in the following section.

\section{Nonlinear manifestation: jets begetting jets.}\label{jets_begetting_jets}
 \begin{figure*}
\begin{center}
\leavevmode
\includegraphics[width=17.5cm]{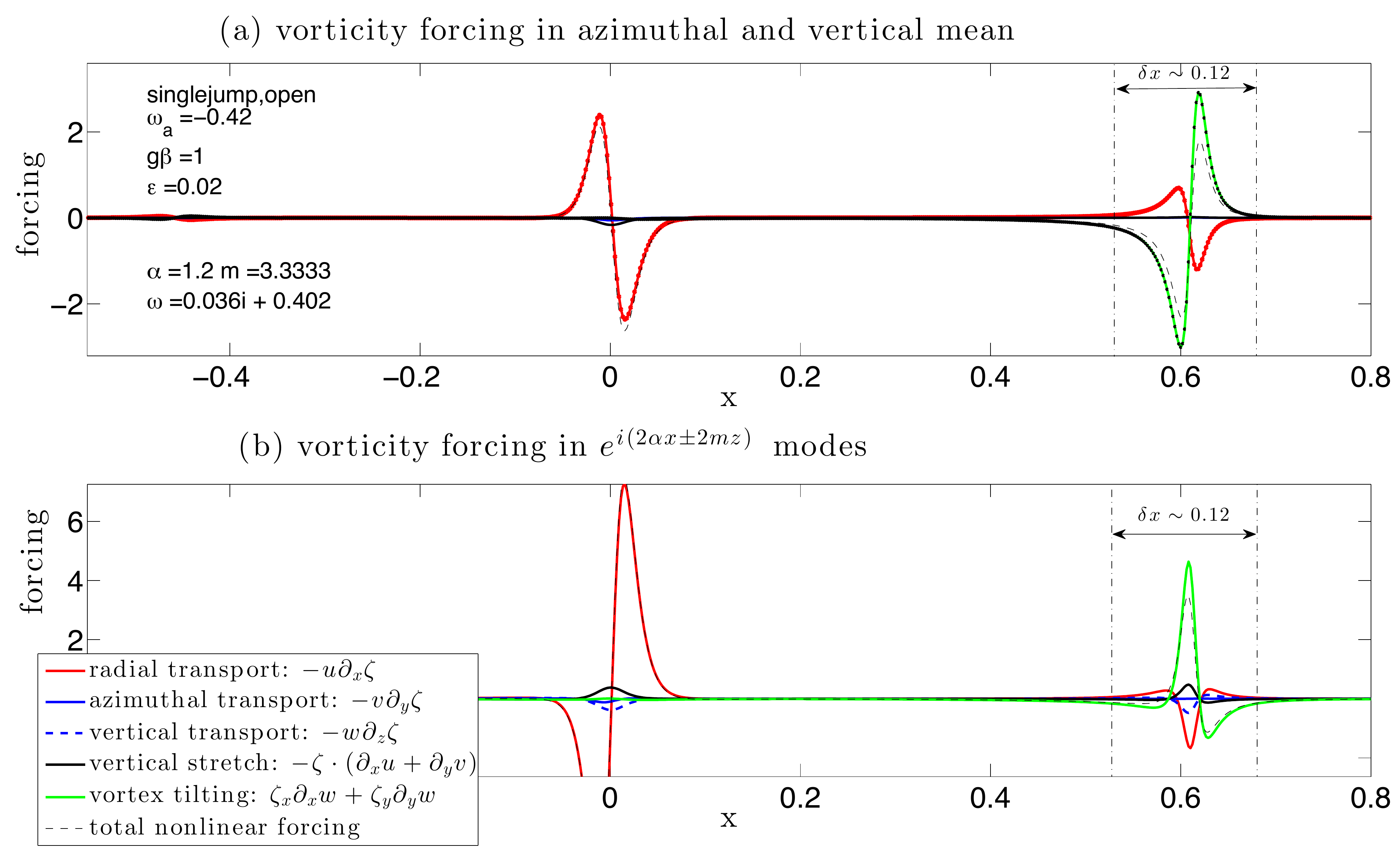}
\end{center}
\caption{Nonlinear vorticity forcing based on the
solutions shown in Figure \ref{General_profiles_1j}.  As
detailed in the text (Section 4), individual and aggregate vorticity
forcing shown for (a) mean-forcing and, (b)2k-forcing.  The dashed black line
shows the aggregate of the six individual forcing terms shown in the boxed
legend of panel (b).  { 	Generated solution values 
are indicated by points superposed on curves
in panel (a) only. } 
The radial transport forcing is strongest at $x=0$, where the location
of the mean vorticity gradient is greatest, but it also has a significant contribution
at the critical layer $x\sub{bg}^+ \approx 0.62$.  
However, the vortex tilting
terms (green) outcompete the radial transport contribution at $x=x\sub{bg}^+ $.
The mean-forcing 
profile indicates that a jet-like feature should develop at $x=x\sub{bg}^+$, while
the strong $2k$ radial transport forcing at $x=0$ suggests that the
vortex profile there should eventually get destroyed. Denoted on the figure is
 the width of the critical layer $\delta x$,
as determined in the linear analysis section.
For parameters shown on this figure and subsequent ones, rates are in units of $\Omega\sub 0$,
$g\beta$ in units of $\Omega\sub 0^2$ while lengths and
wavenumbers in $L$ and $L^{-1}$ respectively.
}
\label{nonlinear_vorticity_forcing_1j}
\end{figure*}
While this is a linear examination, it is instructive to see how
this unstable mode drives nonlinear power.  In particular, we
assess the nonlinear terms in the vorticity equation
(\ref{nonlinear_vorticity_equation}) using the 
linear solutions we have just determined.  There 
are six possible nonlinear forcing terms.  Because
the nonlinearities are quadratic, these will project
power into different wavenumber disturbances.  By example,
let us analyze the transport terms -- familiar
in atmospheric dynamics and meteorology -- and focus
on the radial component expression $-u\partial_x\zeta$.
Given the normal mode form in which
disturbances are $\sim e^{i(\alpha y + mz)}$, the above products give
power in the vertically-azimuthally uniform ``mean" state
\[
=Re\big(\hat u \partial_x \hat \zeta^*\big),
\]
where the star appearing means complex conjugate.  We refer to this
as the {\emph{mean-forcing.}}.
There is also power in the product wave numbers $2\alpha$ and $2m$,
\[
=\Big(\hat u \partial_x \hat \zeta\Big)e^{2i\omega t}e^{2i(\alpha y + mz)}
+ {\rm c.c.}
\]
and we refer to this as the {\emph{$2k$-forcing.}}
We assess these various contributions for all of the nonlinear
forcing terms appearing in equation
for which we call accordingly:
 the azimuthal component of the transport $-v\partial_y \zeta$,
 the vertical component of the transport $-w\partial_z \zeta$,
 the nonlinear vertical stretch $\zeta \partial_z w = -D \zeta$,
and the vortex tilting effects
$\zeta\sub x\partial_x w + \zeta\sub y\partial_y w$.
In the following two sections we examine the nonlinear forcing in the
vorticity-step and jet models.
\par

\subsection{Nonlinear forcing in the vortex step model}
Figure \ref{nonlinear_vorticity_forcing_1j} exhibits 
each of the six individual forcing profiles as well as the aggregate resulting forcing
in the vorticity step model examined in Section \ref{naked_Z-modes}.
As expected, there is strong power in both the mean-forcing and the first harmonic 
(hereafter called {\emph{2k-forcing}})
at $x=0$ and it is dominated by the radial transport term $u\partial_x \zeta$.  This
is understood to represent the primary
vortex step either rolling-up or undergoing a large amplitude undulation 
(e.g., Tamarin et al. 2015).  There is strong amplitude power in both
the mean-forcing and 2k-forcing 
at the critical point $x\sub{bg}^+ \approx 0.62$.
The vorticity forcing there is that of a jet (for comparison
see the vorticity profile
associated with the barotropic jet in Figure \ref{Velocity_profiles}b), which 
means that as the Z-mode instability develops, there will nonlinearly emerge
a jet like structure in the critical layer.\par
What is unexpected is that the amplitude of this vorticity forcing at
$x\sub{bg}^+$ is driven primarily by the vortex tilting terms: perturbations
in the velocity fields
give rise to perturbations in the horizontal vorticity components $\zeta\sub x',\zeta\sub y'$.
These, in turn, nonlinearly couple to the perturbation vertical velocity $w'$ -- in
the critical layer these nonlinear products act as source terms 
generating vertical vorticity.
  The radial transport term also contributes significantly in the critical layer,
but generally acts opposite to the vorticity generation driven by the vertical tilting
of horizontal vorticity.  In all instances we have calculated, the
aggregate vortex forcing is non-zero and always dominated by the vortex tilting terms, and mainly
by the vertical tilting of the radial vorticity $\zeta_x \partial_x w$.  We also note that the vortical forcing
at $x\sub{bg}^-$ is comparatively weak although non-zero.  It is remarkable that
there is strong power in both the mean and 2k forcings indicating that this
instability is quite powerful where it is expressed since many vertical wave numbers
are excited (see the discussion on growth rates and their dependence on the vertical wave number $m$,
found in the previous sections). 
Its expression, however, is constrained to within 
a narrow zone of the critical layer whose
width depends upon the growth rate itself: faster growth
rates mean wider the critical
layer zones (see further below).
% {vortex_tilting}
\par
\bigskip

 \begin{figure*}
\begin{center}
\leavevmode
\includegraphics[width=16cm]{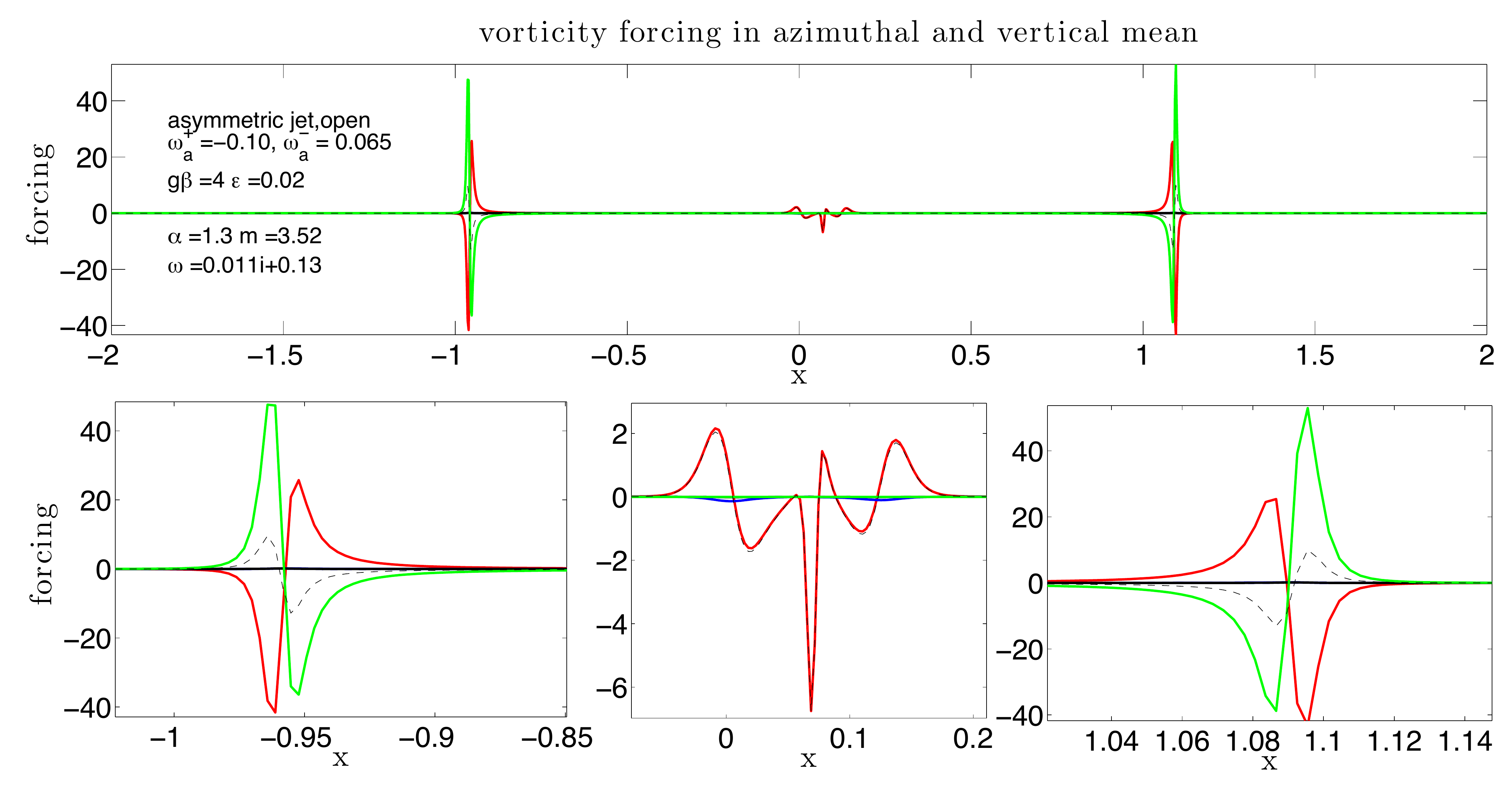}
\end{center}
\caption{
Nonlinear vorticity forcing in the mean for the asymmetric jet profile with $\alpha = 1.3$
$\Delta^+ = 0.13$, $\Delta^- = 0.20$ and $\omega_a^+ = -0.1 $ (with
 $\omega_a^- = 0.065 $) and $g\beta = 4$, the latter
 being comparable
 to conditions examined in M13. 
 The model flow profile has an effective Ro $\sim 0.05$ 
 The vertical wavenumber $m = 3.52$ corresponding
to the fastest growth rate (Im$(\omega) \approx 0.011$) 
of the dispersion curve associated with the diamonds shown
in Figure \ref{Growth rates_profiles_asymmetric_jump}.  Top panel
shows the distributed forcing power over space and the lower row of panels shows
closeups of the active regions.
  Note the relative weakness of the nonlinear
power near the location of the forcing jet for this relatively weak value of $\omega_a^+$.
The instability for this parameter regime is dominated by the Z-mode instability
and the the nonlinear power is concentrated mostly in the critical layer zones ($x\sub{bg}^- \approx -0.96$
and $x\sub{bg}^+ \approx 1.09$) where
vortex tilting and radial transport are the dominant forms of vortex forcing.  1750 grid points were used in generating
these solutions.
All quoted quantities scaled according to convention described
in Figure \ref{nonlinear_vorticity_forcing_1j}.
}
\label{Vorticity_Forcing_lowvamp}
\end{figure*}

 \begin{figure*}
\begin{center}
\leavevmode
\includegraphics[width=16cm]{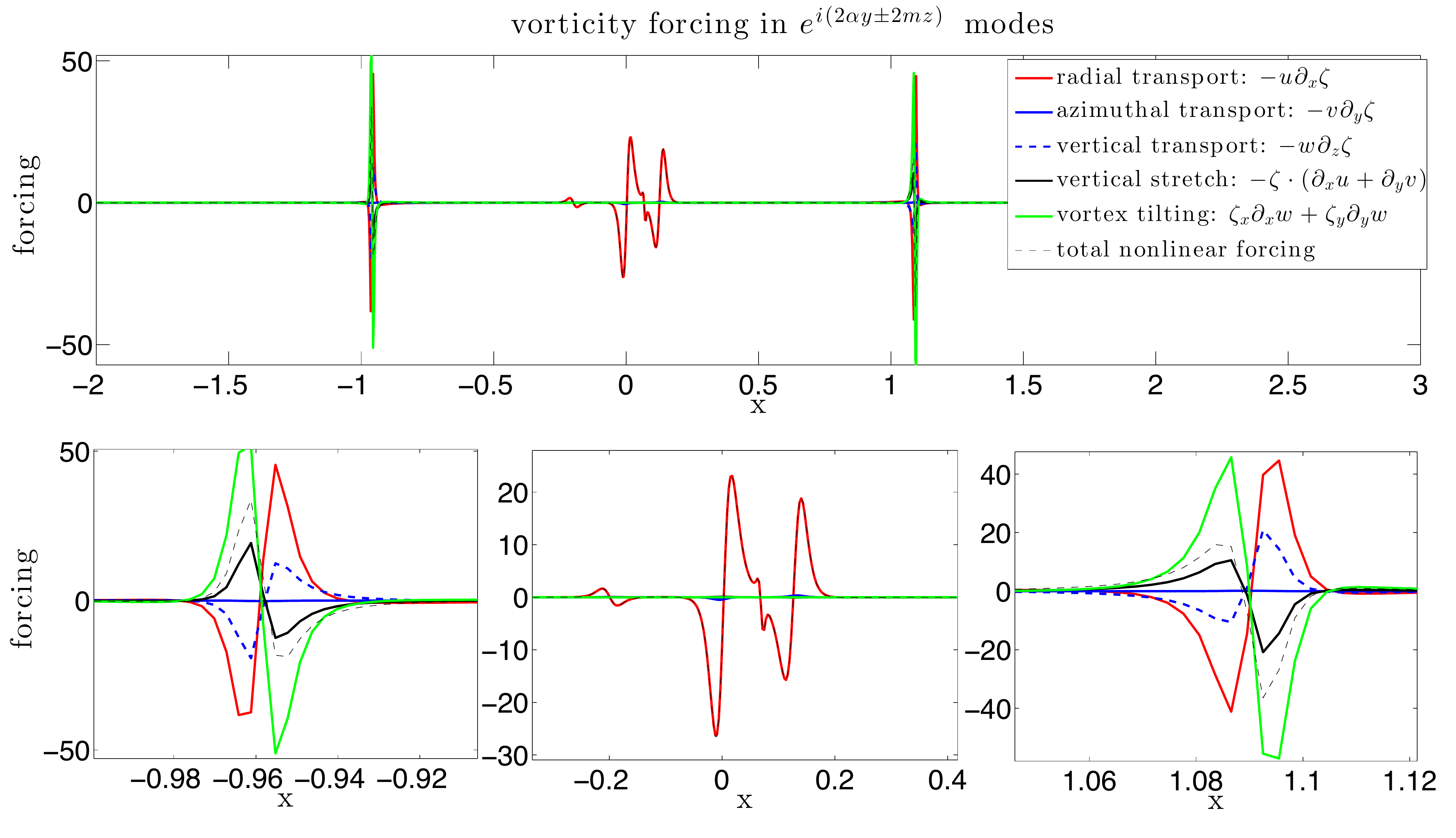}
\end{center}
\caption{Like Figure \ref{Vorticity_Forcing_lowvamp} except the forcing
power in the 2k modes shown.
}
\label{Vorticity_Forcing_lowvamp2}
\end{figure*}

\subsection{Nonlinear forcing in the jet model}

The nonlinear expression of the instability in the jet model field 
(Sections \ref{RWI_in_asymmetric_jet}--\ref{zmodes_asymmetric_jet})
also underscores the implications we have inferred in the previous section.
 For conditions in which 
$\omega^+_{a{{\rm c1}}} < \omega_a^+ < 0$ the nonlinear forcing is
strongest at the model jet's associated critical layers while it is
relatively weak at the location of the jet itself 
as indicated in both Figures \ref{Vorticity_Forcing_lowvamp}--\ref{Vorticity_Forcing_lowvamp2}.
These figures also emphasize how strongly localized the vortical forcing is to the buoyant critical
layer zones. 
As in the vortex step case examined in Section 6.1,
these figures also clearly exhibit how (i) the emergent
vertical vorticity profile is yet another jet and (ii) the two {\emph{main}} competitors
in the nonlinear vertical vorticity driving is between
 the vertical tilting of the radial perturbation vorticity and 
 radial advection of the perturbation vertical vorticity with the former
 outlasting the latter.
 \par
\begin{figure*}
\begin{center}
\leavevmode
\includegraphics[width=15cm]{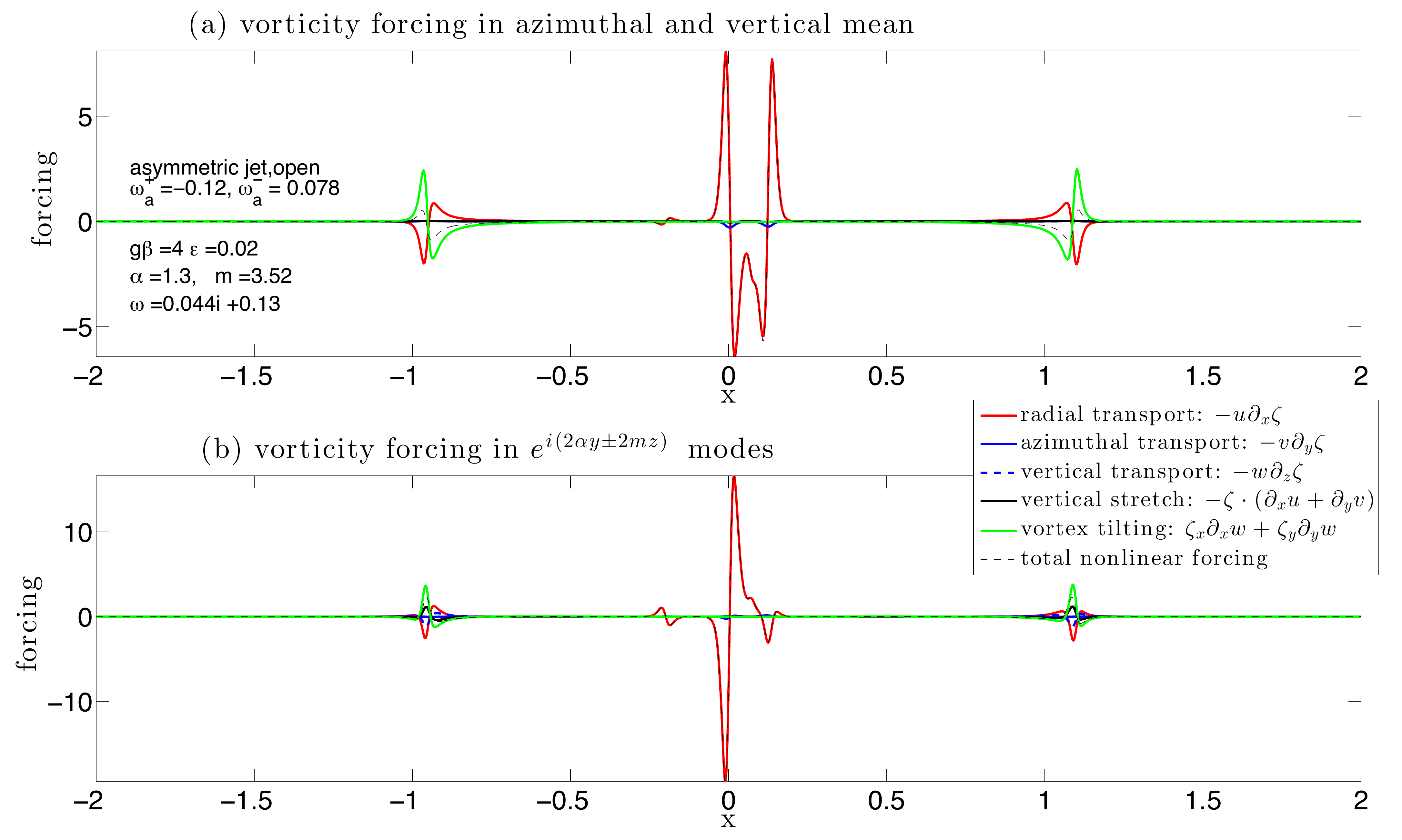}
\end{center}
\caption{Like Figure \ref{Vorticity_Forcing_lowvamp} 
except $\omega\sub a^+ = -0.12$ (and $\omega\sub a^- = 0.078$), which
means this is a model flow profile with Ro $\sim 0.06$: (a)
top panel forcing in the mean and (b) forcing in 2k.  The nonlinear properties
depicted are those for the fastest growing mode shown in Figure \ref{Growth rates_profiles_asymmetric_jump}
(solid line).
The system supports both Rossby and Z-mode instabilities in which the former
is more strongly expressed than the latter.  
The nonlinear forcing, mostly due to the radial advection of the mean vorticity gradient,
is strongest now near the location of the
original jet itself and is relatively weak at the buoyant
critical layers (near $x=-1$ and $x=1.1$).  This strong power, especially in the 2k forcing, 
is indicative of nonlinear roll-up of shear layers.
Scrutiny of the forcing shows that it is focused in the region $0<x<0.2$ which,
given that $\omega\sub a^+ < 0$, corresponds to the anticylonic side of the model
jet profile. All quoted quantities scaled according to convention described
in Figure \ref{nonlinear_vorticity_forcing_1j}.
}
\label{asymmetric_jet_medvamp2_forcing}
\end{figure*}
\begin{figure*}
\begin{center}
\leavevmode
\includegraphics[width=16cm]{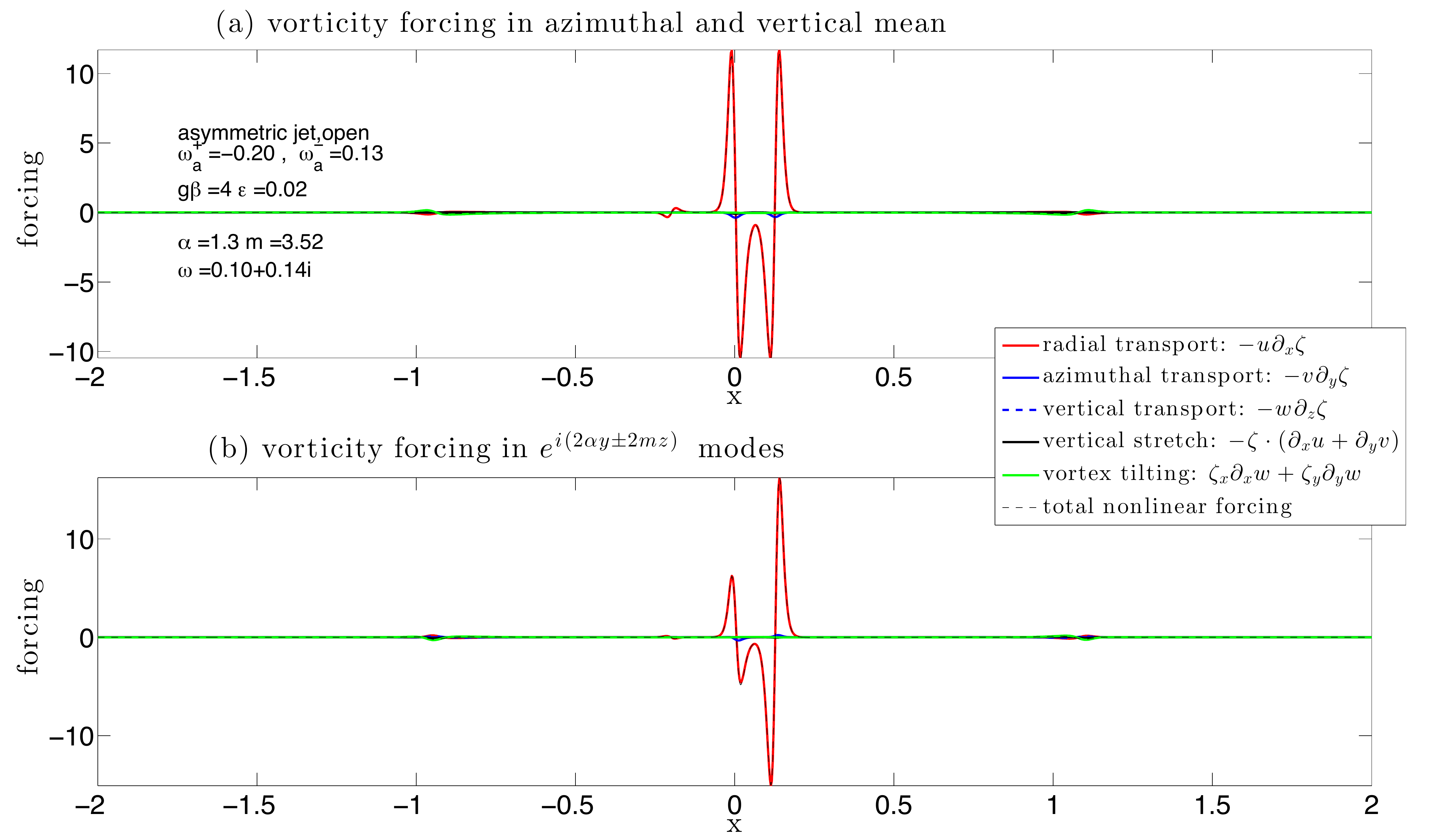}
\end{center}
\caption{Like Figure \ref{Vorticity_Forcing_lowvamp} 
except $\omega\sub a^+ = -0.2$ (and $\omega\sub a^- = 0.13$), which
that this is a model flow profile Ro $\sim 0.1$: (a)
top panel forcing in the mean and (b) forcing in 2k.
The instability, Im$(\omega) \approx 0.1$, 
is almost totally dominated by the Rossby wave instability
with very weak power at the location of the buoyant critical layers.
Similar to the previous figure, most of the nonlinear forcing is concentrated on
the anticyclonic side of the model jet profile.
All quoted quantities scaled according to convention described
in Figure \ref{nonlinear_vorticity_forcing_1j}.
}
\label{asymmetric_jet_highvamp_forcing}
\end{figure*}

 However, when the the magnitude of the model jet increases so that
 we are now in a parameter regime with $\omega_a^+ \le \omega^+_{a{{\rm c1}}}$, then the character begins to change.
 Near the marginal condition for the onset of the RWI ($\omega_a^+ \approx \omega^+_{a{{\rm c1}}}$)
 Figure \ref{asymmetric_jet_medvamp2_forcing} shows how nonlinear vertical vorticity forcing
 at the location of the jet is slightly larger in magnitude than the corresponding forcing
 at the buoyant critical layer zones.  We also note that the nonlinear vortical forcing upon the
 model jet is mainly in the region $0<x<0.2$. Given the value of $\omega_a^+ \approx -0.12\Omega\sub 0$
 it means
 that {\emph{the nonlinear driving is focused mainly on the anti-cyclonic side of the asymmetric jet}} suggesting
 that this will ultimately result in the nonlinear roll-up of that part of the jet - something
 that is seen throughout the numerical experiment reported in M13.
Eventually, when the value of $\omega_a^+ $ is strongly in the RWI instability regime
(i.e. $\omega_a^+ < \omega^+_{a{{\rm c1}}}$)
the nonlinear forcing is almost entirely focused upon the anti-cyclonic side of the model
jet with relatively little corresponding forcing power at the critical layer zones
or the cyclonic side of the jet (Figure \ref{asymmetric_jet_highvamp_forcing}).
\par
Although not shown here, the same qualitative pattern holds if $\omega_a^+$ is sufficiently positive.
For example we have examined the results using the same model 
jet dimensions assumed for the results shown in Figures \ref{Vorticity_Forcing_lowvamp}--\ref{asymmetric_jet_highvamp_forcing}
including the value of $\alpha\Delta^+ \approx 0.16$, $\delta \approx 1.54$ and $g\beta = 4\Omega\sub 0^2$ where, instead, we
examine the response for positive
values of $\omega_a^+$.  Inspection of
 Figure \ref{Barotropic_Growth_rates_profiles_asymmetric_jump}
 shows that the corresponding value of the onset of the RWI for the
 jet is given by $\omega^+_{a{{\rm c2}}} \approx 0.22\Omega\sub 0$.  The same pattern of results
 follows - for $0 < \omega_a^+ < \omega^+_{a{{\rm c2}}}$ only the Z-mode instability is expressed
 and the nonlinear vortical forcing is dominant in the critical layer zones.
 Similarly, as $\omega_a^+  > \omega^+_{a{{\rm c2}}}$ the RWI becomes more important in which
 the nonlinear vortical forcing is dominant in the vicinity of the model jet.
 In particular, the forcing is concentrated in the region $-0.2 < x < 0$ 
 but this also happens to be the anti-cyclonic side of the model jet since $\omega_a^+ > 0$  means
 $\omega_a^- < 0$ according to equation (\ref{restricted_omega_minus}).
 \par

\subsection{Jets begetting jets?}
 M13 consider the fluid response of a stably stratified Keplerian shearing box 
 to an azimuthally aligned tube of surplus vertical vorticity placed in 
 the origin of their experiment.  They find that after some time, the fluid domain
 steadily transitions into similarly confined narrow/tubular regions of deviation jet-like profiles.
 For a surplus vorticity field placed at the origin containing the midplane, the response flow
 generates jet like vertical vorticity above and below the mid plane at the location of buoyant critical points.
 A close inspection of both Figures 1 and 2 of that work shows that the generated jet-like vertical vorticity fields
(the original children)  have
 some vertical extent.  As the simulation marches forward in time, these original child vortices 
 themselves become parents to a new generation of
 child vortices at new critical layers associated with their parents, and so on the process
 continues.\footnote{We observe that in the numerical experiment results presented in 
 M13, the original tube of surplus vertical vorticity located at the origin remains constituted
 as such throughout the duration of the domain and does not get destroyed.  This is probably
 because the the width and girth of this line charge vorticity is one which does not
 go unstable via the RWI (or its equivalent analog for such a model flow).  Further analysis needs to be done to confirm this conjecture.}  
 The results and intuition derived from the linear study
 considered here largely confirms this self-replicating scenario
 proposed in M13. \par
 The insights garnered of the analysis here helps to fill in more of the details of this
 emergent (possibly) turbulent self-sustaining process.  A vorticity field supporting a localized
 Rossby wave will generate a new vorticity field at a far-field buoyant critical layer.
 This buoyant critical layer $x_{bg}$ is found to be the location
 where the Doppler shifted frequency of the Rossby wave is resonant with the local
 Brunt-V\"ais\"al\"a frequency.  The resonance induces a vortex generation mechanism
 which is dominated by
 two mostly opposing processes: (i) the local creation of vertical vorticity 
 in which the local  perturbation radial vorticity
 (i.e. the $\zeta_x$ found in the region around $x=x_{bg}$)
 is converted into vertical vorticity through the vertical tilting process: 
 $\zeta_x \partial_x w$, and (ii) the local (again near $x=x_{bg}$) radial advection of the 
 perturbation vertical vorticity: $u'\partial_x \zeta$.  Both effects act nearly opposite to
 one another, but the vertical tilting mechanism appears to dominate in the jet profiles
 we have tested here in this study.
 \par
 Among other things, this results in an asymmetric jet-like structure which
 grows in amplitude until it either saturates or the jet itself gets destroyed
 because of roll-up due to the RWI.   While the spawned jet grows it can, on its own
 accord, induce new jet-like structures at new critical points associated with the spawned jet
 and the linear analysis done here largely supports the picture sketched in M13 of {\emph{jets begetting jets}}.
 As a ``parent" structure, a jet-like profile can give rise to new ``child" jets only 
 while the parent jet maintains its structural integrity.  This can be inferred from the linear stability
 of the asymmetric jet examined in Section 6.3: as the amplitude of the jet $|\omega_a^\pm|$ is increased
 from zero
 first only Z-modes are unstable.  However, as $|\omega_a^\pm|$ increases eventually 
 the most unstable mode is the RWI which induces transformative/destructive roll up of the
 parent jet.  
 It is only during the low amplitude range of jet amplitudes, prior to the RWI becoming
 important, will a given jet generate into existence a new jet(s) at its
 associated buoyant critical layer(s).
 \par
 In this way, the parent jet must stay as such long enough for child jets to grow into maturity
 so they too can generate the next generation of jets, and so on perpetuating the
  self-replication process.  Critical to this is the time spent
 growing each jet - as it must grow slow enough so that the Z-mode instability (which is relatively
 slow by our current account) may manifest itself and start
 spawning the next generation of jets.  If a jet grows too fast then the Z-mode process has no time
 to birth the new generation and the whole process shuts off.  If the jet grows quite slowly
 it could, in a general sense, give rise to a pattern state like that reported in M13.
 Conditions lying somewhere between these two extremes might give rise to a turbulent flow state that could be either
 decaying or self-sustaining
 - perhaps something in the flavor of a ``chaotic propeller" 
  identified as being operative in subcritical transitions in both plane-Couette 
  and rotating-plane-Couette flows
  (e.g., see discussion in Rincon et al. 2007). 
 \par
   Whether or not this process can be self-sustaining and lead to a turbulent state
   under a wide umbrella of conditions appropriate to Dead Zones
 remains to be determined.  It would seem, however given our considerations,
 that its self-sustainability depends centrally on what way the original disturbances are structured
 since it requires there to be some relatively large amplitude disturbance to start the whole thing off.
 A recent announcement by Marcus et al. (2015) on this process in a shearing sheet disk model with vertically varying gravity and 
 stratification seems to indicate that it can give rise to a strongly turbulent state.

\subsection{Spatially localized instability}
In Section \ref{shear_layer} we examined the properties of the shear layer system and we
identified that there is a critical value of the \BV frequency $N\sub{{\rm cr}}$ for which
the location of the critical layer goes from being found
outside the shear layer ($g\beta > N\sub{{\rm cr}}^2$) 
to appearing inside the shear layer ($0<g\beta < N\sub{{\rm cr}}^2$).  The discussion
in the previous section concerning
 the self-reproductive spatial spreading of jet creation/destruction
 is the expected outcome of the former of the two conditions.  What might happen
 when the critical layer appears inside the shear layer?  
 \par
\begin{figure*}
\begin{center}
\leavevmode
\includegraphics[width=16cm]{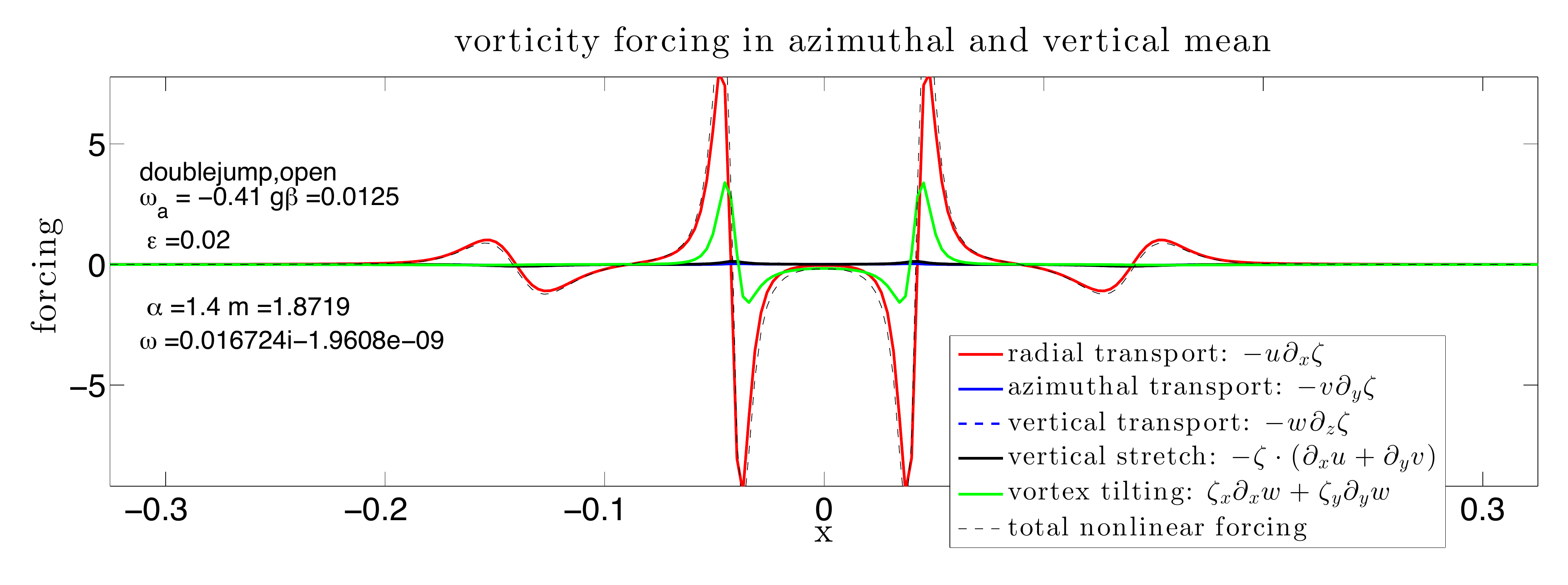}
\end{center}
\caption{The nonlinear vorticity driving for the linear solutions
shown in Figure \ref{omega_Z_Shear_Layer_solutions}b.  In this example
the critical layer exists inside the shear layer near $x=\pm 0.45$.  The edges
of the jet are at $x=\pm 0.14$.  All shears and growth rates in units of $\Omega\sub 0$
while $g\beta$ in units of $\Omega\sub 0^2$.  
}
\label{confined_transition}
\end{figure*}

 Consider the nonlinear
 vorticity forcing arising from the conditions
 shown in Figure \ref{omega_Z_Shear_Layer_solutions}b in which the critical layer
 is clearly inside the shear layer and the shear layer itself is stable against
 the destructive RWI.  We show in Figure \ref{confined_transition} the corresponding 
 mean vertical vorticity forcing profile
 arising under these parameter conditions
 $g\beta < N_{{\rm cr}}^2$.  The imprint of the primary shear layer is  
 evident in the vicinity of the locations $x\approx \pm 0.14$ while
 the nonlinear forcing at the critical layer is also 
 clearly visible near $x\approx \pm 0.045$.  
 The resulting aggregate forcing profile qualitatively resembles forcing profiles
 we have discussed thus far
 for when the critical layers appear outside.  We observe that unlike the examples of the
 vorticity-step and jet profiles, the critical layer jets are primarily
 driven by the radial advection of the perturbation vertical vorticity
 ($-u\partial_x \zeta$)
  while
 the vertical tilting of radial perturbation vorticity
 ($\zeta_x \partial_x w$) is relatively weak by comparison but acts
 with the same sense as the radial advection of the perturbation vertical vorticity.
 \par The implications are interesting: These results suggest that
 under those conditions in which the \BV frequency is small,
 that jets will grow inside the shear layer and, as per our
earlier insights, once the jet amplitude grows large enough it will self-destruct while
generating new jets associated with its critical layers.  If the whole process remains
confined to inside the shear layer, then there are at least two possible outcomes:  In the first
of these,
the parent shear layer structure remains stable 
(also as suggested by linear analysis) while the jet inside grows and once
having reached a sufficient amplitude it will self-destruct having spawned child jets
along the way (presumably contained inside the shear layer as well).  The second possibility
is that while the jet inside grows, the parent shear layer nonlinearly destructs as well.
If an external agent is responsible for the creation and maintenence of the parent shear layer in the first
place -- e.g, by either direct thermally driven relaxation (Les \& Lin, 2015, Lobo Gomes et al., 2015)
or by the VSI \citep{Richard_etal_2016} -- 
then either of the two scenarios envisioned could give rise to a process of 
unsteady jet creation and destruction entirely contained inside the shear layer, where the only
difference in the outcomes of the two possibilities is the degree of the unsteady activity -- possibly
turbulent.  Global numerical calculations of the several processes mentioned (thermally driven
relaxation or the VSI), when sufficiently resolved,
can test whether or not this dynamical scenario indeed manifests itself.

\section{Summary discussion}
{
{  This study independently confirms
and complements the ZVI process and effect uncovered and reported in M13. 
We assert that the Z-mode instability, i.e., the Doppler-shifted
frequency resonance between a distant Rossby wave
and the local \BV frequency, }
is a bonafide instability of a sheared
system not supporting the more familiar centrifugal instability studied
in the context of the Taylor-Couette system.
\par
We have adopted
an alternative mathematical framework to analyze the linearized problem
of perturbations in a strongly sheared and stratified atmosphere.  The relatively sophisticated
approach, which involves converting the underlying linearized differential system instead into a set of coupled
integral equations, is here undertaken because
of the complexities inherent to discerning eigenmode structure 
of PDEs supporting one or more irregular singular points.  In this problem,
those singular points are 
the buoyant critical layers responsible for the instability.
\par
The single-step vorticity jump examined at length in Section
\ref{naked_Z-modes} exhibits the Z-mode instability stripped down
to its bare minimum ingredients, that is, a stably stratified atmosphere
with a shear profile $v_0(x)$ containing (at least) one location with a strong
radial gradient which, in turn, supports a radially localized Rossby wave which is the
basic carrier mode of the Z-mode instability.  Supposing
this Rossby wave is vertically uniform and has a wavelength $\lambda_\alpha = 2\pi/\alpha$ 
with corresponding propagation
speed $c\sub{rw}$,  
on its own accord it will not become unstable. However,
because the atmosphere is stratified with a squared \BV frequency $N^2\sub B \equiv g\beta$, 
sufficiently far from the center of the Rossby wave there exist radial
locations $x\sub{bg}^\pm$ in which the location specific value of the background shear flow plus
the local {\emph{vertically uniform}} inertia-gravity wave propagation speed
$\pm N\sub B/\alpha$ is equal to the Rossby propagation wave speed, 
\beq
c\sub{rw} = v_0\big(x\sub{bg}^\pm\big) \pm \frac{N\sub B}{\alpha},
\eeq
that is to say, the location of the buoyant critical layer.
Said in another way, {\emph{there is a frequency resonance}} in which the locally observed Doppler shifted Rossby wave frequency
equals the absolute value of the local \BV frequency.
This result directly confirms the stated requirements for instability as reported in M13.  
%As
%such, we see this effect as an interaction of a single Rossby wave and a buoyant
%critical layer (see further discussion in Section \ref{label_sri}).  
%The critical layer
%condition says that  between the Doppler-shifted faraway Rossby wave
%frequency
%and the local Brunt-V\"ais\"al\"a frequency (i.e. the frequency of the vertically uniform internal gravity wave associated
%with azimuthal wave vector $\alpha$). In the setting examined here,
%${N\sub B}$ is a constant which means that this internal gravity wave frequency is the same everywhere
%in the domain.  For real disks this is not the case since the vertical component of
%gravity and (where appropriate) the buoyancy gradient are both functions of the vertical
%coordinate $z$.
\par
The growth rates of Z-modes
are shown in all three flow models to be maximal at some non-zero vertical wavenumber which depends upon the amplitude and
type of model shear flow under consideration.  
%We confirm M13's findings that in the shear layer model
Although the shear layer model considered here and the setup considered
by M13 are formally different, we consider them to be similar
enough to compare our results with theirs.  
Using parameter
values that are qualitatively most similar to their setup, we confirm
that the Z-mode instability gets expressed around values of ${\tt Ro} \approx 0.20$.
\par
The jet flow models we have considered show that the
unstable Z-modes exist for values of 
${\tt Ro} \approx 0.05$
thereby pushing down the critical values
of ${\tt Ro}$ needed to
express the instability by a factor of four of that surmised in M13. 
These trends and results emphasises the need
for a general
assessment the Z-mode instability including necessary conditions.
%Among other things, it means examining conditions for the onset of the Z-mode
%instability as a function of azimuthal wavenumber and model shear flow amplitudes
%$\omega\sub a$ (or, as in the case of the assymetric jets, $\omega\sub a^\pm$).
\par
The vertical velocity fluctuations are highly localized
in narrow regions containing the critical layer as, e.g., Figure \ref{Velocity_profiles_1j}(c)
plainly depicts.  The localization scales linearly with the growth rate of the Z-mode 
- a lower growth
rates means a narrower critical layer zone.  
This trend is probably why many previous
numerical simulations (prior to M13),
conducted both in shearing sheet and cylindrical geometries,
have been unable to 
see any evidence of this effect. \par
We have examined with some detail the conditions
in which unstable Z-modes exist both independently of and concurrently with
the RWI in the shear-layer model.  Using conditions similar
to that examined in M13 (namely, fixed horizontal box size and shear layer width) 
we demonstrate for shear layers
whose amplitudes $\omega\sub a$ lie between $\omega\sub Z > \omega\sub a > \omega\sub R$
that Z-modes exist in the absence of the RWI.  The condition $\omega\sub a = \omega\sub R$
is the condition for the onset of the RWI and for values of 
$\omega\sub a < \omega\sub R$ both instabilities are present.  
 $\omega\sub Z$, 
which is the critical value of the onset of Z-modes, 
is a function of the \BV frequency $g\beta$ and the former's
dependence upon the latter
was numerically determined for the special
case considered and summarized in Figure \ref{omega_Z_Shear_Layer}.
It is unexpected that the instability persists even in the limit
where the stratification vanishes ($g\beta \rightarrow 0$) suggesting that plain
inertial modes in the shearing sheet should also be unstable 
under suitable conditions in the presence of any number
of model flows like those considered here.  This remains to be further examined.
\par
The qualitative trends
regarding the existence of Z-modes and their concurrence with the RWI
 reported above for the shear layer carry over to the jet flow model
as well.  However, the same relatively
detailed analysis done for the shear layer model
remains to be done for the jet flow model.\par
%Of course there are subtleties relating to exactly constitutes a Z-mode versus an R-mode
%under various flows.  This is especially relevant to these modes in the shear layer model.
%We have arbitrarily defined a R-mode as being the unstable mode that survives into the
%2D limit where the vertical wavenumber goes to zero, i.e., $m\rightarrow 0$ (with $\epsilon$ fixed).
%Z-modes are defined as the unstable mode first emerging at some non-zero value of $m$ as $\omega\sub a$
%dips below $\omega\sub Z$ (for given $\alpha$ and $g\beta$).  The character of the associated mode
%appears to resemble the an R-mode as $\omega\sub a$ approaches the critical value
%for the RWI transition $\omega\sub R^{-}$ (see Sections 7.2.1 -- 7.2.2).  Indeed, the nonlinear
%driving rate by the Z-mode at the location of jet's edges becomes stronger than the nonlinear jet
%production rate occurring in the associated buoyant critical layer as $\omega\sub a \rightarrow 
%\omega\sub R^-$ (see Figures .  The 

%as the unambiguous demonstration (i.e., free of numerical artifacts) of the VSI requires
%very high resolution numerical experiments to be performed
%(Nelson et al., 2013, Stoll \& Kley, 2015).
%Similar considerations apply for the linear stability analysis
%done herewith and in future examinations forthwith: unless one undertakes a high resolution calculation, this
%effect and other quasi-global ones like it, will go undiscovered.
\par
\medskip
Given the calculated perturbation responses including the derived
velocity and vorticity fields,
one can assess the total nonlinear vertical vortex forcing arising from these modes
according to equation (\ref{vortex_tilting}).  What is remarkable is that 
at the buoyant critical layer $x\sub{bg}$ there are two main competitive effects driving 
the nonlinear response, namely,
that of the radial transport of the perturbation vertical vorticity, $u'\partial_x\zeta'$,
and the vertical tilting  
of the perturbation {radial vorticity}, $\zeta_x'\partial_x w'$.
Figure \ref{nonlinear_vorticity_forcing_1j}, which
serves as a representative depiction of these varied forcings,
shows the vertical tilting of the radial perturbation vorticity generally
winning out over the radial advection of perturbation
vertical vorticity and, ultimately, giving rise to a vertically/azimuthally uniform
jet flow as well as power in higher harmonics in 
the azimuthal and vertical directions.  The emerging jet flow is strongly localized due to the
size of the original critical layer.  This resultant nonlinear forcing
at the buoyant critical layers is the same for all three model flows examined
here.  
\par
We note that the only qualitative difference in the outcome of the instability
in the critical layer is in the sign in the change of vorticity
across the central ridge line of the emerging jet, which
depends upon whether or not $u'\partial_x\zeta'$ 
dominates $\zeta_x'\partial_x w'$ inside the critical layer during the growth
of the instability.  This is because these two forcing terms appear to act
opposite to one another - at least this is so in all of the profiles we have
examined while preparing this study.
\par
We have also shown here
that both jets and shear layers are subject to the RWI if their vorticity
amplitudes get large enough.  We have considered this in detail for
the jet model flow where we find that as the amplitude of the jet approaches
the critical value needed for
the onset of the RWI, the nonlinear forcing upon itself increases
as well (sequenced in Figures  
\ref{Vorticity_Forcing_lowvamp}--\ref{asymmetric_jet_highvamp_forcing}).
Moreover, this nonlinear self interaction, which is separate from
the nonlinear forcing occurring at the buoyant critical layers, 
is focused mainly on the anti-cyclonic side
\footnote{Recall that parts of the shear flow are
termed anti-cyclonic with respect to the background Keplerian shear flow.  See 
discussion formal definition of this found in Section 4.}
 of the jet which
should cause it to undergo a roll-up type of instability.
\par
%\par
Indeed, the study of \citet{Talia2015}
 explicitly examines how the $2k$-forcing\footnote{as defined in Section 8.}
 of an unstable Rayleigh shear layer is most
responsible for the destructive transformation of the shear layer in the nonlinear regime.  
%We similarly infer that for these large values of  $|\omega_a^\pm|$ the jet
%itself gets destroyed.   We also observed that for these sufficiently large values of
%the $|\omega_a^\pm|$ the nonlinear vortical power is concentrated upon the anti-cyclonic
%side of the model jet
%
% which itself indicates that the instability ought to induce roll-up of only that side
%of the model profile.
% Figure \ref{2D_jet_rollup} depicts this
%for a model two-dimensional simulation of an unstable jet 
%of the kind examined in Section \ref{RWI_in_asymmetric_jet}
%(numerical method details discussed in figure caption)
%showing the nonlinear roll-up of
%a jet profile subject to the RWI.
%Only the anticyclonic part
%rolls up while the cylonic portion remains more or less intact.
%The qualitative features of the roll-up sequence of the model asymmetric jet
%shown mirrors that of the
%simulations snapshots shown in M13.  We emphasize that the
%numerical experiment of the model jet shown in Figure \ref{2D_jet_rollup} is two-dimensional and cannot
%represent the later generation of new jets at the model jet's associated buoyant critical layers.
The point we wish to make is
the claim that this part of the process discussed in M13
can be understood independently within the framework of the RWI.
A similar role of the RWI has been recently uncovered in the nonlinear development
of the VSI \citep{Richard_etal_2016}.
\par
The totality of the aforementioned results leads one into considering
the possibility, as originally suggested in M13, that
under suitable conditions the nonlinear outcome of this instability
is that of a self-replicating dynamical
mechanism of jet creation and destruction.
If the circumstances are just right, this could result in widespread turbulence
which could perhaps be sustained. 
A given parent jet flow
can bring into being, through the Z-mode instability, a first generation
of child jets at the buoyant
critical layers of the parent.  Once the amplitude of the child jets grows
large enough they, in turn, bring into existence a second generation of child
jets
at the buoyant critical layers associated with the first generation, and so
on the process continues.  However, once the vorticity amplitudes of the first
generation get large enough, the jets will experience the RWI destroying
the anticyclonic side of their flow profiles and, possibly, severely disrupting the orderly flow of
their cyclonic sides as well.  
A once quiet laminar disk can undergo a transformation into a non-steady
dynamical state as the creation/destruction process described herein replicates
itself and spreads over the entirety of the domain.
\par
Of course, the robustness of the overall scenario described 
and its relevance
to real protoplanetary disk Dead Zones remains to be assessed and should be the focus
of future study.
Such an evaluation should
center on the fact that the zombie vortex instability requires the presence of a large amplitude initial
disturbance (like the three model flows considered in this study) 
to set off the process in the first place.  How does such
a configuration come about in the first place would have to be plausibly
explained and somehow justified.   The recent numerical 
work of \citet{M15},
which examines this process for a more realistic disk model in which, effectively
speaking with respect to our analysis, the \BV frequency is
a function of the vertical coordinate,
reports a similar pattern of eruption of the laminar state
and subsequent transformation into an unsteady dynamical flow.  A
parallel linear analysis like that done here but applied in the more 
physically relevant
setting considered by Marcus et al. (2015) -- in which
the vertical gravity and stratification is coordinate dependent --
 is both challenging and awaits to be done.
 
 \subsection{``Is the ZVI an instance of the SRI?" and other open issues}
{ 
What relationship does the ZVI have to the stratorotational instability (SRI) is difficult to say
without further careful mathematical analysis.  The SRI 
\citep{Molemaker_etal_2001,Yavneh_etal_2001,2005A&A...429....1D},
is an instability shown to arise in Taylor-Couette experiments in which
co-aligned stratification and
rotation serve to destabilize azimuthally propagating waves.  
\citet{2005A&A...429....1D} argued for the relevance of the SRI for sheared Keplerian flows
in the shearing box approximation.
Indeed, the fundamental equations governing the response of disturbances are shared between
our study and that of \citet{2005A&A...429....1D}.  The differences lie in the base flows
and boundary conditions considered: \citet{2005A&A...429....1D} treat a pure Keplerian flow but with
inner and outer impenetrable radial boundaries (i.e., a rotating channel) while in this study
we have no kinematic boundaries in the radial direction other than the decay of all quantities
as $|x| \rightarrow \infty$ but, instead, we examine the linear response
to flows that deviate somewhat from the basic Keplerian flow background.  \citet{2005A&A...429....1D}
report the existence of both exponential modes (e-modes) and oscillatory modes (o-modes).  The e-modes
were identified as being the channel analog of the RWI \citep{Umurhan_2006,Umurhan_2008} but no analog of the o-modes have been identified.  
\par
The o-modes of \citet{2005A&A...429....1D}'s study appear to share the same order of magnitude growth rates as the growth rates associated with the unstable Z-modes examined in section 7.1.  It is also interesting to note that a response of a non-stratified Keplerian flow in the shearing box model
{\emph {with only a single wall boundary condition}} (with the other end being open) 
results in the propagation
of a Rossby wave along the wall \citep{Umurhan_2008,2010A&A...521A..25U}.  In other words, the presence of a wall 
qualitatively mimics the effect of having a vorticity jump at the wall location.  By similar logic, the two walls in a channel flow might be thought of as qualitatively mimicking the effect
of two oppositely signed vorticity jumps, respectively, at the wall locations
\citep{Umurhan_2008}.
\par
In this work we have
shown that all that one needs to see the critical layer instability manifest itself is some
vertical stratification and a localized azimuthally propagating Rossby wave somewhere in the physical domain.
It therefore seems certainly plausible that the presence of channel walls, which brings into existence
individual Rossby waves that propagate along the walls, can simultaneously trigger a buoyant critical
layer response within the channel.\footnote{A possibility suggested to us by an anonymous referee as well.}
Although this is certainly no proof that the o-modes of \citet{2005A&A...429....1D}'s study
and the Z-modes of our study are manifestations of the same mechanism driven by different means,
the aforementioned physical connections and similarities existing between the two systems might serve as a useful clue in resolving
this matter in a future study. 
\par
\bigskip
Other remaining questions and issues are the following:
\begin{enumerate}
\item If the criterion for the critical layer instability is simply the resonance between
the Doppler-shifted frequency of a Rossby wave and the local \BV frequency, then why
is there a minimum amplitude required in the vorticity-step (as in the single vortex jump problem)
in order for the mode to exist at all?  
\item The differences between the vertically uniform shear layer profile 
we consider in this study and that of 
the shear layer line charge considered in M13 must be kept in mind.  
Most importantly, M13 show that jet creation at the critical layer occurs
in a horizontal plane not containing the original shear layer line charge.
Why is this so?  This study
is unable to answer the reasons for this phenomenon.  However, we speculate that
this has something to do with 
the superposition of nonlinear power arising from the windowed range of vertical wavenumber
that can be unstable.  
\par
We note that in the framework of the shear-layer calculation
we examined here and when we adopt parameter values for the shear layer intensity
($\omega_a$)
 and width ($\Delta$)
that best approximates the line charge considered in M13, the vertical wavenumber corresponding
to the fastest growth we find is approximately the scale of the spacing between spawned jets
observed in the critical layers of M13's study.  
We consider this as a possible clue for further understanding. 
\item  How does this instability fare in the face of thermal cooling?  A systematic survey of
the ZVI with the explicit inclusion of, say, optically thin thermal relaxation 
remains to be done.  Since thermal cooling is generally
dampening, a thermal cooling time shorter than the growth rate timescales associated with
the ZVI determined in this study may likely indicate stabilization of the ZVI.  
Based on conservative estimates of the growth rates determined for the flows examined in this study
($\sim 0.02 \Omega\sub 0$), it would
suggest that thermal cooling times shorter than about $50\Omega\sub 0 ^{-1}$, or approximately
8 local orbit times, may stabilize the ZVI (Malygin \& Klahr 2016, personal communication).
\end{enumerate}
}

\subsection{A concluding historical reflection}
 It appears that our conceptions regarding the nature and development of 
 disk activity -- possibly even turbulent -- have returned to some ideas
 that were discussed nearly 25 years ago.  
 In an often overlooked study by Dubrulle \& Knobloch (1992)
 the authors performed a careful analysis of a non-stratified shearing box 
 and showed that all small amplitude disturbances
 are stable.  They went on to argue that a pure Keplerian
 flow is not enough to generate an instability and they went further
 to stress that some other
 additional {\emph{finite}} amplitude profile -- which possesses at least
 one inflection point -- would be necessary to instigate some kind
 of turbulent transition in a (non-magnetized) disk.  
 Two previous studies done right prior
 to that work, namely by Lerner \& Knobloch (1988) and
 Dubrulle \& Zahn (1991), explicitly demonstrated that instability
 was feasible in a plane-Couette setting
 with a flow profile similar to the shear layer profile
 examined in this work. 
 Indeed, one can argue that the RWI is an example of such an inflection point
 instability since the shearing
 box is the rotating version of classical plane-Couette flow
 and the flow profiles examined in the original study of 
 Lovelace et al. (1999) contain at least one inflection point.
  We see the Zombie vortex discovery by M13 and Marcus et al. (2015)
 -- as well as the linear analysis we have conducted here -- as taking the
 vision expounded by Dubrulle \& Knobloch (1992)
 one step further for a stratified shearing box:  
 that a finite amplitude profile without an inflection point
 can lead to a critical layer instability.  When this recipe is further
 expanded to include
 profiles that also have inflection points, the result
 can turn into a self-replicating process and if the conditions are right,
 the shearing box flow might very well transition into a turbulent state.
%%\cite{1974ARA&A..12..279Z}
%and 
%\citep{2015ApJ...808...87M}.
%
%In the recent study of \citet{2015ApJ...808...87M}
}
 
\par

%\begin{thebibliography}
%%	
%
%\bibitem{}
%Richard, S., Nelson, R. P., Umurhan, O. M., 2015,
%MNRAS (in press)
%
%\bibitem{}
%Rincon, F., Ogilvie, G. I.,  Cossu, C., 2007, A\&A, 463, 817
%
%      

%\end{thebibliography}

\appendix

\section{Two dimensional analytical solutions}\label{assymetric_jet_analysis}
For the extreme limiting case that $\epsilon = 0$ the asymmetric jet profile can
be rendered into piecewise linear profiles.  We have, four regions in which
the perturbation vorticity field $\hat\zeta = 0$:  ``Zone I" for $x < -\Delta^-$,
 ``Zone II" for $ -\Delta^-< x < 0$,
 ``Zone III" for $0< x < \Delta^+$,
 ``Zone IV" for $x > \Delta^+$.
 In each of these separate zones the normal mode stream function
 is given by
 \beqa
 \hat\psi\sub{{\rm I}} &=& A e^{\alpha (x+\Delta^-)}, \\
 \hat\psi\sub{{\rm II}} &=& A_+ e^{\alpha x} + A_- e^{-\alpha x}, \\
 \hat\psi\sub{{\rm III}} &=& B_+ e^{\alpha x} + B_- e^{-\alpha x}, \\
 \hat\psi\sub{{\rm IV}} &=& B e^{-\alpha (x-\Delta^+)}.
 \eeqa
 The velocity fields in each region may be immediately inferred from these
 above forms.  Additionally, the normal mode pressure field $\hat\Pi$ may also
 be determined from the normal mode reexpression of equation (\ref{v_prime_eqn})
 which is
 \beq
 \hat\Pi = \left(\frac{\omega}{\alpha} - v\sub 0\right)\frac{d\hat\psi}{dx}
 + \big(2\Omega\sub 0 + v\sub{0x}\big) \hat \psi
 \eeq
 Solutions in each zone must be matched to one another subject
 to the continuity of the normal (perturbation) velocities and
 pressures at each transition zone
 $x=0,\pm\Delta^\pm$ which amounts to six conditions
 \beqa
 & & \hat\psi\sub{{\rm I}}(-\Delta^-) = \hat\psi\sub{{\rm II}}(-\Delta^-),
 \qquad
  \hat\psi\sub{{\rm II}}(0) = \hat\psi\sub{{\rm III}}(0), \nonumber 
  \\
  & & \hskip 3.0cm
  \hat\psi\sub{{\rm III}}(\Delta^+) = \hat\psi\sub{{\rm IV}}(\Delta^+),
 \eeqa
 and
  \beqa
 & & \hat\Pi\sub{{\rm I}}(-\Delta^-) = \hat\Pi\sub{{\rm II}}(-\Delta^-),
 \qquad
  \hat\Pi\sub{{\rm II}}(0) = \hat\Pi\sub{{\rm III}}(0), \nonumber 
  \\
  & & \hskip 3.0cm
  \hat\Pi\sub{{\rm III}}(\Delta^+) = \hat\Pi\sub{{\rm IV}}(\Delta^+),
 \eeqa
 for the six unknowns $A,A_\pm, B, B_\pm$.  Nontrivial solutions exist
 only if the determinant of the resulting matrix system is equal to zero
 which imposes a condition on the eigenvalue $\omega$ quoted in
 text equation (\ref{eigenvalue_for_barotropic_jet}).  For the special
 restricted condition upon $\omega_a^-$ expressed in equation (\ref{restricted_omega_minus})
 we have
 \beq
  \ \ a\left(\omega_a^+,\alpha\Delta^+,\delta\right) = 
 \frac{1}{2} \alpha\Delta^+(4\omega_a^+ + 3\delta - 3),
 \eeq 
 and
 \beqa
 & & b\left(\omega_a^+,\alpha\Delta^+,\delta\right) = \nonumber \\
& & \ \  \frac{1}{4}\Big(\omega_a^+\Big)^2\bigg[
-1 + e^{-2\alpha\Delta^+} - \frac{1}{\delta^2}  + \frac{e^{-2\alpha\delta\Delta^+}}{\delta^2}
- \frac{1}{\delta} + \nonumber \\
& & \ \ \ \frac{e^{-2\alpha\delta\Delta^+}-e^{-2\alpha(1+\delta)\Delta^+}}{\delta}
+ 2\alpha\Delta^+ \frac{\delta+1}{\delta}
 + 4 \left(\alpha\Delta^+\right)^2\bigg] +  \nonumber \\
 & & \ \   \Big(\alpha\Delta^+\Big)^2\left(-\frac{9}{4}\delta -\frac{3}{2}\omega_a^+ + \frac{3}{2}\omega_a^+\delta\right),
 \eeqa
 and
 \beqa
 & & c\left(\omega_a^+,\alpha\Delta^+,\delta\right) = e^{-2\alpha(1+\delta)\Delta^+}\frac{(1+\delta)\omega_a^+ }{8\delta^2}\times
 \nonumber \\
 & & \ \ \ \left[\omega_a^+ + e^{2\alpha\Delta^+}\left(-\omega_a^+ + (\alpha\Delta^+)(-3 + 2\omega_a^+)\right)\right]\times
 \nonumber \\
 & &  \ \ \ \left[\omega_a^+ + e^{2\alpha\delta\Delta^+}\left(-\omega_a^+ + \delta(\alpha\Delta^+)(3\delta + 2\omega_a^+)\right)\right].
 \eeqa

\end{document}